\documentclass[aps,prl,twocolumn,groupedaddress]{revtex4-1}

\usepackage{graphicx, graphics}
\usepackage{amsmath, amssymb, amstext, amsfonts}
\usepackage{rotating}
\usepackage{subcaption}
\usepackage{multirow}

\newcommand{\enum}[2]{\ensuremath{#1\times10^{#2}}} 

\newcommand{\qty}[2]{\ensuremath{#1\,\mathrm{#2}}}

\newcommand{\bfunc}{$\beta$-function}
\newcommand{\bstar}{\ensuremath{\beta^*}}
\newcommand{\bstarval}[1]{$\bstar = #1\,\mbox{m}$}

\newcommand{\emittn}{\ensuremath{\epsilon_n}}

\newcommand{\sigs}{\ensuremath{\sigma_s}}
\newcommand{\sigp}{\ensuremath{\sigma_p}}
\newcommand{\kb}{\ensuremath{k_b}}

\newcommand{\Nb}{\ensuremath{N_b}}
\newcommand{\Eb}{\ensuremath{E_b}}

\newcommand{\etev}[1]{\ensuremath{\Eb=\qty{#1}{TeV}}}
\newcommand{\VRF}{\ensuremath{V_{\mathrm{RF}}}}

\newcommand{\aibs}{\ensuremath{\alpha_{\mathrm{IBS}}}}
\newcommand{\aibss}{\ensuremath{\alpha_{\mathrm{IBS},s}}}
\newcommand{\aibsx}{\ensuremath{\alpha_{\mathrm{IBS},x}}}
\newcommand{\aibsy}{\ensuremath{\alpha_{\mathrm{IBS},y}}}
\newcommand{\aibsxy}{\ensuremath{\alpha_{\mathrm{IBS},x,y}}}
\newcommand{\aradd}{\ensuremath{\alpha_{\mathrm{rad}}}}
\newcommand{\aradds}{\ensuremath{\alpha_{\mathrm{rad},s}}}
\newcommand{\araddx}{\ensuremath{\alpha_{\mathrm{rad},x}}}
\newcommand{\araddy}{\ensuremath{\alpha_{\mathrm{rad},y}}}
\newcommand{\araddxy}{\ensuremath{\alpha_{\mathrm{rad},x,y}}}

\newcommand{\A}{\ensuremath{A_\text{ion}}}
\newcommand{\Circ}{\ensuremath{C_\text{ring}}}
\newcommand{\lumi}{\ensuremath{\mathcal{L}}}

\begin{document}


\title{Potential performance for Pb-Pb, p-Pb  and p-p collisions  in a future circular collider }

\author{Michaela Schaumann}
\email{Michaela.Schaumann@cern.ch}


\affiliation{CERN, Geneva, Switzerland and RWTH Aachen University, Aachen, Germany}



\begin{abstract}
The hadron collider studied in the Future Circular Collider (FCC) project could operate with protons and lead ions in similar operation modes as the LHC. In this paper the potential performances in lead-lead, proton-lead and proton-proton collisions are investigated. Based on average lattice parameters, the strengths of intra-beam scattering and radiation damping  are evaluated and their effect on the beam and luminosity evolution is presented. Estimates for the integrated luminosity per fill and per run are given, depending on the turnaround time. Moreover, the beam-beam tune shift and bound free pair production losses in heavy-ion operation are addressed.
\end{abstract}


\maketitle


\section{Motivation}

The Future Circular Collider (FCC) is a recently proposed collider study in a new 80--100~km tunnel at CERN in the Geneva area \cite{VHELHC}. The design study includes three collider options: FCC-ee (formerly known as TLEP), a $e^+e^-$ collider with a center-of-mass energy of  90--400~GeV, seen as a potential intermediate step; FCC-hh, a hadron collider with a centre-of-mass energy of the order of \qty{100}{TeV} in proton-proton collisions as a long-term goal; and FCC-he, combining both as a hadron-electron collider.

The beam energy of the hadron machine is expected to be $E_b= \qty{50Z}{TeV}$, where $Z$ is the charge number of the circulating nuclei. Its main purpose will be to search for new physics in energy regimes which have never been reached before.
The FCC-hh will therefore spend most of its physics time  providing proton-proton collisions to its experiments. Nevertheless, operating this machine with heavy ions is being considered. It would provide, for example, Pb-Pb and p-Pb collisions at $\sqrt{s_{\rm NN}}=39$ and \qty{63}{TeV}, respectively.
From the heavy-ion physics point of view, using the FCC-hh as a heavy-ion collider would open a whole new regime of research opportunities \cite{ADainese_QM2014}.

This paper discusses potential FCC-hh beam parameters for heavy-ion operation. The dominating beam dynamic effects and estimates for the time evolution of luminosity, intensity, emittances and bunch length by analytic equations and  Collider Time Evolution (CTE) \cite{cte} simulations are presented. An approximated smooth lattice model is assumed. Lead-lead (Pb-Pb) and proton-lead (p-Pb) operation are considered. We close with a short discussion of proton-proton (p-p) operation, based on the same techniques.

\section{General Assumptions}
It is foreseen to operate the FCC-hh with different types of particles, e.g., protons (p) and lead-ions (Pb), but potentially also other ion species. The choice of certain parameters and hardware components has to  ensure the compatibility with all potential beams. As mentioned, the production of p-p collisions will be the main task, restricting the heavy-ion run time to a few weeks per year, similar to the current Large Hadron Collider (LHC) schedule. In order to optimise time and cost, the operation with different species should share mostly the same equipment and machine settings should be kept as similar as possible. For this reason, the parameters to be chosen for the heavy-ion operation are in line with those for p-p operation documented in \cite{FCC1}, where possible. 
This work focuses on the baseline option of a ring  with $C_\text{ring}=\qty{100}{km}$ circumference requiring \qty{16}{T} Nb$_3$Sn dipoles to provide a maximum beam energy of \mbox{$E_b= \qty{50Z}{TeV}$}.

\subsection{Pre-Accelerator Chain}
The study of this new hadron collider began only recently and the requirements for the pre-accelerator chain are still undefined. 
Assuming the same ratio of injection to full energy as for the LHC, the injection energy of the FCC-hh would be \mbox{$E_{b,\text{inj}} = \qty{3.3Z}{TeV}$}.

Taking the existing CERN infrastructure into account, reference~\cite{FCC1} tentatively suggests three options for the last accelerator injecting into the FCC: a machine built either in the SPS, the LHC or the FCC tunnel. The magnet strength required for an injection energy of \qty{3.3Z}{TeV} would be \qty{1}{T}, for an injector with normal conducting magnets in the \qty{100}{km} FCC tunnel. \qty{3.6}{T}, using superconducting LHC-type magnets (Nb-Ti) in the existing LHC tunnel. \qty{13.5}{T}, using Nb$_3$Sn magnets replacing the SPS. A choice has not been made, but using the existing superconducting LHC  magnets seems to be the most favoured and cost effective option today. Equipping the LHC magnets with new power converters and ramping to only about half their maximum field could reduce the ramp time to an acceptable value of a few minutes. 

Based on this, it will be assumed here that the existing pre-accelerator complex, including the LHC, is used to accelerate the particles up to \qty{3.3Z}{TeV} before injection into the new ring. Both LHC rings are filled and the beams are injected in opposite direction into the FCC. This is a reliable but conservative assumption. Major upgrades are essential in the injector chain to satisfy the requirements of the FCC experiments and to obtain a realistic filling time. The heavy-ion programme will benefit from the efforts made. It can be expected that the performance and turnaround time will be significantly improved compared to the current situation, but the amount of improvement would be speculative today.

\subsection{Smooth Lattice Approximation}
\label{s_smoothLatticeApprox}

At the time of this study, the lattice design is still preliminary \cite{BHolzer_fcclattice}. However, for the calculation of many parameters and effects, the knowledge of certain lattice properties is required. 
In the design of a new machine, one has to respect some constraints, from which at least a first approximation of the range of these quantities can be derived.

As a baseline it is assumed that the lattice would be a similar FODO design as in the LHC. The maximum (and minimum) $\beta$-function in a FODO cell is directly proportional to the cell length, $L_c$ \cite{HandbookAccelPhys}:
\begin{eqnarray}
\beta^{\pm} = \frac{L_c(1 \pm \sin\frac{\mu}{2})}{\sin\mu} \propto L_c,
\label{eq_fodobeta}
\end{eqnarray}
where $\mu$ is the phase advance per cell. 
To keep the beam size in the arcs at a reasonable value, $L_c$ should not exceed twice the LHC value of $L_{c,\text{LHC}}= \qty{106.9}{m}$. It seems adequate to investigate cell lengths between one and two times the LHC value.
A tendency to the upper range, close to $2\, L_{c,\text{LHC}}$, seems to be favoured as a compromise between magnet aperture and strength.
 
The horizontal dispersion is produced in the bending magnets and is therefore proportional to the bending angle per cell, $\theta_c$, times $L_c$. The average dispersion in a FODO cell, $\langle D_x \rangle$, is given by \cite{HandbookAccelPhys}:
\begin{eqnarray}
\langle D_x \rangle = \frac{L_c \theta_c}{4}\left( \frac{1}{\sin^2\frac{\mu}{2}}-\frac{1}{12}\right) \propto L_c \theta_c.
\label{eq_fododispersionaverage}
\end{eqnarray}
The total bending angle of the ring, the sum over $\theta_{c,i}$ of all cells, is $2\pi$: 
\begin{eqnarray*}
2\pi &=& \Sigma \theta_{c,i} = N_c \theta_c \\
\Rightarrow \theta_c &=&\frac{2\pi}{N_c},
\end{eqnarray*}
where $N_c$ is the total number of FODO cells in the ring. The length of the circumference, filled by the arcs, is: 
\begin{eqnarray*}
 L_\text{arcs}= N_c L_c = \frac{2 \pi}{\theta_c}L_c.
\end{eqnarray*}
Of this length, the dipoles themselves only occupy the fraction $F_\text{arc}$, giving:
\begin{eqnarray*}
L_\text{dipole} = 2 \pi \rho_0 = F_\text{arc} L_\text{arcs} = F_\text{arc} \frac{2 \pi}{\theta_c}L_c,
\end{eqnarray*}
with $\rho_0$ as the dipole bending radius. It follows that the average horizontal dispersion is related to the cell length as:
\begin{eqnarray}
\theta_c L_c &=&  L_c^2 \frac{F_\text{arc}}{\rho_0} \propto L_c^2 
\label{eq_fodothetapropto}
\\
 \Leftrightarrow \langle D_x \rangle &\propto& L_c^2.
 \label{eq_fododispersionpropto}
\end{eqnarray}

The vertical dispersion is in general very small and corrected for. Therefore, it is assumed to be zero:
\begin{equation*}
\langle D_y \rangle =0.
\end{equation*}

Assuming a phase advance of $\mu=\pi/2$ per cell and an arc filling factor of $F_\text{arc}=0.79$, as in the LHC, Eq.~\eqref{eq_fodobeta}, \eqref{eq_fododispersionaverage} and \eqref{eq_fodothetapropto} can be used to express the dispersion and \bfunc s in terms of the cell length $L_c$.

The momentum compaction factor, $\alpha_c$, and the relativistic gamma factor at transition energy, $\gamma_T$, can be approximated via the average horizontal dispersion:
\begin{eqnarray}
\alpha_c \equiv \frac{1}{\gamma_T^2} = \frac{1}{\Circ}\oint \frac{D_x}{\rho_0} \text{d}s \approx \frac{2\pi \langle D_x \rangle}{\Circ}.
\label{eq_fodogammataverage}
\end{eqnarray}

\subsection{Beam Parameters}
The potential beam parameter space is constrained by many different limitations, including the injector performance and dynamic effects in the whole operational cycle. The beam parameters presented in the following are an example of what could be possible from today's knowledge. Further studies should be performed to confirm their validity and to determine the optimum parameter set.

Using the existing pre-accelerator chain, it can be expected that  beam parameters at least as good as in the LHC can be achieved.
For the moment, the bunch-by-bunch differences observed in LHC operation \cite{IPAC13BbB} are neglected. Average bunch parameters measured in the 2013 proton-lead run \cite{JJowett_IPAC13, MSchaumann_phd} are taken as a conservative baseline. The assumed beam parameters for the lead and proton beams for heavy-ion operation of the FCC-hh are given in Table~\ref{t_fcc_beamparameters}.

\begin{table}
\caption[Assumed beam parameters for heavy-ion operation.]{\label{t_fcc_beamparameters} Assumed beam parameters for heavy-ion operation in Pb-Pb and p-Pb collisions.}
\begin{ruledtabular}
\begin{tabular}{rcccc}
Parameter  & Symbol & Unit& Lead & Proton \\
\colrule
No. of particles per bunch & \Nb & [$10^8$]&	1.4 & 115 \\
 Normalised transv. emittance & \emittn & [$\mu$m] & 1.5 & 3.75\\
RMS bunch length & $\sigma_s$ & [m] & 0.08 & 0.08 \\
No. of bunches per beam & \kb & - & 432 & 432 \\
\bfunc\ at IP & \bstar & [m] & 1.1 & 1.1 \\
\end{tabular} 
\end{ruledtabular}
\end{table}

For the number of bunches per beam, \kb, given in  Table~\ref{t_fcc_beamparameters}, one injection per beam from the LHC is assumed. The LHC filling is assumed to be the planned "baseline" filling scheme after LS2 \cite{IPAC14Model}.
One shot from the LHC fills only about one quarter of the total circumference of the FCC. This implies that either only one experiment, clusters of experiments or two experiments, placed at opposite positions in the ring, could be provided with collisions. The reason for this choice is related to the turnaround time of the LHC as an injector, which will be explained in the discussion of the luminosity evolution. The \bstar-values are the same as during p-p operation.

Intensity losses and emittance growth at injection, during the ramp and while preparing collisions are neglected.

\subsection{RF System and Longitudinal Parameters}
\label{s_longitudinalParameters}
An RF system similar to the one currently used in the LHC, which has a frequency of $f_\text{RF} = \qty{400.8}{MHz}$, gives an harmonic number of
\begin{eqnarray*}
h = \frac{f_\text{RF}}{f_\text{rev}} 
= 133692 ~(=2^2 \times 3 \times 13 \times 857)
\end{eqnarray*}
in a ring with a circumference of  exactly $\Circ=\qty{100}{km}$. In reality, the circumference  will be adjusted to give an $h$ with more small factors, but this is not important in the following.

\subsubsection{Injection}
When the beam is injected, assuming bunch to bucket transfer, the longitudinal beam parameters, i.e., the relative RMS momentum spread, $\sigma_p$, the RMS bunch length, $\sigma_s$, and the longitudinal emittance, $\epsilon_s$, are defined by the previous accelerator. To conserve the beam quality, the RF bucket has to be matched to the arriving beam. Assuming an injected bunch length of $\sigma_s=\qty{0.1}{m}$, the corresponding $\sigma_p$ and  $\epsilon_s$ arriving from the LHC can be calculated as
\begin{eqnarray}
\sigma_{p} &=& 2\pi \frac{f_s \sigma_s}{c |\eta|} = 1.9 \times 10^{-4},
\label{eq_momentumSpread}\\
\epsilon_s &=& 4\pi \sigma_p \sigma_s \beta_\text{rel} E_b/(Z c) = \qty{2.6}{eVs/charge},
\label{eq_longEmittance}
\end{eqnarray}
where $f_s$ is the synchrotron frequency given by
\begin{eqnarray}
f_s &=& f_\text{rev}\sqrt{\frac{|\eta| V_\text{RF} h Z e }{2\pi \beta_\text{rel} E_b}},
\label{eq_longRevFreq}
\end{eqnarray}
with $f_\text{rev}$ as the revolution frequency, $\beta_\text{rel} = v/c$ and $E_b$ as the energy of the synchronous particle. $\eta = \frac{1}{\gamma_T^2} - \frac{1}{\gamma^2}$ is the slip factor with $\gamma$ as the relativistic Lorentz factor, $Z e$ is the particles' charge. At $E_b=\qty{3.3Z}{TeV}$, an RF voltage of $V_\text{RF} = \qty{12}{MV}$ was used   in the LHC.

From Eq.~\eqref{eq_longEmittance}, it follows that $\epsilon_s$ is constant, if $\sigma_s$ and $\sigma_p$ are constant.  If $\sigma_s$ can be preserved during the transfer, Eq.~\eqref{eq_momentumSpread} and \eqref{eq_longRevFreq} show that for a given lattice the RF voltage is the only free parameter to match the momentum spread.

\begin{figure}
\includegraphics[width=0.45\textwidth]{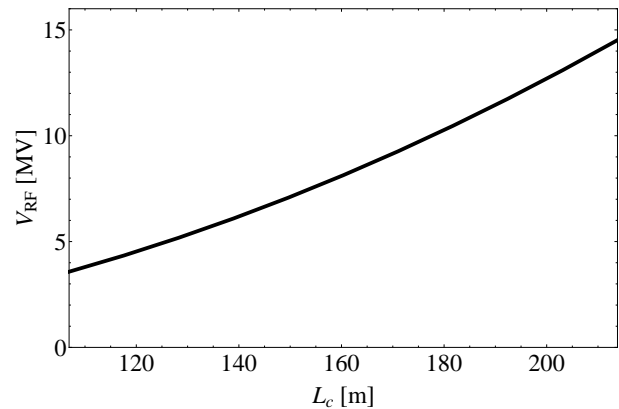}
\caption[RF voltage dependence on lattice at injection in the FCC for matched bucket condition.]{\label{f_fcc_VoltageVsCellLength} RF voltage dependence on lattice at injection in the FCC for matched bucket condition.}
\end{figure}

Because of the preliminary stage of the lattice design, the effect of a varying cell length should be investigated. $\gamma_T$ is the only parameter in Eq.~\eqref{eq_momentumSpread} depending on the lattice. From Eq.~\eqref{eq_fodogammataverage}, \eqref{eq_fododispersionaverage} and \eqref{eq_fododispersionpropto}  follows 
\begin{eqnarray}
\gamma_T &\propto& \frac{1}{L_c} \nonumber \\
\Rightarrow \sigma_p &\propto&  \gamma_T  \sqrt{V_\text{RF}}\propto \frac{\sqrt{V_\text{RF}}}{L_c}  
\label{eq_momentumSpreadPropto}
\end{eqnarray}
for $\gamma \gg \gamma_T$. To obtain a matched distribution with $\sigma_p$ equal to the injected value, $V_\text{RF}$ has to be increased proportionally to the square of the cell length as shown in Fig.~\ref{f_fcc_VoltageVsCellLength}.

We define a baseline FCC-hh lattice with a FODO cell length of $L_c\approx \qty{203}{m}$ for the calculations in the following. With this, Eq.~\eqref{eq_fododispersionaverage} and \eqref{eq_fodogammataverage} estimate $\gamma_T \approx 103$. Figure~\ref{f_fcc_VoltageVsCellLength} shows that for this baseline lattice, an RF voltage of about $V_\text{RF} =\qty{13}{MV}$ is required at injection in the FCC.

\subsubsection{Top Energy}
To counteract the adiabatic damping of the bunch length during the energy ramp, white RF noise is applied to keep $\sigma_s$ at a constant value of \qty{0.08}{m}. This value is taken from the p-p parameter list and is based on the resolution limits of the experiments,  imposing  a minimum length of the luminous region.

Using  an RF voltage of $V_{RF}=\qty{32}{MV}$, twice the LHC design value \cite{LHCDesignReport}, at top energy of the FCC-hh, the synchrotron frequency, the relative RMS momentum spread and the longitudinal emittance are
\begin{eqnarray*}
f_s &=& \qty{3.4}{Hz},\\
\sigma_p &=& 0.6\times 10^{-4},\\
\epsilon_s &=& \qty{10.1}{eVs/charge}.
\end{eqnarray*}

The bucket height, $(\Delta p/p)_\text{max}$, and area, $A_\text{bucket}$, evaluate to \cite{HandbookAccelPhys}
\begin{eqnarray*}
\left(\frac{\Delta p}{p}\right)_\text{max} &=& \sqrt{\frac{2 Z e V_\text{RF}}{\pi h |\eta| \beta_\text{rel} E_b}} = 1.8\times 10^{-4},\\
A_\text{bucket} &=& \frac{8 \Circ}{h \pi c} \sqrt{\frac{Ze V_\text{RF} E_b}{2 \pi h |\eta|}}=\qty{28.6}{eVs/charge}.
\end{eqnarray*}
At injection energy these values are $(\Delta p/p)_\text{max}=4.5\times 10^{-4}$ and $A_\text{bucket}=\qty{4.7}{eVs/charge} $. The calculation is based on the baseline lattice defined in the previous paragraph.

\begin{figure}[tb]
\includegraphics[width=0.45\textwidth]{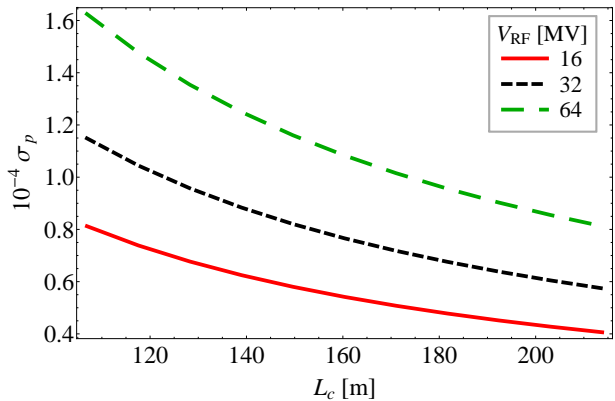}
\caption[RMS momentum spread dependence on lattice and RF voltage.]{\label{f_fcc_SimgapVsCellLength} RMS momentum spread dependence on lattice and RF voltage at top energy, as described by Eq.~\eqref{eq_momentumSpreadPropto}.}
\end{figure}

An energy spread of $0.6 \times 10^{-4}$ seems small and it has to be investigated in detail, if this would cause instabilities. 
As Eq.~\eqref{eq_momentumSpreadPropto} states and Fig.~\ref{f_fcc_SimgapVsCellLength} visualises, increasing the RF voltage could be advantageous, but the gain in $\sigma_p$ is small for $L_c$ on the order of twice the LHC cell length.
In the design stage of the machine, it could as well be an option to increase $\gamma_T$ by decreasing the cell length to obtain a higher $\sigma_p$. Nevertheless, the benefit has to be weighed against other design criteria relying on the cell length. For a chosen bunch length, the longitudinal emittance will behave proportionally to the momentum spread.

In general, it seems reasonable to aim for a similar momentum spread as in the LHC, around $\sigma_p = \enum{1.1}{-4}$. This however would require an unrealistically high RF voltage of about $V_\text{RF}\approx \qty{100}{MV}$.

\section{Lead-Lead Operation}
Based on the assumptions made above, approximations of relevant beam properties and effects are calculated in the following section. Because of the preliminary state of the accelerator design, simplifying assumptions had to be made in several places, therefore the study presented here can only give a first indication of what could be expected from  heavy-ion operation of such a machine.

\subsection{Intra-Beam Scattering}
\label{s_fcc_ibs}
Intra Beam Scattering (IBS) is a dynamic effect within a bunch of charged particles, where multiple small-angle Coulomb scattering leads to particle losses and emittance growth. This effect can become very strong and reduce the potential luminosity. 

\subsubsection{Formalism and Scaling}
Several formalisms are available describing the physical effects derived by Piwinski, Bjorken and Mitingwa, Bane, Nagaitsev or Wei \cite{APiwinski_IBS, JBjorken_IBS, KBane_IBS, SNagaitsev_IBS, JWei_IBS}, based on different assumptions and suitable for different situations. To estimate the effect in the FCC-hh, the methods of Piwinski \cite{APiwinski_IBS} and Wei \cite{JWei_IBS} are used. 

Piwinski's equations for the IBS emittance growth rates, \aibs, can be found in \cite{APiwinski_IBS1}. In his formalism the \aibs\ are proportional to 
\begin{eqnarray}
A_\text{p} &=& \frac{2 r_0^2 N_b m_0 c^2}{16 \pi \gamma \epsilon_{n,x} \epsilon_{n,y} (Z\epsilon_s)},
\label{eq_IBSPiwiA}
\end{eqnarray}
where $\epsilon_s$ is the invariant longitudinal emittance per charge given by Eq.~\eqref{eq_longEmittance}, $\epsilon_{n,xy} = \beta_\text{rel} \gamma \epsilon_{xy} $ are the transverse normalised emittances, $r_0$ the classical particle radius, which relates to the classical proton radius, $r_{p0}$, as $r_0 = Z^2/A_\text{ion} r_{p0}$, and $m_0$ is the rest mass of the particle.
This factor gives an indication of the scaling and quantities most important for the IBS strength. Equation~\eqref{eq_IBSPiwiA} scales inversely with the energy, 
meaning the IBS growth is strongly suppressed at higher energies. On the other hand, the rates increase with bunch intensity and decrease with growing emittances ($ \aibs \propto \Nb/(\epsilon_{n,x} \epsilon_{n,y} \epsilon_s)$), implying that the higher the bunch brightness, desired for luminosity production, the stronger the IBS. A third relevant proportionality is the relation to $r_0$, which depends on the particles' mass and charge ($\aibs \propto Z^2/A_\text{ion}$), hence the effect is stronger for heavy ions compared to protons. 
The remaining factors in Piwinski's equations are complicated and depend mainly on lattice parameters, like the dispersion and \bfunc s, and the beam divergences in all dimensions.

In a simplified formalism J. Wei derived analytical equations of the IBS emittance growth rates of hadron beams  \cite{JWei_IBS}, provided that the lattice of the accelerator mainly consists of regular FODO cells.
For full coupling between the horizontal and vertical motion, the growth rates average in the transverse dimension. For round beams ($\epsilon= \epsilon_x = \epsilon_y$) and if the motion is fully coupled, Wei's formulae for the IBS emittance growth rates are
\begin{eqnarray}
\aibsxy &=& \frac{C_1 \Nb}{\sigma_s \epsilon^2 \sqrt{\epsilon + C_2 \sigma_p^2}}
\label{eq_IBSgrowthRateWeixy}\\
\aibss &=& \frac{C_3 \,\epsilon}{\sigma_p^2} \aibsxy,
\label{eq_IBSgrowthRateWeis}
\end{eqnarray}
where $C_1$, $C_2$ and $C_3$ are constant during operation:
\begin{eqnarray*}
C_1 &=& \frac{5 \sqrt{2} c  Z^4 r_{p0}^2}{ 8  \A^2 \gamma^5 
\beta_\text{rel}^3} \frac{ 2 D_x^2 \gamma^2 - \beta_x( \beta_x +\beta_y)}{
 \beta_x  \sqrt{\beta_x + \beta_y}}\\
 C_2 &=& \frac{D_x^2}{\beta_x}\\
 C_3 &=& \frac{4 \gamma^2 \beta_x}{2 D_x^2 \gamma^2 - \beta_x (\beta_x +\beta_y)}.
\end{eqnarray*}
Following the smooth lattice approximation, the average of the dispersion and \bfunc s around the ring are used in the equations.
In this form the longitudinal and transverse growth rates are directly related.

\subsubsection{Calculation of IBS Growth Rates}

The large parameter space, originating from the uncertainties of the lattice design, defines a range of IBS growth rates.  
Equation~\eqref{eq_fodobeta} and \eqref{eq_fododispersionaverage} are used to estimate the average dispersion, $\langle D_x \rangle$, and \bfunc s, $\langle \beta_{x,y} \rangle$, required to approximate the IBS growth rates. 

\begin{figure*}
	\begin{subfigure}[h]{0.45\textwidth}
		\includegraphics[width=1\textwidth]{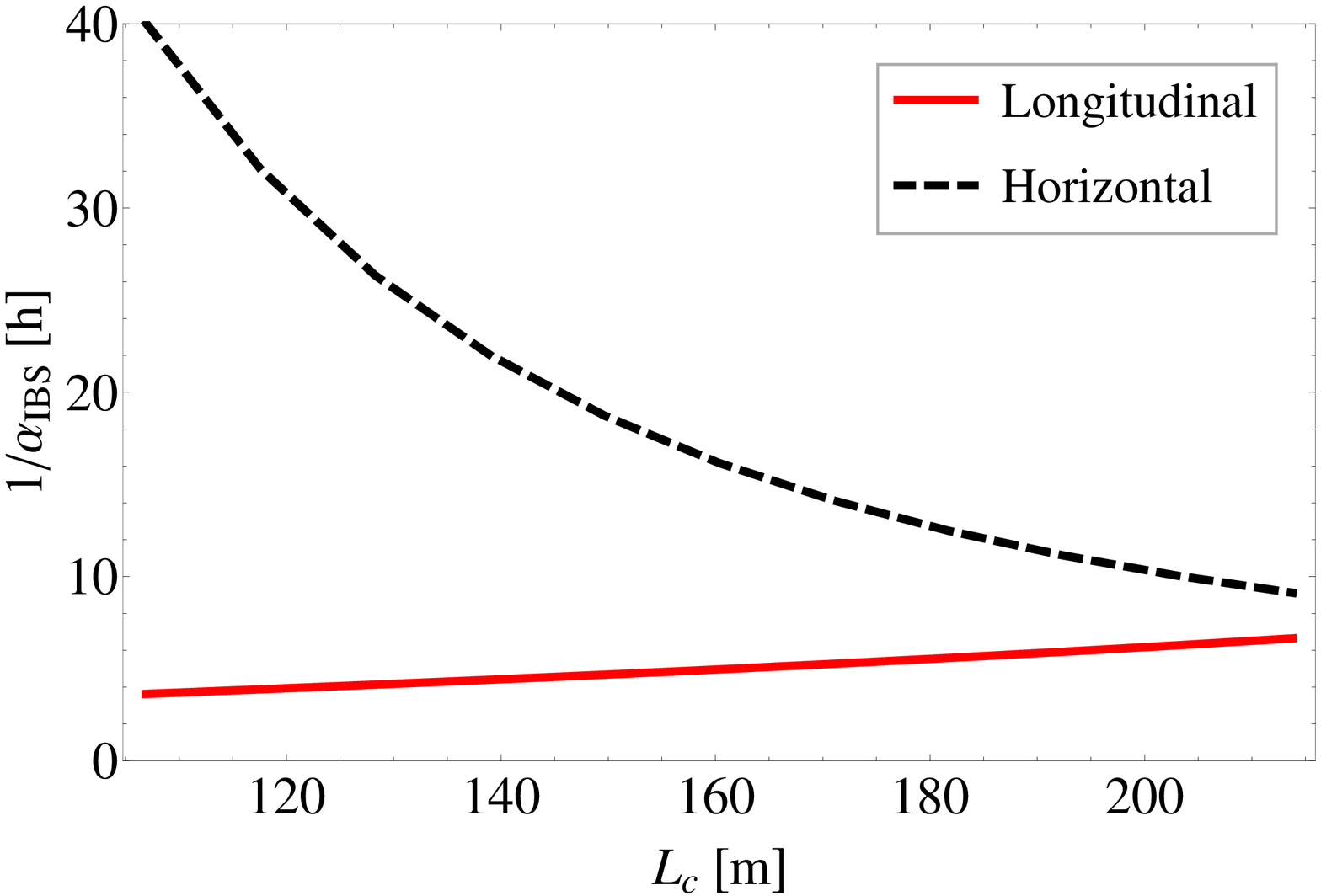}
	\caption{\label{f_fcc_ibsgrowthtimesInjection} IBS at \qty{3.3Z}{TeV} and $V_\text{RF}=\qty{13}{MV}$.}
	\end{subfigure}
	\begin{subfigure}[h]{0.45\textwidth}
		\vspace*{0.3cm}
		\includegraphics[width=1\textwidth]{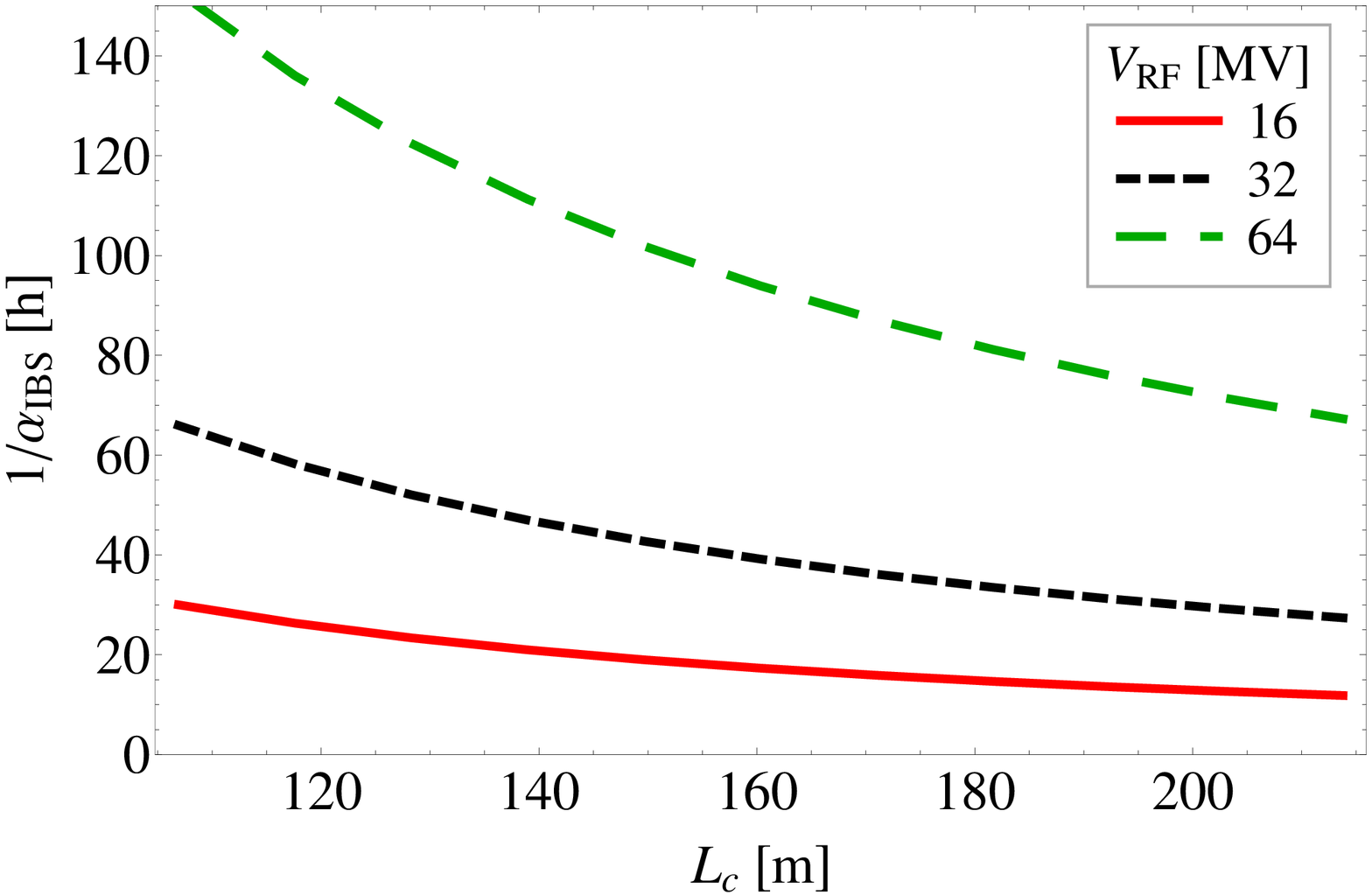}
		\caption{\label{f_fcc_ibsgrowthtimesCollisionLong} Longitudinal IBS at \qty{50Z}{TeV}.}
	\end{subfigure}
	\begin{subfigure}[h]{0.45\textwidth}
		\vspace*{0.3cm}
		\includegraphics[width=1\textwidth]{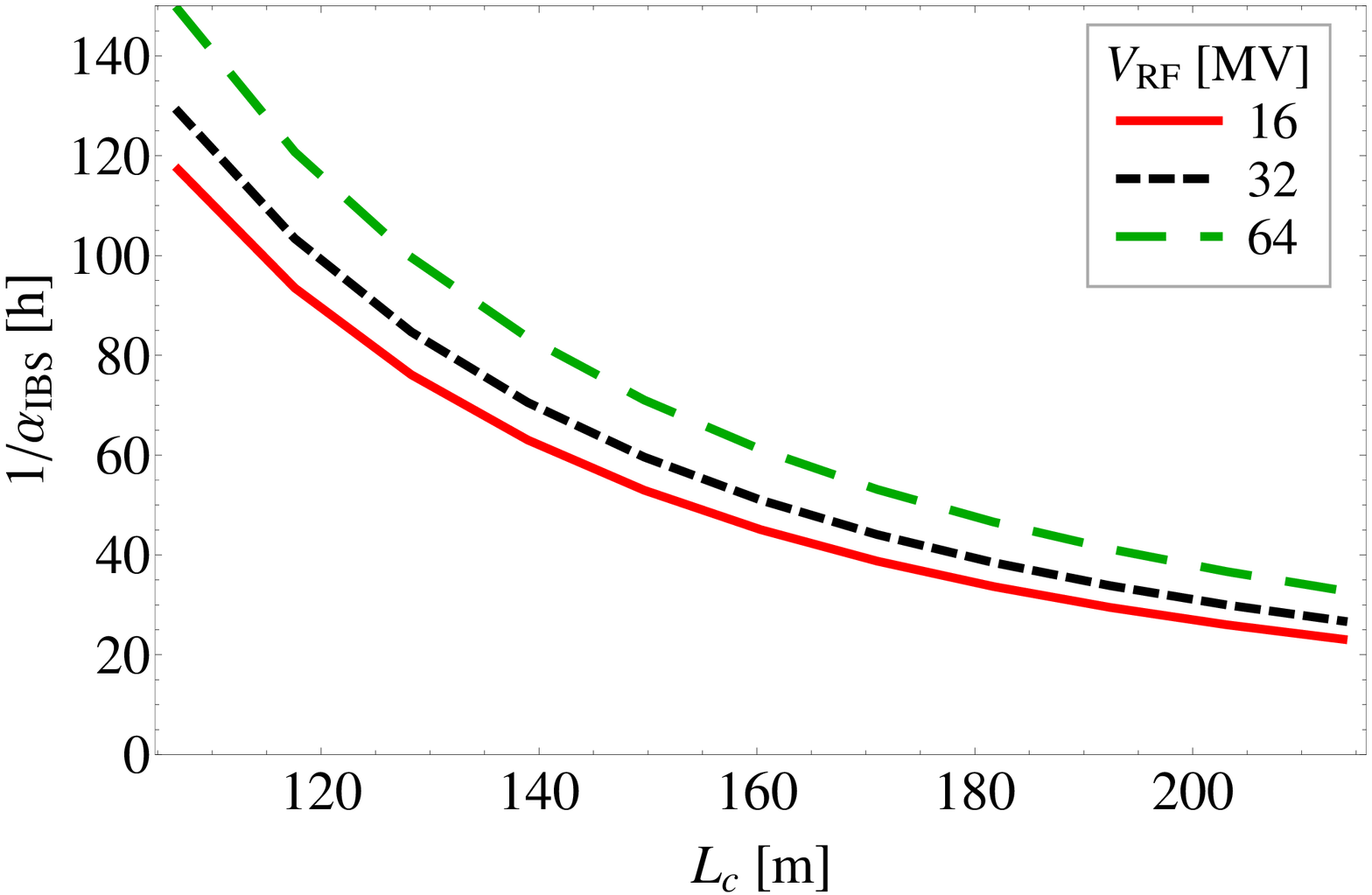}
	\caption{\label{f_fcc_ibsgrowthtimesCollisionHor} Horizontal IBS at \qty{50Z}{TeV}.}
	\end{subfigure}
\caption[Pb initial IBS growth times.]{\label{f_fcc_ibsgrowthtimes} Range of initial IBS  growth times as a function of the cell length, $L_c$, at injection (\subref{f_fcc_ibsgrowthtimesInjection}) and collision (\subref{f_fcc_ibsgrowthtimesCollisionLong}, \subref{f_fcc_ibsgrowthtimesCollisionHor}) energy, evaluated with Piwinski's equations \citep{APiwinski_IBS1}. (\subref{f_fcc_ibsgrowthtimesCollisionLong}) longitudinal, (\subref{f_fcc_ibsgrowthtimesCollisionHor}) horizontal growth times. For given values of the total RF voltage, \VRF. No transverse coupling assumed. }
\end{figure*}

Figure~\ref{f_fcc_ibsgrowthtimes} shows 1/\aibs\ as a function of $L_c$ at (\subref{f_fcc_ibsgrowthtimesInjection}) \qty{3.3Z}{TeV} (injection energy) and  at (\subref{f_fcc_ibsgrowthtimesCollisionLong}, \subref{f_fcc_ibsgrowthtimesCollisionHor}) \qty{50Z}{TeV} (collision energy) for the initial bunch parameters given in Table~\ref{t_fcc_beamparameters}. Only the longitudinal and horizontal plane are shown. IBS in the vertical plane is negligible without coupling, as assumed in the calculations.  For the plots at top energy the dependence of \sigp~on $L_c$ is taken into account, while \sigp~is constant at injection energy. The results are calculated for a set of RF voltages.  \sigp~can become very small for long cells (Fig.~\ref{f_fcc_SimgapVsCellLength}) and larger RF voltage can mitigate this effect. 

Note that the horizontal IBS strength increases (= decreasing $1/\aibs$) and the longitudinal decreases with increasing $L_c$ at injection, but at top energy both, \aibss~and \aibsx, become stronger for longer cells. The factors in the IBS calculation depending on the cell length  are  $\langle D_x \rangle \propto L_c^2$, $\langle \beta_x \rangle \propto L_c$ and $\sigp \propto 1/L_c$. If \sigp\ is independent of $L_c$ (as in the case of the injected beam), \aibss\ only has a weak dependence on the lattice, while \aibsx\ features a second term $\propto D_x^2 \propto L_c^4$ \cite{APiwinski_IBS1}. At top energy, \sigp\ is influenced by the lattice conditions and becomes a function of $L_c$.
Thus the strong dependence of \aibss\ on \sigp\ takes over and the longitudinal IBS growth is enhanced for long cells.

In general, IBS could lead to longitudinal emittance growth at injection energy, while the transverse growth rates are moderate. At collision energy IBS should still be modest, with growth times above \qty{20}{h}. The situation even improves, if the energy spread could be kept at the LHC design value, see Table~\ref{t_fcc_ibsgrowthtimes}. 

However, as it will be shown in the next section, this is only true for the initial beam parameters right after arriving at top energy. Because of the strong radiation damping, the beam emittances will shrink and the IBS will become strong enough to balance the damping.

Table~\ref{t_fcc_ibsgrowthtimes} summarises the IBS growth times for a bunch with initial parameters, assuming (a) $\sigma_p = \enum{0.6}{-4}$ (obtained with $\gamma_T$ of baseline lattice) and (b) $\sigma_p = \enum{1.1}{-4}$ (LHC design) at collision energy, $\sigma_p = \enum{1.9}{-4}$ at injection energy. The dispersion and \bfunc s are taken as calculated from the baseline lattice with $L_c = \qty{203}{m}$. The comparison of the formalisms by Piwinski and Wei shows that Wei estimates a systematically slightly stronger IBS rate. The overall agreement is better than 10\% at high energy and 20\% at injection energy for the given parameters.

\begin{table*}[tb]
\caption[Pb initial IBS growth times.]{\label{t_fcc_ibsgrowthtimes} 
Initial IBS growth times for Pb-ions calculated with the Piwinski \cite{APiwinski_IBS1} and Wei \cite{JWei_IBS} formulae, assuming baseline lattice ($L_c = \qty{203}{m}$) and no transverse coupling. Assumption for momentum spread: injection $\sigp = \enum{1.9}{-4}$ at $V_\text{RF}=\qty{13}{MV}$, collision (a) $\sigp = \enum{0.6}{-4}$ (obtained with $\gamma_T$ of baseline lattice), (b) $\sigp = \enum{1.1}{-4}$ (LHC design) at $V_\text{RF}=\qty{32}{MV}$. }
\begin{ruledtabular}
\begin{tabular}{rccccccc}
\multirow{3}{*}{Growth Times}   & \multirow{3}{*}{Unit}& 
\multicolumn{2}{c}{Injection} & \multicolumn{4}{c}{Collision}  \\

				&&	\multirow{2}{*}{Piwinski} & \multirow{2}{*}{Wei}	& \multicolumn{2}{c}{(a)} & \multicolumn{2}{c}{(b)} \\
				&& & & Piwinski & Wei &Piwinski &Wei \\
\colrule
 $1/\aibss$ & [h] & 6.3  	& 5.1 &	29.1   	& 27.3 &  141.4 & 132.0\\
 $1/\aibsx$ & [h] & 10.0 	& 8.2 & 30.0  	& 28.0 &43.9 & 41.0\\
 $1/\aibsy$ & [h] & $-10^4$ & $-10^3$& $-10^6$ &$-10^6$ & $-10^6$ & $-10^6$\\ 
\end{tabular} 
\end{ruledtabular}
\end{table*}

\subsection{Radiation Damping}
\label{s_fcc_raddamp}

A charged particle travelling in a storage ring will radiate energy, when it is bent on its circular orbit. Because of the average energy loss into this synchrotron radiation, the betatron and synchrotron oscillation amplitudes are damped like $A_i = A_{0,i} e^{-\alpha_{i} t}$, where $i=x,y,s$, with the radiation damping rates $\alpha_{i}$ given in Chapter 3.1.4 of \cite{HandbookAccelPhys}.

For a flat, isomagnetic ring with separated function magnets and zero vertical dispersion, the radiation emittance damping rates can be approximated by 
\begin{eqnarray}
\aradds = 2\araddxy \approx  2E_b^3  C_{\alpha} \frac{4\pi}{\rho_0 \Circ},
\label{eq_raddampRateSimplifyed}
\end{eqnarray}
where  $C_\alpha = r_0 c/(3(m_\text{ion} c^2)^3)$. Those quantities do not depend on the beam parameters. The strongest dependence is on the third power of the energy, the machine size and the particle type. Note that the longitudinal damping is twice as fast as the transverse.

\begin{table}
\caption[Pb radiation damping times.]{\label{t_fcc_raddamprates} Emittance radiation damping times for Pb-ions.}
\begin{ruledtabular}
\begin{tabular}{rccc}
Damping Times   & Unit& Injection & Collision \\
\colrule
 $1/\aradds$ & [h] & 852 &	0.24  \\
 $1/\araddx$ & [h] & 1704 & 0.49 \\
 $1/\araddy$ & [h] & 1704 & 0.49  \\ 
\end{tabular} 
\end{ruledtabular}
\end{table}

To get an impression how strong the radiation damping will be in the FCC, the damping rates are compared to the LHC design values:
\begin{eqnarray*}
\frac{\alpha_\text{rad,FCC}}{\alpha_\text{rad,LHC}} 
 \approx \frac{ E_\text{FCC}^3/C_\text{FCC}^2}{ E_\text{LHC}^3/C_\text{LHC}^2} \approx \frac{7^3}{4^2} \approx 22.
\end{eqnarray*}
This scaling is valid for all planes, because of relation~\eqref{eq_raddampRateSimplifyed}. The circumference of the accelerator was chosen such that the required dipole field does not exceed the expected technical limits. Therefore, the bending radius can be approximated to be proportional to the circumference, $\rho_0 \propto \Circ$. The new machine will be about a factor 4 longer than the LHC. Moreover, the energy will be increased by about a factor 7 ($=\qty{50Z}{TeV}/\qty{7Z}{TeV}$).
Table~\ref{t_fcc_raddamprates} quotes the radiation damping times at injection and collision energy.
Note that the horizontal equilibrium emittance from quantum excitation for lead beams  at top energy is of the order of \qty{10^{-5}}{\mu m}, the effect is thus still negligible.

\subsection{Luminosity}
The quantity that measures the ability of a particle accelerator to produce the required number of interactions is the \textit{luminosity}. It represents the proportionality factor between the number of produced events per unit of time, $\text{d}R/\text{d}t$, and the production cross-section of the considered reaction, $\sigma_{c}$:
\begin{equation}
\frac{\text{d}R}{\text{d}t}= \sigma_{c} \,\mathcal{L} .
\label{eq_LumiEventRate}
\end{equation}

In the specific case of a circular collider and when the particle density distribution can be approximated to a Gaussian, the luminosity of two beams, colliding exactly head-on, is given by:
\begin{equation}
\mathcal{L} = \frac{N_{b1} N_{b2} f_\text{rev}\kb}{2\pi \sqrt{\sigma_{x1}^2+\sigma_{x2}^2} \sqrt{\sigma_{y1}^2+\sigma_{y2}^2}}  =\frac{\Nb^2 f_\text{rev} \kb \gamma}{4\pi\epsilon_n \beta^*}, 
\label{eq_Luminosity}
\end{equation}
where $N_{b1}$ and $N_{b2}$ are the two colliding bunch intensities, \kb~the number of colliding bunches per beam, $\sigma_{xi}$ and $\sigma_{yi}$ are the transverse beam-sizes in the horizontal and vertical direction, respectively. The second equality follows in the approximation of equal and round distributions and optics of both beams: $\Nb = N_{b1} = N_{b2}$, $\sigma_{xy} = \sigma_{xi} = \sigma_{yi} = \sqrt{\epsilon_n \beta^* / \gamma}$.  With $\epsilon_n$ as the normalised emittance and $\beta^*$ as the \bfunc\,\,at the interaction point (IP).

Using Table~\ref{t_fcc_beamparameters} and Eq.~\eqref{eq_Luminosity} the initial luminosity at the beginning of collisions computes to
\begin{equation*}
\mathcal{L}_\text{initial} = \qty{\enum{2.6}{27}}{cm^{-2} s^{-1}}.
\end{equation*}
Which is, due to the higher intensity and energy, already 2.6 times higher than the design luminosity for Pb-Pb of the LHC.

The total event cross-section, $\sigma_{c,\text{tot}}$, is given by the sum over the cross-sections of all possible interactions removing particles from the beam in collision (burn-off). Apart from the inelastic hadronic interactions, the effects of Bound Free Pair Production (BFPP) and Electromagnetic Dissociation (EMD) are very important for Pb-Pb collisions: 
\begin{eqnarray}
\sigma_{c,\text{tot}} &=& \sigma_{c,\text{BFPP}} + \sigma_{c,\text{EMD}} + \sigma_{c,\text{hadron}}
\label{eq_fcc_crosssection} \\
			&\approx& \qty{354}{b} +\qty{235}{b}  + \qty{8}{b} = \qty{597}{b}. \notag
\end{eqnarray}
The numerical values in Eq.~\eqref{eq_fcc_crosssection} are estimated for $E_b=\qty{50Z}{TeV}$ with the aid of References~\cite{HMeier_BFPP, JJowett_EDM}.

\subsubsection{Luminosity Evolution}
While the beams are in collision, the instantaneous value of the luminosity will change, through intensity losses and emittance variations,
\begin{equation}
\mathcal{L}(t) = A \frac{\Nb^2(t)}{\sqrt{\epsilon_{x} (t) \epsilon_{y} (t)}},
\label{eq_LumiTimedependence}
\end{equation}
where all time independent factors are merged in $A = f_\text{rev} \kb /(4\pi \beta^*)$. For simplification, equal beam populations and sizes of both beams are assumed. To obtain the beam evolution with time, a system of four differential equations for the intensity, emittances and bunch length evolution has to be solved. The solutions can be inserted into Eq.~\eqref{eq_LumiTimedependence} to obtain the luminosity evolution. 
\begin{eqnarray}
\frac{\text{d}\Nb}{\text{d}t} &=&  -\sigma_{c,\text{tot}} A \frac{\Nb^2 }{\sqrt{\epsilon_{x} \epsilon_{y} }}
\label{eq_intensityEvolutionDiff}\\
\frac{\text{d}\epsilon_{x}}{\text{d}t}& =& \epsilon_{x} (\aibsx  -\araddx) 
\label{eq_horemittanceEvolutionDiff}\\
\frac{\text{d}\epsilon_{y}}{\text{d}t} &=& \epsilon_{y}( \aibsy  -\araddy)
\label{eq_veremittanceEvolutionDiff}\\
\frac{\text{d}\sigs}{\text{d}t}& =& \frac{1}{2} \sigs(\aibss -  \aradds)
\label{eq_bunchlengthEvolutionDiff}
\end{eqnarray}

The factor 1/2 in Eq.~\eqref{eq_bunchlengthEvolutionDiff} was introduced because the emittance growth rates are twice the amplitude growth rates.
The change in particle number with time, $\text{d}\Nb/\text{d}t$, is linked to the luminosity production rate described in Eq.~\eqref{eq_LumiEventRate}. Now, $\sigma_c =\sigma_{c,\text{tot}}$ is the sum of cross-sections for all processes that remove particles. A minus sign is introduced, since for each collision event generated one particle is lost: $\text{d}R/\text{d}t = -\text{d}\Nb/\text{d}t$.
The time evolution of the emittances and bunch length is influenced by dynamic IBS growth and constant radiation damping. The total emittance growth rate $\alpha_\epsilon = \aibs - \aradd$, thus varies dynamically in time and it is impossible to find an analytic solution of this system. 

In the given form, Eq.~\eqref{eq_intensityEvolutionDiff} assumes that all beam losses are from luminosity burn-off. 
In LHC p-p operation, a large fraction of particles is lost on the collimators. The amount, however, strongly depends on the collimator settings, which in past runs (2012) were tight in order to clean the beam halo. 
Nevertheless, experience from RHIC shows that, owing to the applied stochastic cooling, it is possible to achieve very low loss rates from non-luminous processes \cite{WFischer_UUcrosssection}. The strong radiation damping at FCC energies will have a similar effect as the stochastic cooling in RHIC, supporting the assumption of negligible non-luminous losses made in Eq.~\eqref{eq_intensityEvolutionDiff}.

In the following, approximations will be made for which an analytic description is possible.  To simplify the situation, round beams and fully coupled transverse motion is assumed, such $\epsilon(t) = \epsilon_x(t) =\epsilon_y(t)$ at all times, reducing the ordinary differential equation (ODE) system to three equations.

(I) In the first case, $\epsilon(t) =\epsilon_{0} = \text{const.}$ should be considered, which is achieved when $\aibs-\aradd =0$ and thus $\text{d}\epsilon/\text{d}t=0$. For zero crossing angle, the bunch length evolution is (in first order) decoupled from the luminosity.  Eq.~\eqref{eq_intensityEvolutionDiff} simplifies to 
\begin{eqnarray*}
\frac{\text{d}N}{\text{d}t} = -\sigma_{c,\text{tot}} A \frac{\Nb^2}{\epsilon_{0}}
\end{eqnarray*}
This can easily be solved for the intensity evolution and, in combination with Eq.~\eqref{eq_LumiTimedependence}, for the luminosity evolution with time:
\begin{eqnarray*}
\Nb(t) &=& \frac{N_{b0}}{A N_{b0} \sigma_{c,\text{tot}} t/\epsilon_{0} +1}\\ 
\Rightarrow \mathcal{L}(t) &=& \mathcal{L}_0 \left( \frac{1}{A N_{b0} \sigma_{c,\text{tot}} t/\epsilon_{0} +1} \right)^2.
\end{eqnarray*}
By investigating those equations, it becomes clear that the only non-constant factor is the time $t$, which appears only in the denominator, i.e. the intensity and with it the luminosity can only decay.

(II) In the second case, the total emittance damping rate should be constant, $\alpha_\epsilon = \text{const.}$, with \mbox{$\aibs \ll \aradd$}. It is implicitly approximated that IBS is independent of the beam parameters, decoupling the bunch length and emittance evolutions.   Simultaneously solving the two remaining differential equations \eqref{eq_intensityEvolutionDiff} and \eqref{eq_horemittanceEvolutionDiff} gives
\begin{eqnarray}
\epsilon_n(t) &=&\epsilon_{0} \exp[-\alpha_{\epsilon} t]
\label{eq_emittanceEvolutionConstTau}\\
\Nb (t) &=& \frac{N_{b0} \epsilon_{0}}{\epsilon_{0} + A N_{b0} \sigma_{c,\text{tot}} (\exp[\alpha_{\epsilon}t]-1)/\alpha_{\epsilon}  }
\label{eq_intensityEvolutionConstTau}\\
\mathcal{L}(t) &=&\mathcal{L}_0 \frac{\epsilon_{0}^2 \exp[\alpha_{\epsilon} t]}{(\epsilon_{0} +A N_{b0} \sigma_{c,\text{tot}} (\exp[\alpha_{\epsilon} t]-1)/ \alpha_{\epsilon})^2 }.
\label{eq_luminosityEvolutionConstTau}
\end{eqnarray}
Again $t$ is the only non-constant parameter. As expected, $\epsilon_n(t)$ (Eq.~\eqref{eq_emittanceEvolutionConstTau}) and $\Nb(t)$ (Eq.~\eqref{eq_intensityEvolutionConstTau}) can only decay. However, the combination of both, the luminosity evolution (Eq.~\eqref{eq_luminosityEvolutionConstTau}), features the exponentially growing factor $\exp[\alpha_{\epsilon}t]$ in the numerator and denominator. This means, as long as the numerator \mbox{$ \epsilon_{0}^2 \exp[\alpha_{\epsilon}t]$} predominates the denominator \mbox{$(\epsilon_{0} +A N_{b0} \sigma_{c,\text{tot}} (\exp[\alpha_{\epsilon}t]-1)/ \alpha_{\epsilon})^2 $} a growth of the initial luminosity to a higher peak is possible. It should be noted that the assumption of a constant damping leads to emittances asymptotically approaching zero, which is non-physical. Because of this effect, the luminosity peak computed with Eq.~\eqref{eq_luminosityEvolutionConstTau} is overestimated.

(III) In reality the IBS growth rate changes dynamically with the intensity and emittance, thus it will become stronger, while the emittances shrink due to radiation damping. 
Since the total emittance growth rate is given by $\alpha_\epsilon = \aibs - \aradd$, the  emittance will approach a value where the growth from IBS balances the damping. This balance is not a real equilibrium, where the emittance and bunch length would be constant. But the IBS strength keeps decreasing due to intensity burn-off, leading to a slowly shrinking emittance and bunch length to maintain the balance.

\begin{figure}
		\includegraphics[width=0.45\textwidth]{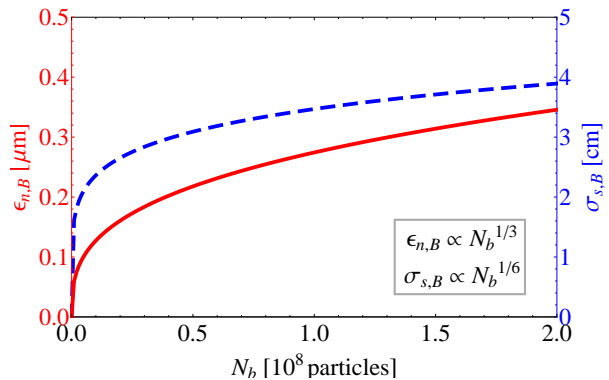}
\caption[Emittance and bunch length for balanced IBS and radiation damping.]{\label{f_fcc_equilibriumEmitLength} Normalised emittance (red) and bunch length (blue) for balanced IBS and radiation damping.}
\end{figure}

An analytical expression for the balance value of the emittance and bunch length can be derived from Wei's IBS formalism given by Eq.~\eqref{eq_IBSgrowthRateWeixy} and \eqref{eq_IBSgrowthRateWeis}. Even in this simplified form, the transverse growth rate shows a rather complicated dependence on $\epsilon$, providing only a numerical solution. Both factors under the square root in the denominator of Eq.~\eqref{eq_IBSgrowthRateWeixy} depend on evolving beam properties. $\epsilon \propto 10^{-6} /\gamma \approx 10^{-11} $ and $C_2 \sigp^2 \approx D_x^2/\beta_x (10^{-4})^2 \propto 10^{-10} $ are in the same order of magnitude, therefore we approximate  $\sqrt{\epsilon + C_2 \sigp^2} \longrightarrow \sqrt{2 C_2 \sigp^2}$.  Eq.~\eqref{eq_IBSgrowthRateWeixy} can be set equal to $\araddx$ to satisfy the balance condition and be solved for the emittance  $\epsilon_B =\epsilon_{n,B}/\gamma$: 
\begin{eqnarray}
 \epsilon_{n,B} \cong \gamma \sqrt{\frac{C_1 \Nb}{\sqrt{2 C_2} \araddx D_p \sigma_{s,B}^2}}.
 \label{eq_equilibriumEmit1}
\end{eqnarray}
$D_p$ is the proportionality factor between the momentum spread and the bunch length given by Eq.~\eqref{eq_momentumSpread}. $\epsilon_{n,B}$ still depends on the balance value of the bunch length, $\sigma_{s,B}$, which is determined by replacing $\epsilon \longrightarrow \epsilon_{n,B}/\gamma$ in Eq.~\eqref{eq_IBSgrowthRateWeis} and applying $\aibss = \aradds$:
\begin{eqnarray}
\sigma_{s,B} \cong \left(\frac{C_3 \sqrt{C_1 \Nb \araddx}}{D_p^{5/2} (2 C_2)^{1/4} \aradds}   \right)^{1/3} \propto \Nb^{1/6}.
 \label{eq_equilibriumBunchLength}
\end{eqnarray}
Inserting  Eq.~\eqref{eq_equilibriumBunchLength} into Eq.~\eqref{eq_equilibriumEmit1} leads to an equivalent equation for the emittance:
\begin{eqnarray}
\epsilon_{n,B} \cong \gamma \left(   \frac{C_1 \Nb D_p \aradds}{C_3 \sqrt{2 C_2} \araddx^2}  \right)^{1/3} \propto \Nb^{1/3}.
 \label{eq_equilibriumEmit}
\end{eqnarray}
When a balance between IBS and radiation damping is reached, the emittance and bunch length depend only on the bunch intensity. The higher the number of particles, the larger the beam dimensions as shown in Fig.~\ref{f_fcc_equilibriumEmitLength}.  The balanced normalised emittance (red) and the bunch length (blue) are plotted as a function of the intensity. The plot shows that those quantities become small in the expected range of bunch charge. Longitudinal, and potentially transverse, blow-up might become necessary to keep the beam sizes in a reasonable range.

The intensity evolution, where $\aibs = \aradd$, can be obtained by inserting Eq.~\eqref{eq_equilibriumEmit} into Eq.~\eqref{eq_intensityEvolutionDiff}:
\begin{widetext}
\begin{eqnarray}
N_{b,B}(t) \cong 3 \sqrt{3} N'_{b0} \left(3+ 2^{5/6} A \sigma_\text{c,tot} {N'_{b0}}^{2/3} t \left[\frac{C_3 \sqrt{C_2} \araddx^2  }{C_1 D_p \aradds}\right]^{1/3}  \right)^{-3/2}.
\label{eq_equilibriumN}
\end{eqnarray}
\end{widetext}
Note that $N'_{b0} \neq N_{b0}$ is not the initial intensity at the beginning of the fill, but should be the number of particles left when the balanced regime is reached. The luminosity evolution can then be calculated by inserting Eq.~\eqref{eq_equilibriumN} and \eqref{eq_equilibriumEmit} into \eqref{eq_LumiTimedependence}.

\begin{figure*}
	\begin{subfigure}[h]{1\textwidth}
	\centering
		\begin{minipage}{0.45\textwidth}
		\includegraphics[width=1\textwidth]{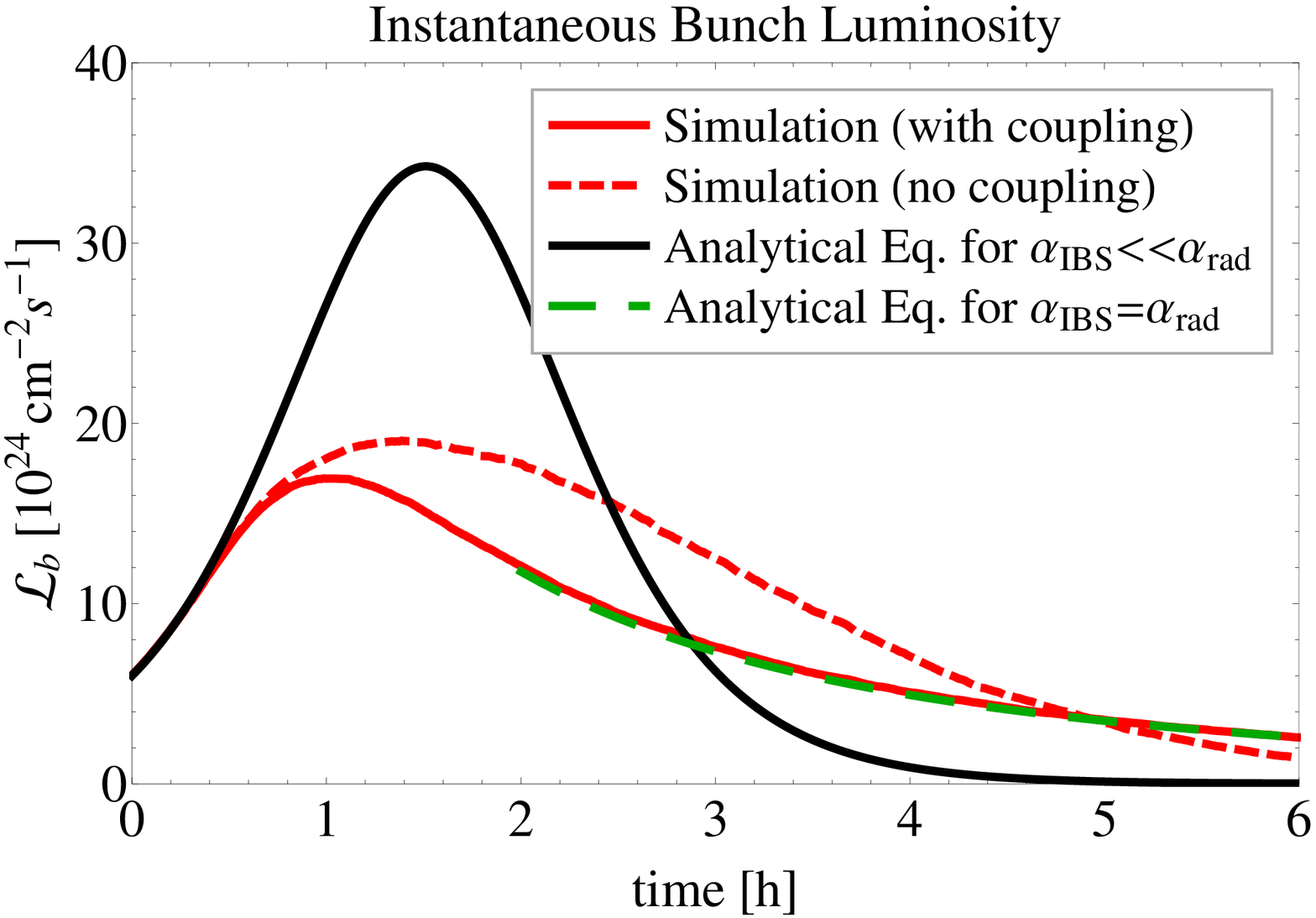}
		\end{minipage}
		\begin{minipage}{0.45\textwidth}
		\includegraphics[width=1\textwidth]{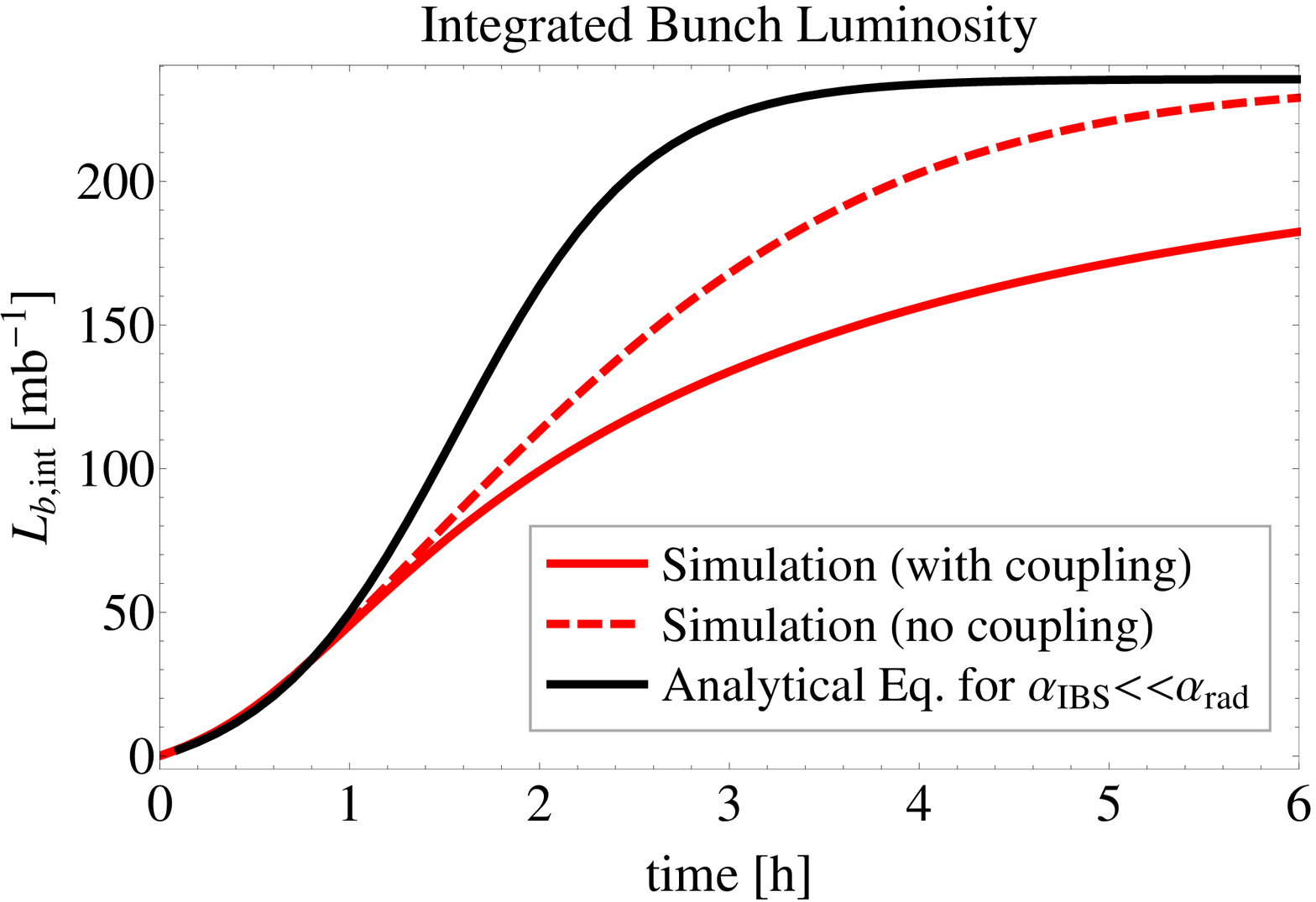}
		\end{minipage}
			\vspace*{0.5cm}
	\end{subfigure}

	\begin{subfigure}[h]{1\textwidth}
	\centering
		\begin{minipage}{0.45\textwidth}
		\includegraphics[width=1\textwidth]{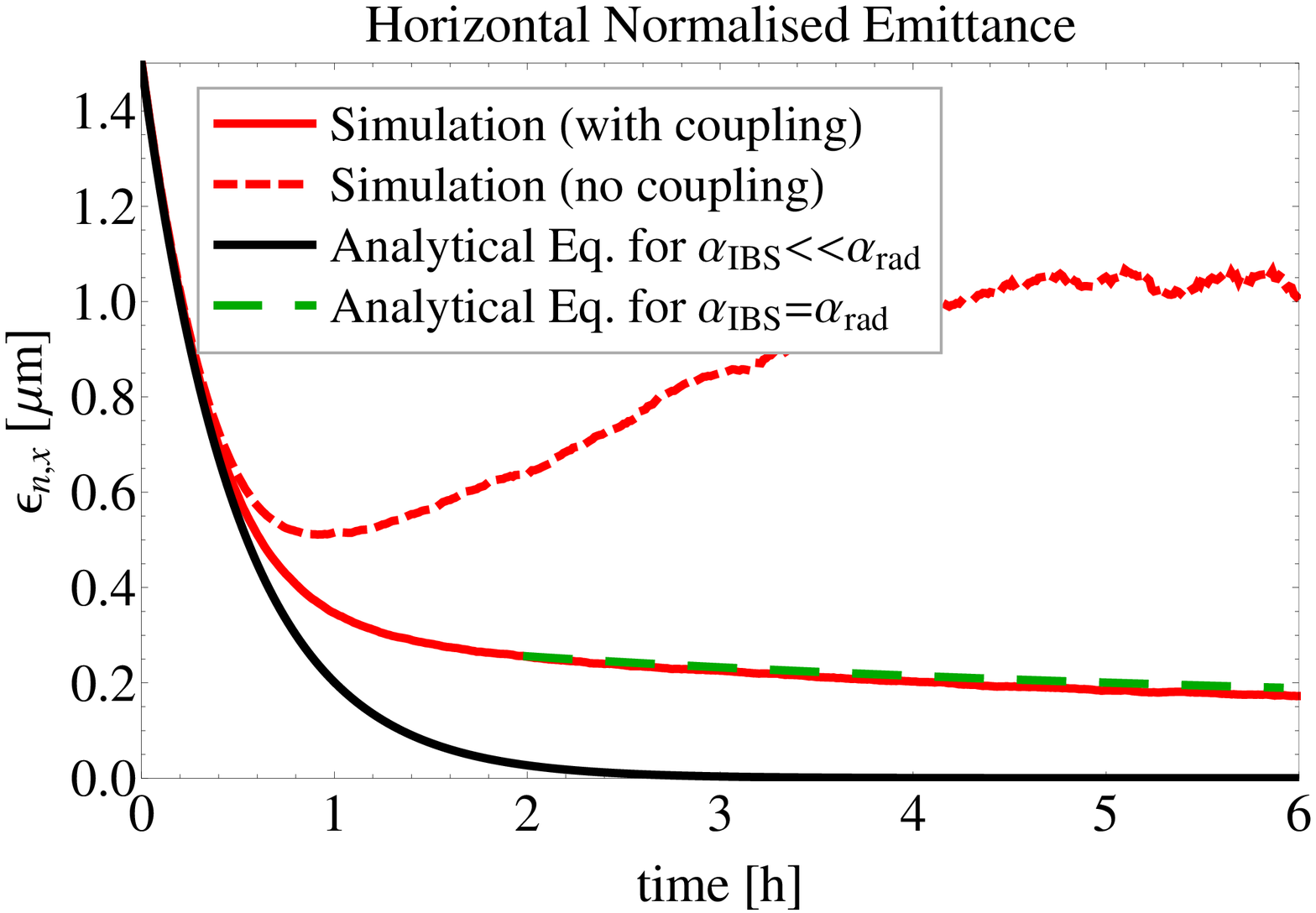}
		\end{minipage}
		\begin{minipage}{0.45\textwidth}
		\includegraphics[width=1\textwidth]{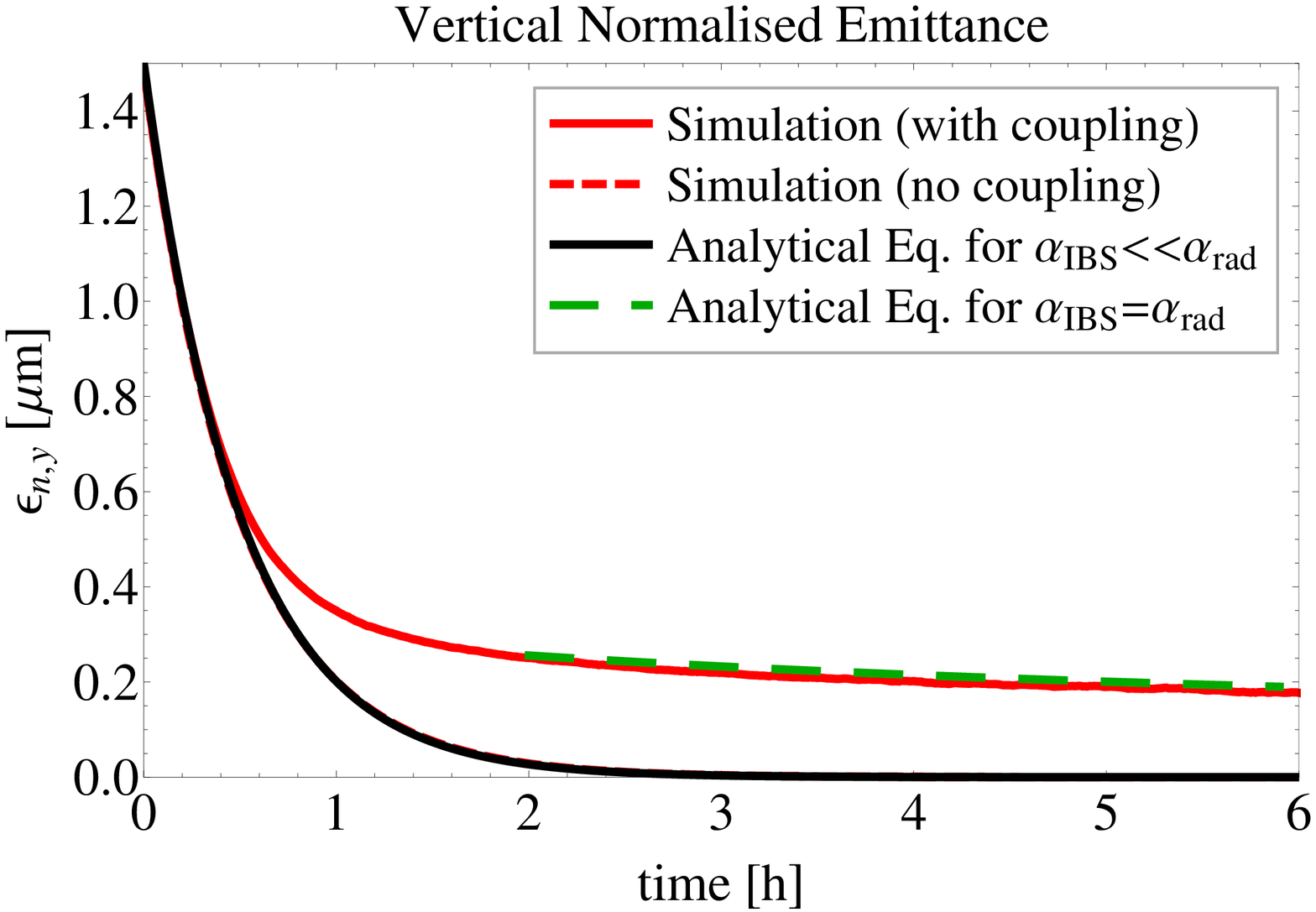}
		\end{minipage}
	\end{subfigure}

	\vspace*{0.5cm}
	\begin{subfigure}[h]{1\textwidth}
		\begin{minipage}{0.45\textwidth}
		\includegraphics[width=1\textwidth]{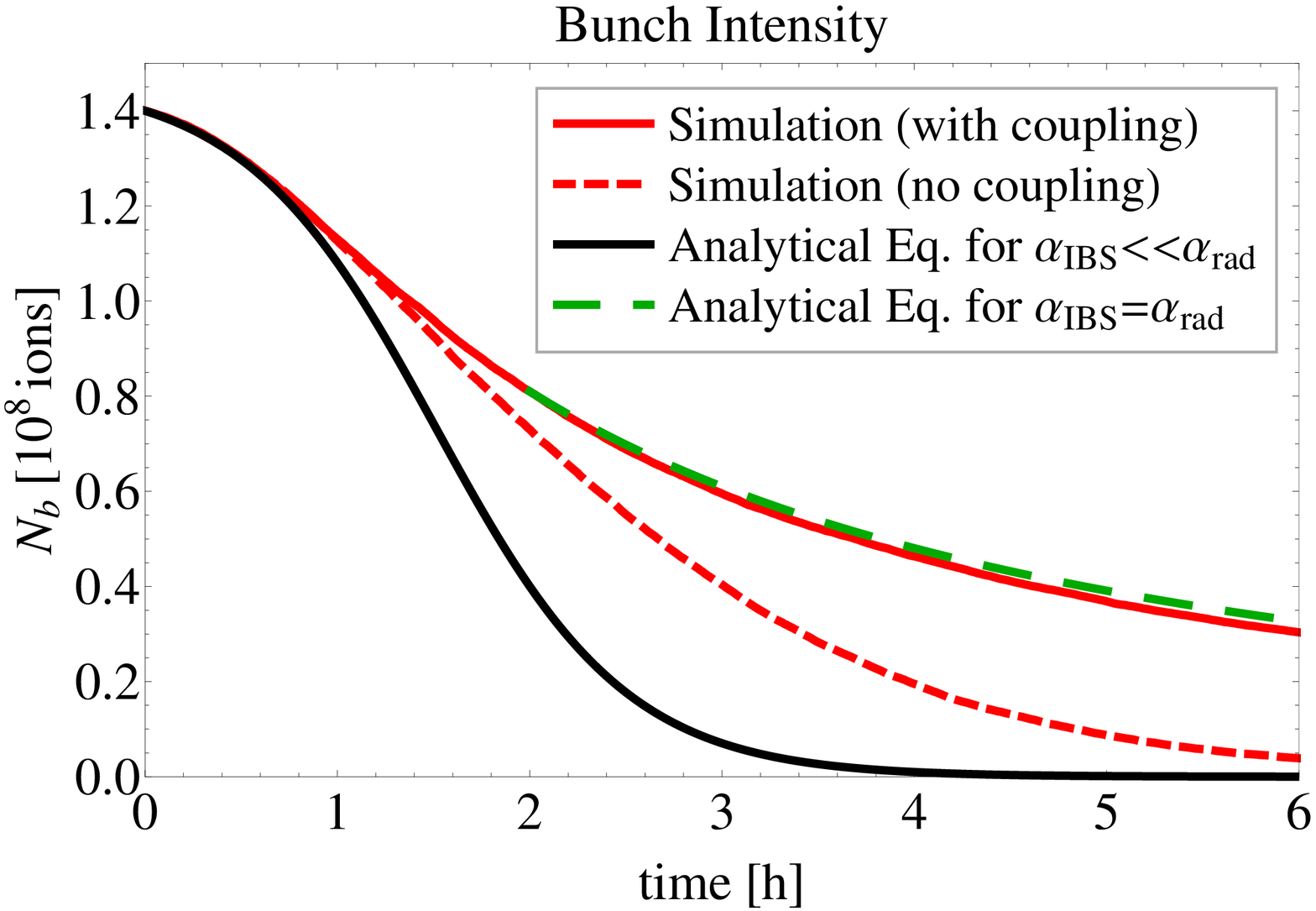}
		\end{minipage}
		\begin{minipage}{0.45\textwidth}
		\includegraphics[width=1\textwidth]{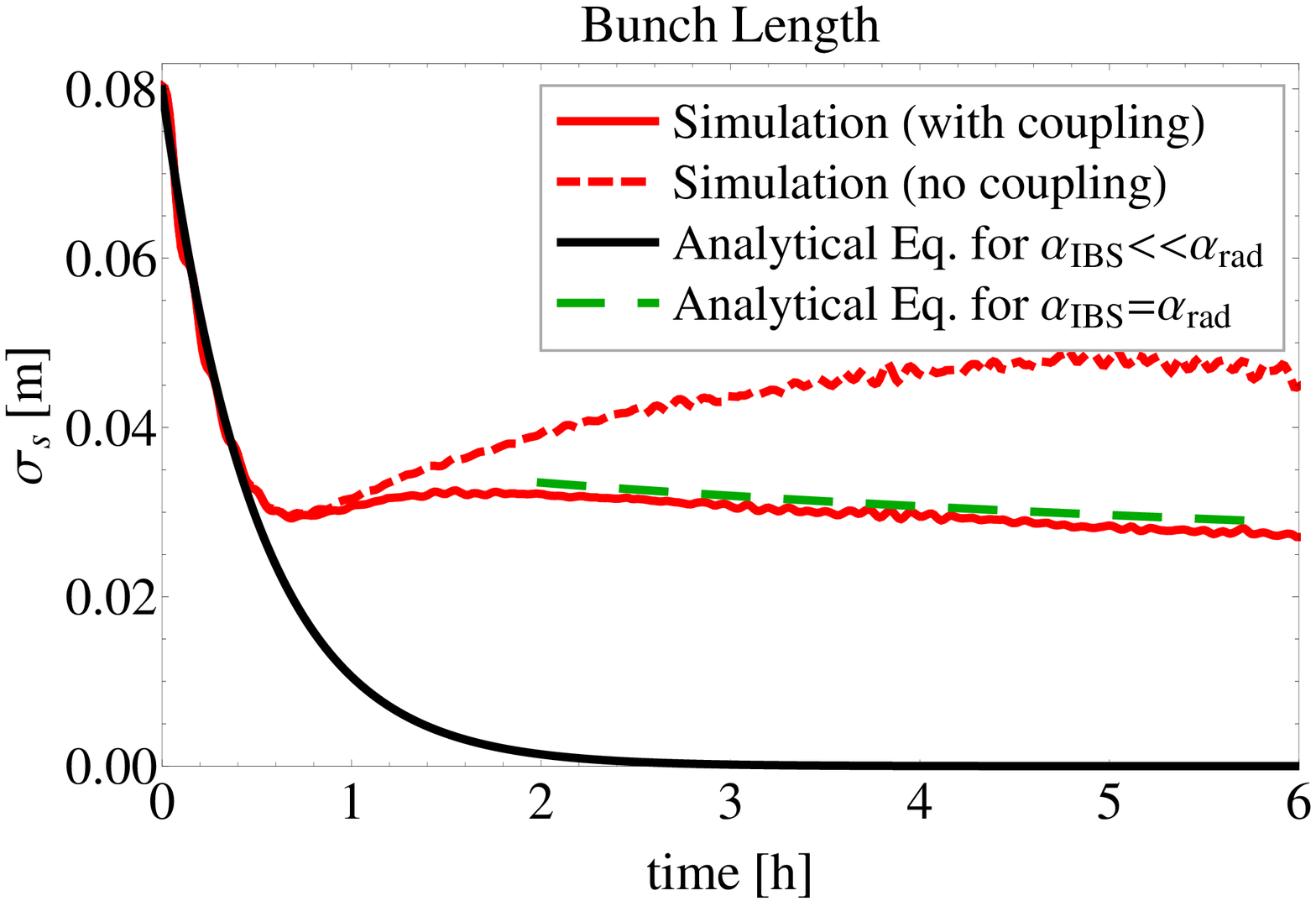}
		\end{minipage}
	\end{subfigure}
\caption[Pb-Pb beam and luminosity evolution.]{\label{f_fcc_beamEvolution} Pb-Pb beam and luminosity evolution. Top: instantaneous (left) and integrated luminosity (right), middle: horizontal (left) and vertical (right) normalised emittance, bottom: intensity (left) and bunch length (right). One experiment is in collisions. The black lines show the calculations done with Eq.~\eqref{eq_emittanceEvolutionConstTau}-\eqref{eq_luminosityEvolutionConstTau} for $\aibs \ll \aradd$, the dashed green lines show the calculations done with Eq.~\eqref{eq_equilibriumBunchLength}-\eqref{eq_equilibriumN} in the regime where IBS and radiation damping balance each other ($\aibs = \aradd$), the two red lines are CTE simulations with (solid) and without (dashed) IBS coupling. Note that the dashed red line in the middle right plot is hidden behind the black line.}
\end{figure*}

Figure~\ref{f_fcc_beamEvolution} shows the beam and luminosity evolution for case (II) and (III) as discussed above in comparison with tracking simulations done with the CTE program \cite{cte}. 
The results are displayed for two colliding lead bunches featuring the beam parameters given in Table~\ref{t_fcc_beamparameters}. One experiment is taking data.
The black line shows the calculations with Eq.~\eqref{eq_emittanceEvolutionConstTau}-\eqref{eq_luminosityEvolutionConstTau} for the approximation where $\alpha_\epsilon = \text{const.}$ and $\aibs \ll \aradd$. The dashed green line shows the calculations done with Eq.~\eqref{eq_equilibriumBunchLength}-\eqref{eq_equilibriumN} in the regime where IBS and radiation damping balance each other ($\aibs = \aradd$). The two red lines are CTE simulations with (solid) and without (dashed) IBS coupling. The simulations are based on the assumption of a smooth lattice  
and Piwinski's IBS formalism.

It is clearly visible that the bunch length and emittances of the analytical calculations for $\aibs \ll \aradd$ (black) asymptotically approach zero, which is non-physical, leading to a strong over-estimation of the luminosity. While the simulation with uncoupled planes (dashed red line) shows a realistic horizontal and longitudinal behaviour, the vertical emittance still damps to zero. In the coupled simulation (solid red lines) all three beam dimensions settle at a balanced value above zero. The transverse normalised emittance reaches around \qty{0.2}{\mu m}, corresponding to a beam size  of $\sigma^*\approx \qty{3}{\mu m}$ at the IP for \bstarval{1.1}.  The bunch length damps twice as fast as the transverse planes, before IBS kicks in and stabilises the bunch length around $\sigs\approx\qty{3}{cm}$. The derivation of the balanced state equations (green dashed) assumes as well coupled transverse motion. The calculation is in very good agreement with the corresponding simulation.

Because of the small beam sizes, problems with instabilities might appear, apart from the fact that it could become difficult to find the collisions. Blow-up might become necessary in the longitudinal but maybe also in the transverse planes. A transverse emittance blow-up could also act as a luminosity levelling method.

\begin{figure*}
	\begin{subfigure}[h]{1\textwidth}
	\centering
		\begin{minipage}{0.45\textwidth}
		\includegraphics[width=1\textwidth]{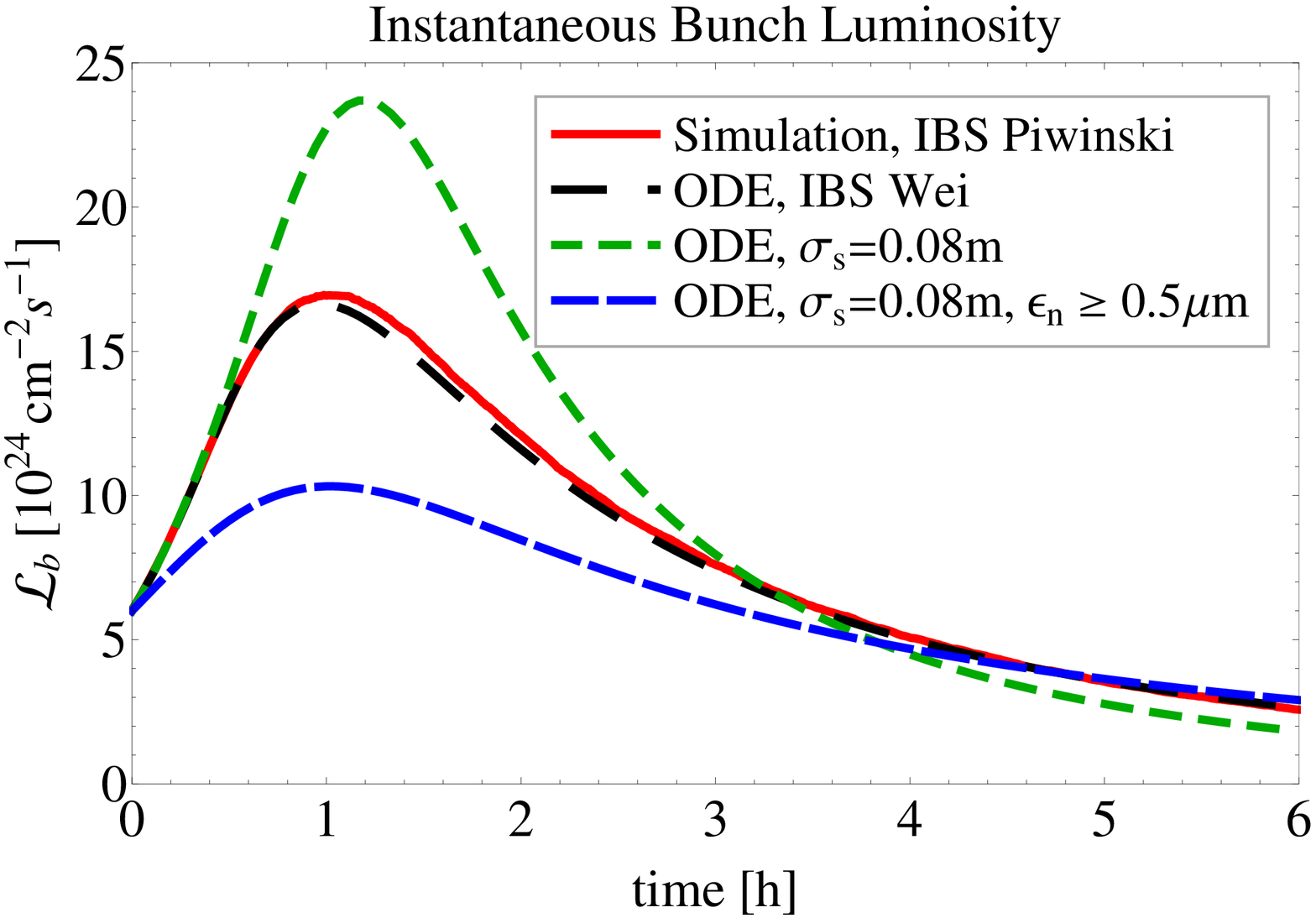}
		\end{minipage}
		\begin{minipage}{0.45\textwidth}
		\includegraphics[width=1\textwidth]{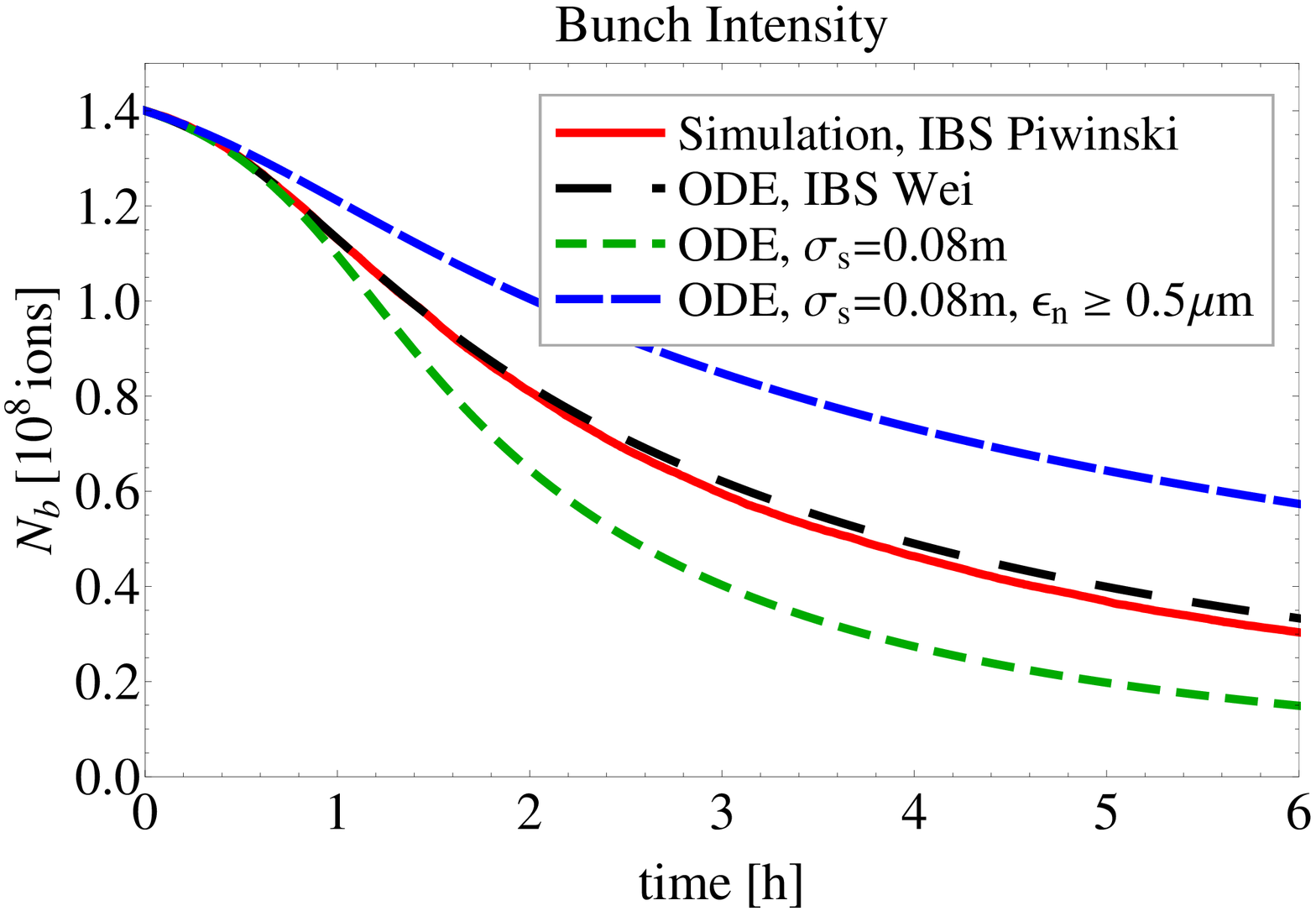}
		\end{minipage}
	\end{subfigure}

	\vspace*{0.5cm}
	\begin{subfigure}[h]{1\textwidth}
		\begin{minipage}{0.45\textwidth}
		\includegraphics[width=1\textwidth]{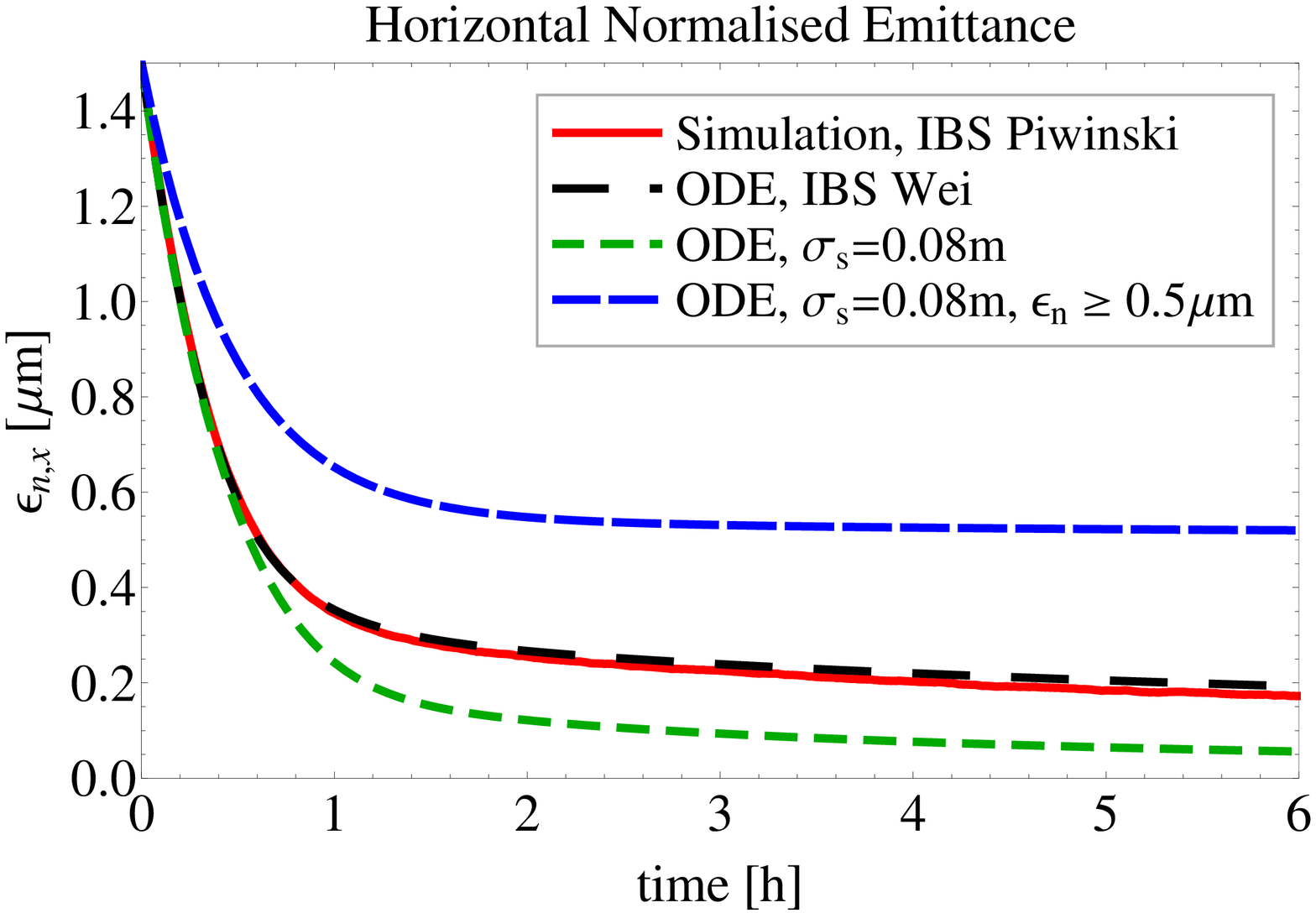}
		\end{minipage}
		\begin{minipage}{0.45\textwidth}
		\includegraphics[width=1\textwidth]{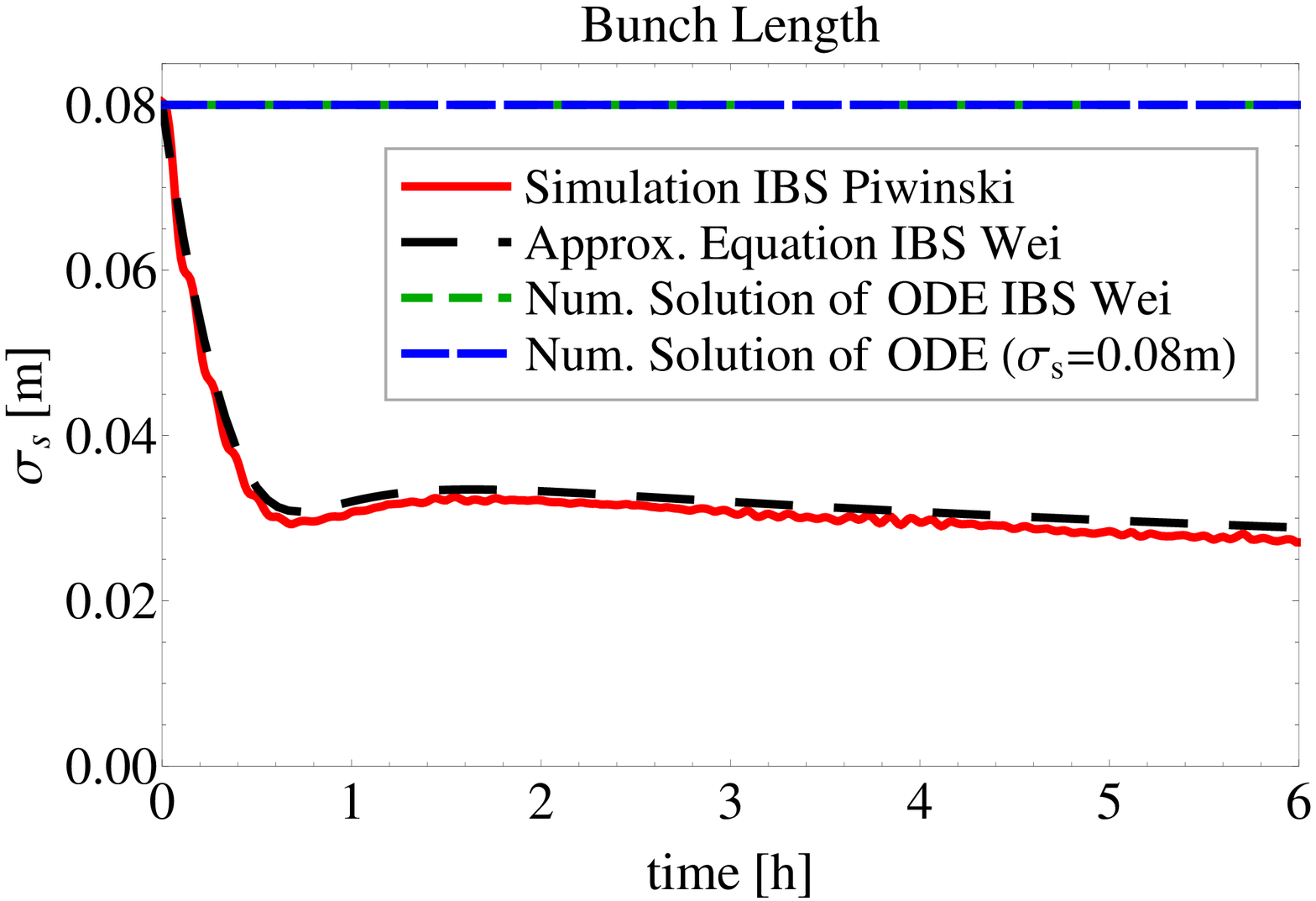}
		\end{minipage}
	\end{subfigure}
\caption[Pb-Pb beam evolution from ODE.]{\label{f_fcc_beamEvolutionODE} Pb-Pb beam evolution derived from ODE in comparison with simulation result. Top: Luminosity (left) and intensity (right), bottom: normalised emittance (left) and bunch length (right).}
\end{figure*}

Without further approximations it is not possible to solve the differential equation system of Eq.~\eqref{eq_intensityEvolutionDiff}-\eqref{eq_bunchlengthEvolutionDiff} analytically. But by using Wei's analytic IBS expressions the system can be solved numerically. Figure~\ref{f_fcc_beamEvolutionODE} presents numerical solutions of the ODE system (dashed lines) obtained with \textit{Mathematica}. Coupled transverse motion and round beams are assumed. The solid red line indicates again the CTE simulation shown in Fig.~\ref{f_fcc_beamEvolution}. The black dashed line shows the corresponding solution of the ODE system. The agreement between the numerical solution and the tracking result is excellent. Hence, the analytic calculation in the balanced regime (with coupling) is in excellent agreement with the ODEs. The small differences, are explained  by the difference in IBS growth rates calculated with Piwinski's and Wei's algorithms for the same beam conditions.  To prevent the bunch length from shrinking to too low values and  to model the evolution under longitudinal blow-up, the ODEs are solved for constant bunch length ($\text{d}\sigs/\text{d}t =0$, green dashed line). This enhances the intensity burn-off and the luminosity peak, since the IBS is weakened, reducing further the balance value of the emittance. Introducing an additional constant term in Eq.~\eqref{eq_horemittanceEvolutionDiff} can constrain the emittance above a certain value, $\epsilon_\text{min}$, similar to the equilibrium between radiation damping and quantum excitation in lepton machines \cite{HandbookAccelPhys}:
\begin{eqnarray*}
\frac{\text{d}\epsilon}{\text{d}t}& =& \aibsx~\epsilon   -\aradd (\epsilon - \epsilon_\text{min}). 
\end{eqnarray*}
Solving the equations for both, constant bunch length and a minimum emittance of e.g. $\epsilon_\text{min}=\qty{0.5}{\mu m}$, results in the blue dashed-dotted curve. As intended, the emittance stops decaying at about \qty{0.5}{\mu m}, naturally coming along with a luminosity reduction.

Looking back at Fig.~\ref{f_fcc_beamEvolution},  the intensity decay is very fast, because of the high burn-off rates going along with the small emittances. 
In the analytical case (black) the total beam intensity is converted into luminosity in about \qty{4}{h}. In the simulation the finite emittances reduce the peak luminosity and spread out the luminosity events over a longer period, however, the event production is still very efficient: only about 20\% of the initial particles are left after \qty{6}{h} collision time. 

For comparison, in a normal LHC fill, the natural cooling from radiation damping is much weaker and not sufficient to increase the luminosity above its initial value. After about \qty{6}{h}, the luminosity has decayed so much that it is necessary to refill. At that time, the beam population has only decreased to 40 or 50\% of its initial value. Those particles have to be thrown away to be replaced with fresh beam. To maximise the integrated luminosity, the time in collisions has to be optimised. 

In a very high energy hadron collider, the event production efficiency will be close to its optimum, where all particles are converted into luminosity. Under equal operational conditions, this will lead to a constant fill length. In this regime the integrated luminosity per fill is given by
\begin{eqnarray}
L_\text{int} = \frac{\Nb \kb}{\sigma_{c,\text{tot}}}.
\label{eq_maxLint}
\end{eqnarray}

\begin{figure}
	\begin{subfigure}[h]{0.45\textwidth}
			\vspace*{-0.4cm}
		\includegraphics[width=1\textwidth]{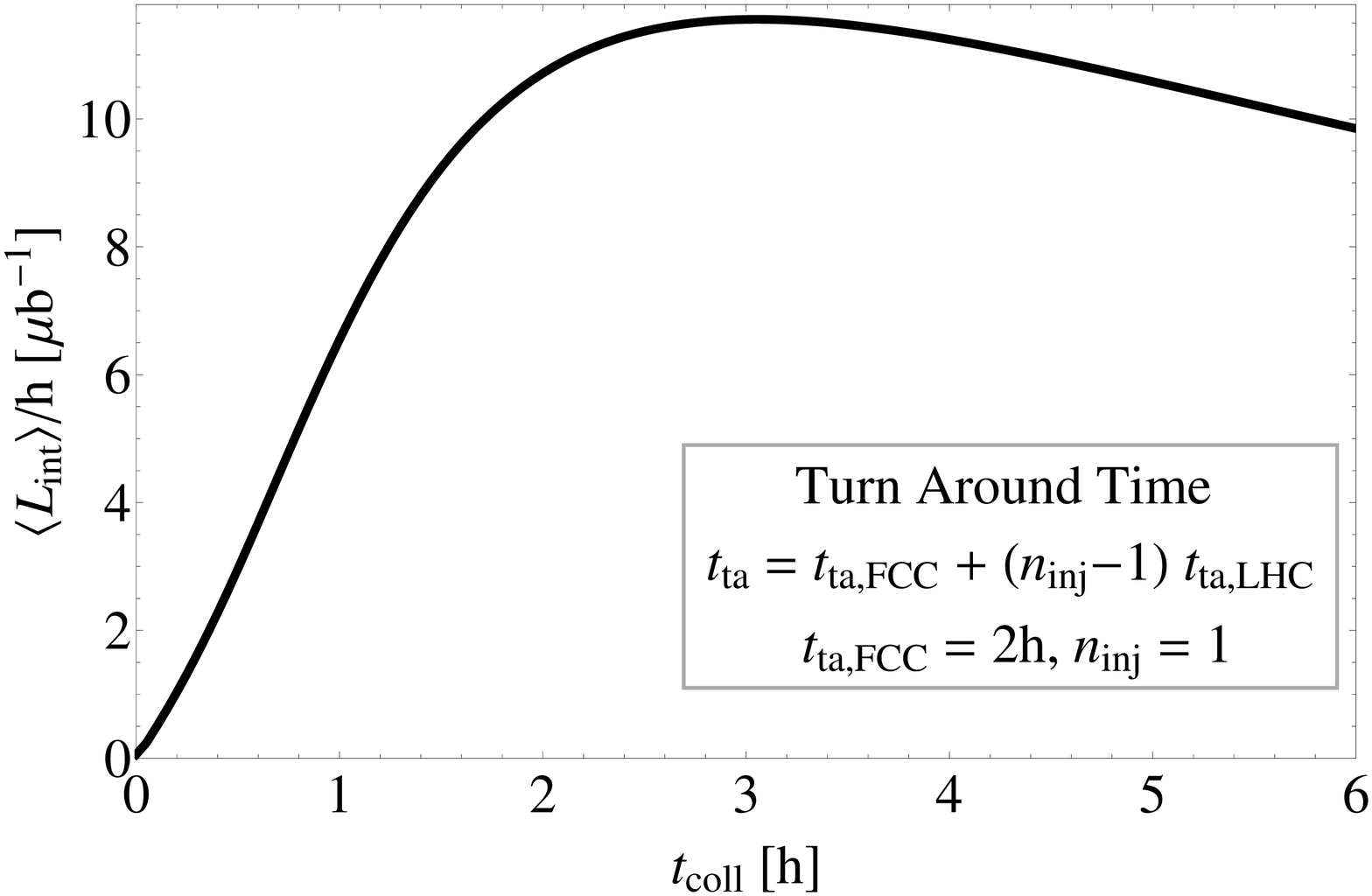}
		\caption{\label{f_fcc_optimisingLumiIntA} Average integrated luminosity per h. }
		\vspace*{0.3cm}
	\end{subfigure}	
	\begin{subfigure}[h]{0.45\textwidth}
		\includegraphics[width=1\textwidth]{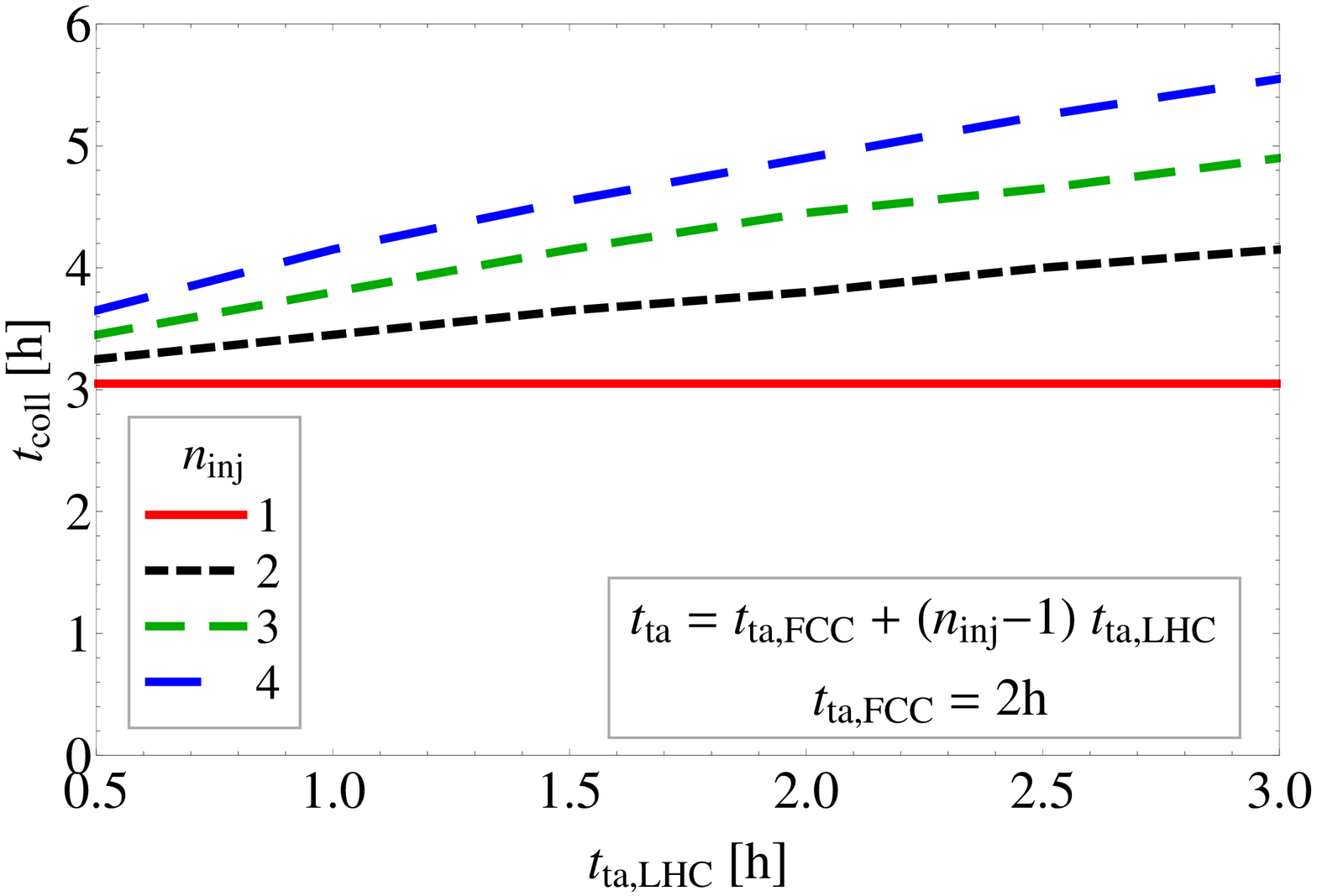}
		\caption{\label{f_fcc_optimisingLumiIntB} Optimal time in collision. }
		\vspace*{0.3cm}
	\end{subfigure}	
	\begin{subfigure}[h]{0.45\textwidth}
		\includegraphics[width=1\textwidth]{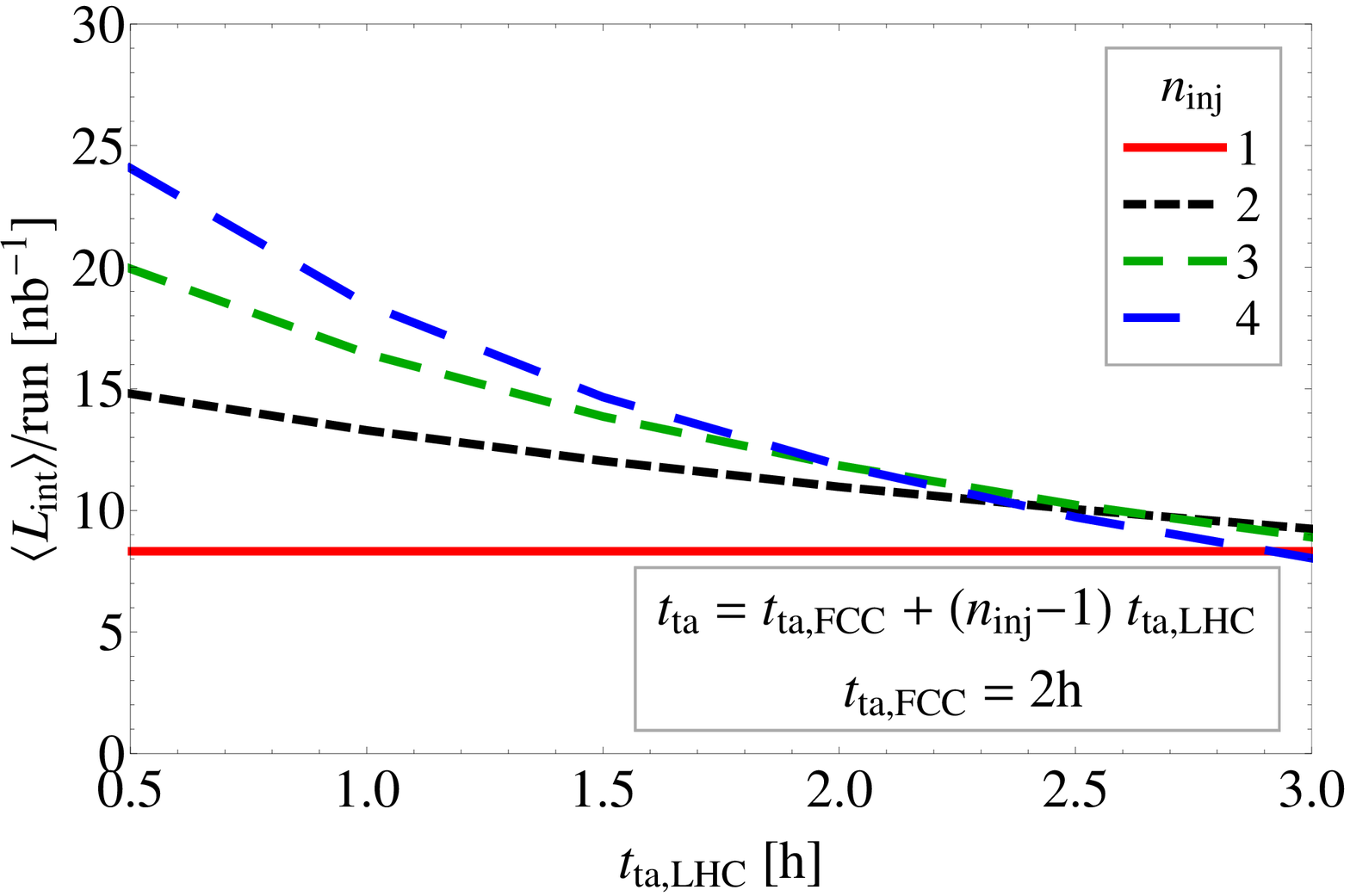}
		\caption{\label{f_fcc_optimisingLumiIntC} Optimised integrated luminosity. }
	\end{subfigure}
\caption[Pb-Pb optimising integrated luminosity.]{\label{f_fcc_optimisingLumiInt} (\subref{f_fcc_optimisingLumiIntA}) Average integrated luminosity per hour, (\subref{f_fcc_optimisingLumiIntB}) optimal time in collision, assuming different number of LHC injections, $n_\text{inj}$, (\subref{f_fcc_optimisingLumiIntC}) optimised integrated luminosity in a 30 days Pb-Pb run.}

\end{figure}

The simulations show that the luminosity evolution is not symmetric to the maximum, but  it drops rather slowly once the balanced regime is reached. Depending on the turnaround time, $t_\text{ta}$, it is advantageous to dump the beams before all particles are burned-off and refill. The turnaround time is the time required to go back into collision after a beam abort. 
The average integrated luminosity defined as
\begin{eqnarray}
\langle L_\text{int} \rangle = \frac{1}{t_\text{coll} + t_\text{ta}} \int_0^{t_\text{coll}} L_\text{int}(t) \text{d}t
\label{eq_fcc_averageLumi}
\end{eqnarray}
can be used to estimate the luminosity outcome per hour, depending on the expected turnaround time and time in collision, $t_\text{coll}$. In fact, for a given $t_\text{ta}$ this equation can be used to find the duration $t_\text{coll}$ for which $\langle L_\text{int} \rangle$ is maximized. The Fig.~\ref{f_fcc_optimisingLumiIntA} shows $\langle L_\text{int} \rangle$/h as a function of $t_\text{coll}$. For $t_\text{ta}=\qty{2}{h}$ the maximum is reached after around $t_\text{coll}= \qty{3}{h}$, which is about the time when the luminosity has decreased back to its initial value. Under optimal running conditions, without failures and early beam aborts, from this point on it is more efficient to dump and refill, rather than collecting at low rates.
As Fig.~\ref{f_fcc_optimisingLumiIntC} displays, around \qty{8}{nb^{-1}} (red solid line) could be collected during such an idealised 30 days Pb-Pb run. It is assumed that only one injection with two beams of 432 bunches each is taken from the LHC.

In general, a maximum of four injections would fit into the FCC.  The total turnaround time consists of two components,
\begin{eqnarray*}
t_\text{ta} = t_\text{ta,FCC} + (n_\text{inj} -1) t_\text{ta,LHC}
\end{eqnarray*}
 firstly $t_\text{ta,FCC}$, including everything  done in the FCC (cycling to go back to injection energy, ramp, preparing collisions etc.), and secondly $t_\text{ta,LHC}$, being the time between injections. $n_\text{inj}$ is the number of LHC injections. The current LHC turnaround time is on average about $t_\text{ta,LHC}=\qty{3}{h}$. Consequently, the already injected bunches would have to wait many hours at the FCC injection plateau. At this energy, the Pb bunches lose about $R_\text{loss}=5\%$ of their intensity per hour from IBS. For more intense bunches, the loss rate is enhanced. Approximating the intensity loss at the injection plateau as linear and neglecting losses during $t_\text{ta,FCC}$, the total colliding beam intensity can be estimated with
\begin{eqnarray*}
N_\text{beam} = \kb \Nb \sum_{i=1}^{n_\text{inj}} (1-R_\text{loss} t_\text{ta} (i-1)).
\end{eqnarray*}
Dividing this by the injected beam intensity, $n_\text{inj}\kb\Nb$, gives the fractional part of the intensity surviving until collision. Taking into account that $\mathcal{L} \propto N^2$, one can approximate that the potential luminosity is reduced by a factor $(N_\text{beam}/n_\text{inj}\kb\Nb)^2$ due to IBS at the injection plateau. 

Multiplying $\langle L_\text{int} \rangle$ by this factor leads to the estimates of $\langle L_\text{int} \rangle$/run shown in Fig.~\ref{f_fcc_optimisingLumiIntC} for up to $n_\text{inj} =4$. The corresponding optimal time in collision is displayed in Fig.~\ref{f_fcc_optimisingLumiIntB}. The total luminosity per run is shown as a function of the LHC turnaround time. This in an essential quantity to be improved for FCC, as it significantly influences the operation strategy. $t_\text{ta,FCC} =\qty{2}{h}$ is assumed. The plot makes clear, that the longer $t_\text{ta,LHC}$ the less attractive it becomes to inject more than once. It has to be considered that the larger $n_\text{inj}$, the higher the risk of losing an LHC fill during the injection process. This would lengthen the injection plateau by several hours and hence reduce the achievable luminosity. Moreover, for shorter $t_\text{ta,FCC}$, the crossing point of the curves shifts to the left, meaning that even for faster LHC cycles the potential luminosity outcome might be higher for fewer injections per fill. The unknown turnaround time imposes a large uncertainty on the estimates of $\langle L_\text{int} \rangle$ per hour and run. Any operational problems leading to delays will reduce the overall efficiency and reduce the estimated performance.

Table~\ref{t_fcc_lumiPb} collects the numerical values for the initial, peak and integrated luminosity per fill in Pb-Pb operation. The values quoted are taken from the simulation including coupling, to treat the most realistic case. The optimisation is taken into account and the values are given for $n_\text{inj}=1$, $t_\text{coll}=\qty{3}{h}$, $t_\text{ta}=\qty{2}{h}$ and $t_\text{run}=\qty{30}{days}$. The initial luminosity value is already 2.6 times over the nominal LHC, the peak could go up to around 7 times nominal LHC, which would be of the order of the requested LHC Pb-Pb luminosity for Run~3.

\begin{table}
\caption[Pb-Pb luminosity.]{\label{t_fcc_lumiPb} Pb-Pb luminosity. The maximum integrated luminosity per bunch calculated with Eq.~\eqref{eq_maxLint} is $L_\text{int,fill}=\qty{0.235}{\mu b^{-1}}$.}
\begin{ruledtabular}
\begin{tabular}{lccc}
   & Unit& per Bunch & \kb\ Bunches \\
\colrule
 $\mathcal{L}_\text{initial}$ & [Hz/mb] & 0.006 &	2.6  \\
 $\mathcal{L}_\text{peak}$ & [Hz/mb] &   0.017 & 7.3 \\
 $L_\text{int,fill}$ & [$\mu$b$^{-1}$] & 0.13 & 57.8  \\
 $L_\text{int,run}$ & [nb$^{-1}$] & 0.02 & 8.3  \\ 
\end{tabular} 
\end{ruledtabular}
\end{table}

\subsubsection{Luminosity Lifetime}
The luminosity lifetime is defined as the time at which the luminosity has decreased to $1/e$ of its initial value. This time can easily be extracted from the simulated data, by searching for the first time where \mbox{$\lumi(t=\tau_{\mathcal{L}}) \leq \lumi(t=0)/e$}:
\begin{equation*}
\tau_{\mathcal{L}} = \qty{6.2}{h},
\end{equation*}
with one experiment in collisions including burn-off, radiation damping and IBS. In case of two exactly opposite experiments, taking data under the same conditions, the luminosity lifetime will decrease accordingly, since the particle losses per turn are doubled.

\subsection{Beam-Beam Tune Shift}
The two beams travel in separated beam pipes. Only in the interaction regions they do pass through a common pipe to bring them into collisions in the local experiment. In those regions of interaction the beams exert electromagnetic forces on each other, the so-called beam-beam force. 
Especially during the passage of one bunch through the other in the IP, during a so-called \textit{head-on} interaction, those forces can be very strong. In a simplified picture, each single particle of one bunch receives a kick from the opposite beam and is deflected by a certain angle. In a linear approximation this kick acts as a quadrupole lens and thus introduces a tune shift, which can be approximated with the linear beam-beam parameter $\xi$ \cite{HandbookAccelPhys}:
\begin{eqnarray}
\xi_{i,u} &=& \frac{N_{b,j} r_{p0} Z_{i} Z_{j} \bstar}{2\pi A_{\text{ion},i} \gamma_i \sigma_{j,u}(\sigma_{j,u} + \sigma_{j,v})},
\label{eq_BeamBeamParameterLong}
\end{eqnarray}
where the beam receiving the kick is labelled with $i$ and the beam exerting the force is labelled $j$. $u$ and $v$ describe the two transverse planes. $r_{p0}$ is the classical proton radius, $Z$ and $A_\text{ion}$ the charge and atomic mass number of the corresponding beams, $\sigma$ the beam size in the corresponding plane.

For equal and round beams Eq.~\eqref{eq_BeamBeamParameterLong} simplifies to
\begin{eqnarray}
\xi &=& \frac{\Nb r_0 \bstar}{4\pi \gamma \sigma^2} =  \frac{\Nb r_0}{4\pi \epsilon_n} = \enum{3.7}{-4},
\label{eq_BeamBeamParameterShort}
\end{eqnarray}
with $r_0$ as the classical radius of the considered particle and $\epsilon_n$ the normalised emittance. As it is easy to see, this equation only depends on the beams themselves and is independent of energy and lattice parameters. 
Equations~\eqref{eq_BeamBeamParameterLong} and \eqref{eq_BeamBeamParameterShort} describe the tune shift introduced due to one head-on collision per turn, if the beams collide in more than one place, $\xi$ has to be multiplied by the number of experiments in which the investigated bunch is colliding.
The numeric value in Eq.~\eqref{eq_BeamBeamParameterShort} was obtained with the initial parameters given in Table~\ref{t_fcc_beamparameters}.

\begin{figure}
		\includegraphics[width=0.45\textwidth]{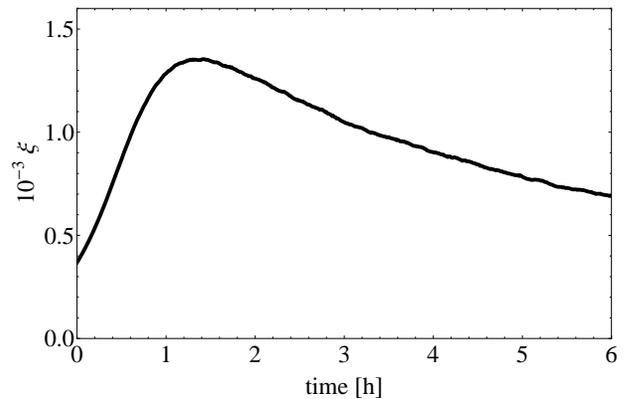}
\caption[Beam-beam tune shift evolution for one experiment.]{\label{f_fcc_beambeamtuneshift} Evolution of the beam-beam tune shift for one experiment in collision.}
\end{figure}

The beam-beam tune shift can be a limiting factor for the luminosity, since, if it becomes too large, the particles could cross resonances and get lost. If this is the case, the intensities have to be reduced or the emittances blown-up to force the tune shift below its limit, consequently the luminosity will be reduced simultaneously.  \Nb~and $\epsilon_n$ change during the fill and thus the beam-beam tune shift. This is especially important in the case discussed here, since with the damped emittance, the tune shift increases. From the simulation results displayed in Fig.~\ref{f_fcc_beamEvolution} the intensity and emittance evolutions are combined to determine the variation of the beam-beam tune shift during a fill with one experiment in collisions, see Fig.~\ref{f_fcc_beambeamtuneshift}.

The peak value reaches $\xi \approx \enum{1.4}{-3}$. If more than one experiment is taking data, this tune shift should be multiplied by the number of experiments. However, this is not exactly true for the curve in Fig.~\ref{f_fcc_beambeamtuneshift}, since it was obtained from the simulated beam evolution considering one active IP. The curve would change slightly (to lower values), due to the faster intensity burn-off and thus the beam evolution for two or more experiments would be different.

Only during operation does it become certain where the beam-beam limit of a collider exactly is. For the p-p operation in the LHC, for instance, a beam-beam limit of 0.015 was expected, based on the Sp$\bar{\rm{p}}$S experience. Nevertheless, the tune shifts achieved in p-p in dedicated experiments exceeded the nominal value by almost a factor of 5 and the value reached in normal operation by already a factor of 2 \cite{MSchaumann_MasterThesis}. 

Comparing to those factors, and taking into account that the usual tune stability in the LHC is in the order of $10^{-3}$, the beam-beam tune shift in Pb-Pb operation for FCC is not negligible, but probably also not at the limit.

\subsection{\label{s_fcc_bfpp} Bound-Free Pair Production Power}

Ultraperipheral electromagnetic interactions dominate the total cross-section during heavy-ion collisions, see Eq.~\eqref{eq_fcc_crosssection}, and cause the initial intensity to decay rapidly \cite{ABaltz_EMinteractions}. The most important interactions in Pb collisions are Bound-Free Pair-Production (BFPP)
\begin{eqnarray*}
^{208}\text{Pb}^{82+} + ^{208}\text{Pb}^{82+} \longrightarrow ^{208}\text{Pb}^{82+} + ^{208}\text{Pb}^{81+} + \text{e}^+
\end{eqnarray*}
 and Electromagnetic Dissociation (EMD)
\begin{eqnarray*}
^{208}\text{Pb}^{82+} + ^{208}\text{Pb}^{82+} \longrightarrow ^{208}\text{Pb}^{82+} + ^{207}\text{Pb}^{82+} + \text{n}.
\end{eqnarray*}
Those interactions change the charge state or mass of one of the colliding ions, creating a secondary beam emerging from the collision point. The resulting momentum deviation of the secondary beam lies outside the momentum acceptance of the ring, resulting in an impact on the beam screen in a localised position (depending on the lattice) most probably around a superconducting magnet downstream the IP. This occurs on each side of every IP where ions collide.

Following Eq.~\eqref{eq_LumiEventRate}, the production rate of those processes is proportional to the instantaneous luminosity and will thus change during the fill. Nevertheless, the magnets would suffer from a continuous high exposure. Already under LHC conditions, the risk of quenching a superconducting magnet due to these losses is high \cite{RBruce_BFPP}. In the FCC the peak luminosity could be an order of magnitude higher, increasing the risk even further. The power, $P$, in those secondary beams can be calculated as the production rate times the particle energy:
\begin{eqnarray*}
P =  \sigma_c \mathcal{L} \gamma m_\text{ion} c^2.
\end{eqnarray*}

\begin{figure}
		\includegraphics[width=0.45\textwidth]{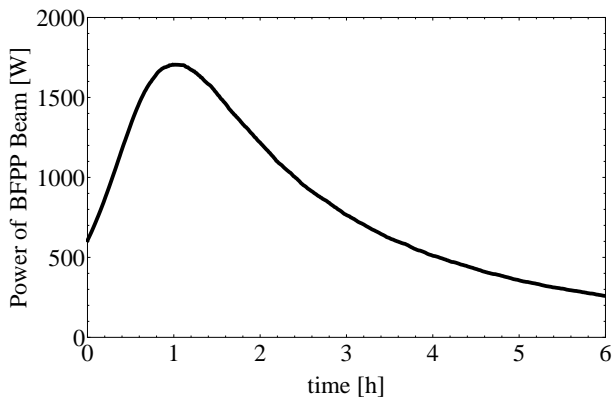}
\caption[BFPP beam power evolution in Pb-Pb operation.]{\label{f_fcc_bfppBeamPowerEvolution} BFPP1 beam ($^{208}$Pb$^{81+}$ ions) power evolution in Pb-Pb operation.}
\end{figure}

Figure~\ref{f_fcc_bfppBeamPowerEvolution} shows the power evolution of the BFPP1 beam ($^{208}\rm{Pb}^{81+}$ ions, capture of one e$^-$), which has the highest cross-section and accordingly the highest intensity and damage potential.
For the calculation the total BFPP cross-section, $\sigma_\text{BFPP} = \qty{354}{b}$ at \qty{50Z}{TeV}, estimated with \cite{HMeier_BFPP}, was used. 
The probability of higher order interaction, i.e., capturing two or more electrons and leading to a charge state of $\leq 80^+$ is much smaller and ignored for the purpose of estimating the upper limit of the stored power.

For the computation of the beam power, the simulated luminosity, shown in Fig.~\ref{f_fcc_beamEvolution}, was used. The maximum power goes up to $P \approx \qty{1.7}{kW}$, but already the initial value of \qty{600}{W} would lead to quenches and prevent from operating the machine.  Depending on the aperture and optics in the FCC, the EMD1 beam ($^{207}\rm{Pb}^{82+}$ ions, emission of one neutron) might as well hit the beam screen, depositing additional energy. 
For comparison, the BFPP1 beam power in the nominal LHC is about \qty{26}{W}, which is already expected to cause operational problems and, possibly, long-term damage. Countermeasures would definitely be required to absorb those particles before they can impact on the superconducting magnets. It has to be studied, if a highly resistant collimator in the dispersion suppressor region, as discussed for HL-LHC heavy-ion operation \cite{JJowett_RLIUP, AZlobin_11Tstatus}, would be sufficient to stop the beams produced in the collisions at the highest energy of the FCC.

\section{Proton-Lead Operation}
\subsection{Beam and Luminosity Evolution}
IBS approximately scales with \mbox{$r_0^2 \propto (Z^2/A_\text{ion})^2$} and is therefore weaker for protons than for lead ions. In fact, IBS is negligible for the (initial) proton beam parameters used in p-Pb operation at top energy. The radiation damping rates in Eq.~\eqref{eq_raddampRateSimplifyed} show a dependence on the particle type as $(\Eb Z)^3 r_0/m_\text{ion}^3 \propto Z^5/A_\text{ion}^4 $. Calculating this ratio shows that the radiation damping for lead is about twice as fast as for protons at the same equivalent energy. Thus, the emittances of both beams evolve with different time constants. Consequently, eight differential equations, four per beam, have to be solved simultaneously to describe the beam and luminosity evolution for p-Pb collisions. Those could be reduced to six equations by assuming fully coupled transverse motion and round beams, in this case $\epsilon(t) = \epsilon_x(t) = \epsilon_y(t)$ holds at all times.
Rewriting Eq.~\eqref{eq_Luminosity} under this approximation leads to the instantaneous luminosity for p-Pb
\begin{eqnarray}
\mathcal{L} = A\frac{\Nb(\text{p})\Nb(\text{Pb})}{ \epsilon(\text{p}) +\epsilon(\text{Pb}) },
\label{eq_LuminosityPPb}
\end{eqnarray}
with $A= f_\text{rev} \kb /(2 \pi \bstar)$. With this, the differential equation system follows
\begin{eqnarray}
\frac{\text{d}\Nb(i)}{\text{d}t} &=&  -\sigma_{c,\text{tot}} A \frac{\Nb(j)\Nb(i)}{\epsilon(j)+\epsilon(i)}
\label{eq_intensityEvolutionDiffppb}\\
\frac{\text{d}\epsilon(i)}{\text{d}t}& =& \epsilon(i) (\aibsxy(i) -\araddxy(i)) \\
\label{eq_emittanceEvolutionDiffppb}
\frac{\text{d}\sigs(i)}{\text{d}t}& =& \frac{1}{2} \sigs(i)(\aibss(i) -  \aradds(i)),
\label{eq_bunchlengthEvolutionDiffppb}
\end{eqnarray}
where only the three equations of beam $i$ (either Pb or p) are noted.  The equations for beam $j$ have an equivalent form with different initial conditions and growth rates. The dependences of the IBS growth rates on \Nb, $\epsilon$ and \sigs~couple the three equations for each beam. The dependence of the luminosity on both beams' emittances and intensities couple the Pb and p beam evolution. An exact analytic solution of this coupled differential equation  system does not exist. Unfortunately, the CTE program does not feature simulations with different particle species, so only approximated analytical and numerical solutions of the ODE system are available to perform estimates.

At the beginning of the fill, $\aibs\ll \aradd$ and it can be approximated that $\alpha_{\epsilon}=\alpha_{\epsilon}(Pb) =2 \alpha_{\epsilon} (p) = \text{const.}$ in all three planes.
This constant emittance decay rate, neglects the dynamically changing IBS with damped emittance. As seen in the Pb-Pb analysis, the peak and integrated luminosity estimates done under those assumptions will be overestimated, due to the emittances asymptotically approaching zero.

In general, the proton beams are more intense compared to lead. In the LHC proton-proton operation, bunches with $10^{11}$ particles are regularly used. Lead bunches have in the order of $10^8$ particles. In proton-lead operation, it is possible to increase the initial luminosity by increasing the proton intensity (the lead intensity is assumed to be at the limit). Nevertheless,  the higher the proton intensity, the stronger the beam-beam effects in those strong-weak interactions. Therefore, it was chosen for the LHC proton-lead run in 2013 \cite{JJowett_IPAC13} to use proton intensities around 10\% of the nominal value used in p-p operation. This should also be the baseline for p-Pb collision mode in FCC-hh. 

In each collision of a proton with a lead ion, those two particles are removed from their beams. Therefore, the maximum integrated luminosity is reached when each lead ion has found a collision partner in the more intense proton beam. The number of lead ions is only about 1\% of the number of protons. In the limit of burning-off all the lead, the proton intensity is hardly changed and could be considered as roughly constant through the whole fill. 

To find an approximated analytical equation for the proton-lead luminosity evolution, the following assumptions are made:
\begin{eqnarray}
\Nb(\text{Pb}) \ll \Nb(\text{p}) &=& N_{b0}(\text{p}) = \text{const.}
\label{eq_fcc_ppB_approxN}\\
\alpha_{\epsilon} = \alpha_{\epsilon}(\text{Pb}) &=& 2 \alpha_{\epsilon} (\text{p}) \approx -\aradd(\text{Pb})
\label{eq_fcc_ppB_approxIBS} \\
\aradds &=& 2 \araddxy
\label{eq_fcc_ppB_approxRad}
\end{eqnarray}
where Eq.~\eqref{eq_fcc_ppB_approxIBS} is assumed for all three planes and Eq.~\eqref{eq_fcc_ppB_approxRad} follows from Eq.~\eqref{eq_raddampRateSimplifyed}.
Applying those constrains to the differential equations~\eqref{eq_intensityEvolutionDiffppb} - \eqref{eq_bunchlengthEvolutionDiffppb} leads to an exponential behaviour of the emittance and bunch length of both beams with related time constants
\begin{eqnarray}
\epsilon(\text{Pb},t)& =& \epsilon_{0}(\text{Pb}) \exp[\alpha_{\epsilon}t]
\label{eq_emittanceEvolutionppbPb}\\
\sigs(\text{Pb},t) &=& \sigma_{s0}(\text{Pb}) \exp[\alpha_{\epsilon}t]\\
\epsilon(\text{p},t)& =& \epsilon_{0}(\text{p}) \exp[\alpha_{\epsilon}t/2]
\label{eq_emittanceEvolutionppbP}\\
\sigs(\text{p},t) &=& \sigma_{s0}(\text{p}) \exp[\alpha_{\epsilon}t/2],
\end{eqnarray}
where the emittance growth rate of the Pb beam is taken as the reference, $\alpha_\epsilon \approx -\araddxy(\text{Pb})$. This value is negative, hence those are exponential decays. 
The proton intensity was assumed to be time independent, thus
\begin{eqnarray}
\Nb(\text{p}, t)= N_{b0}(\text{p}). 
\label{eq_intensityEvolutionppbP}
\end{eqnarray}
To solve the last equation for the Pb intensity evolution, Eq.~\eqref{eq_emittanceEvolutionppbPb}, \eqref{eq_emittanceEvolutionppbP} and \eqref{eq_intensityEvolutionppbP} are inserted into Eq.~\eqref{eq_intensityEvolutionDiffppb}, followed by applying the method of separation of variables. The solution of the arising integral is
\begin{eqnarray*}
\ln \left( \Nb(\rm{Pb},t) \right) = \int \frac{\text{d}x}{x^2 (ax +b)} = -\frac{1}{bx}+\frac{a}{b^2}\ln \left(\frac{ax+b}{x}\right)
\end{eqnarray*}
with $x=\exp[\alpha_\epsilon t/2]$. The final result is
\begin{widetext}
\begin{eqnarray*}
\Nb(\text{Pb},t) = N_{\text{Pb}} e^{- \frac{2 \sigma_{c,\text{tot}} A N_{\text{p}}}{\alpha_\epsilon \epsilon_\text{p}^2} \left( \epsilon_\text{p} (\exp[-\alpha_\epsilon t/2]-1) +   \epsilon_{\text{Pb}} \ln[\epsilon_\text{p}+\epsilon_{\text{Pb}}]- \epsilon_{\text{Pb}}\ln[\epsilon_\text{p}\exp[-\alpha_\epsilon t/2] + \epsilon_{\text{Pb}}] \right) }.
\end{eqnarray*}
\end{widetext}

\begin{figure*}
	\begin{subfigure}[h]{1\textwidth}
	\centering
		\begin{minipage}{0.45\textwidth}
		\includegraphics[width=1\textwidth]{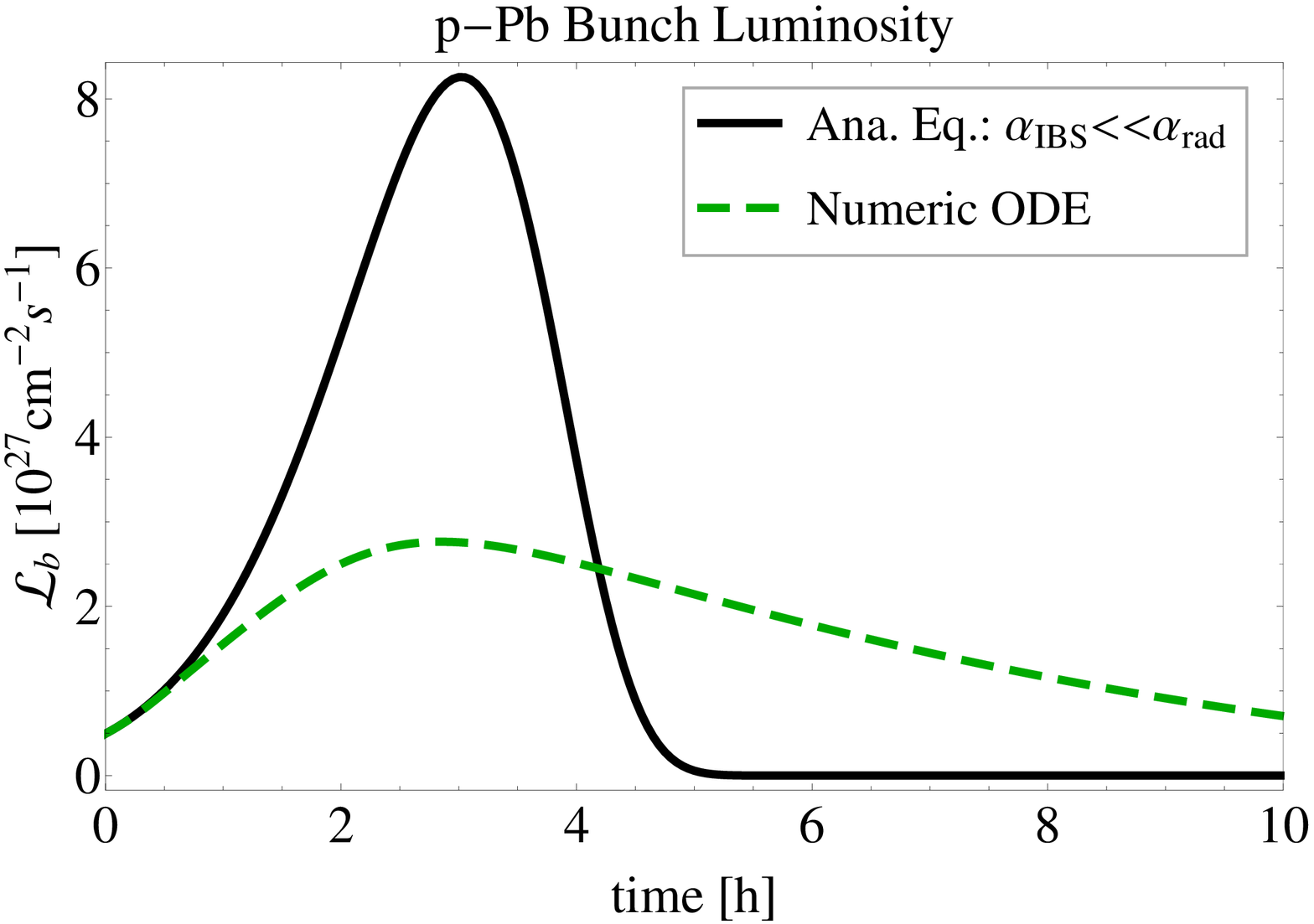}
		\end{minipage}
		\begin{minipage}{0.45\textwidth}
		\includegraphics[width=1\textwidth]{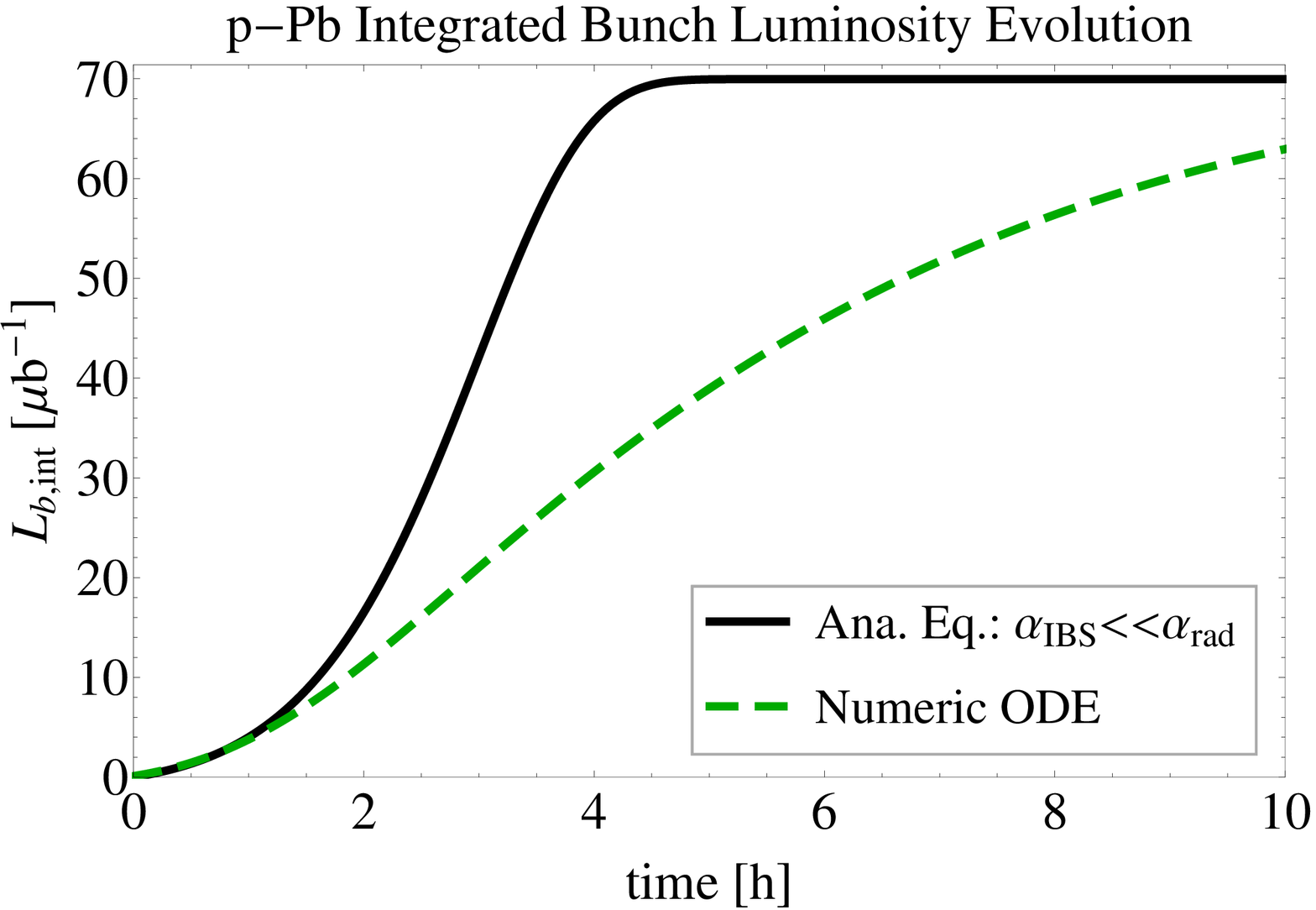}
		\end{minipage}
			\vspace*{0.5cm}
	\end{subfigure}

	\begin{subfigure}[h]{1\textwidth}
	\centering
		\begin{minipage}{0.45\textwidth}
		\includegraphics[width=1\textwidth]{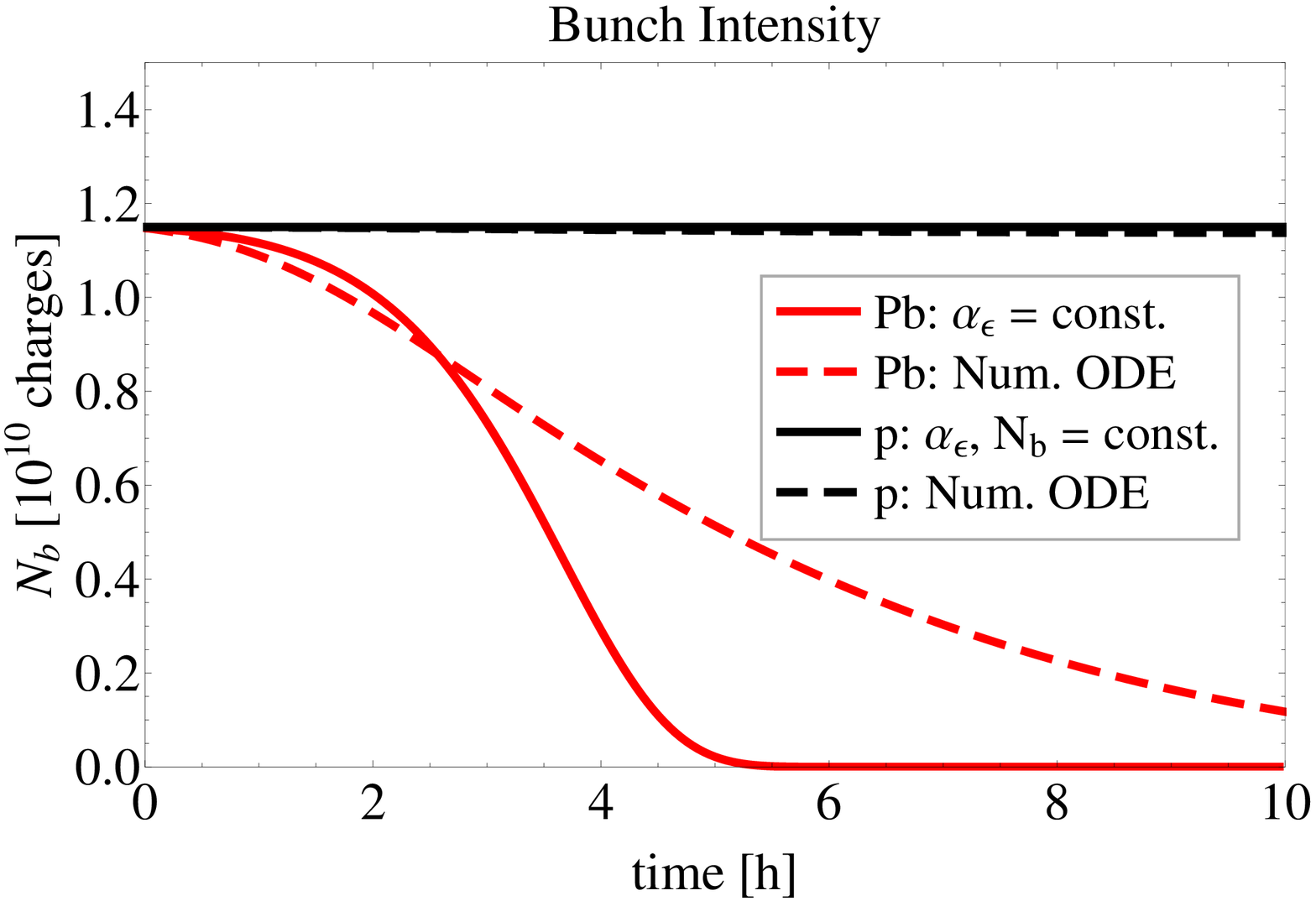}
		\end{minipage}
		\begin{minipage}{0.45\textwidth}
		\includegraphics[width=1\textwidth]{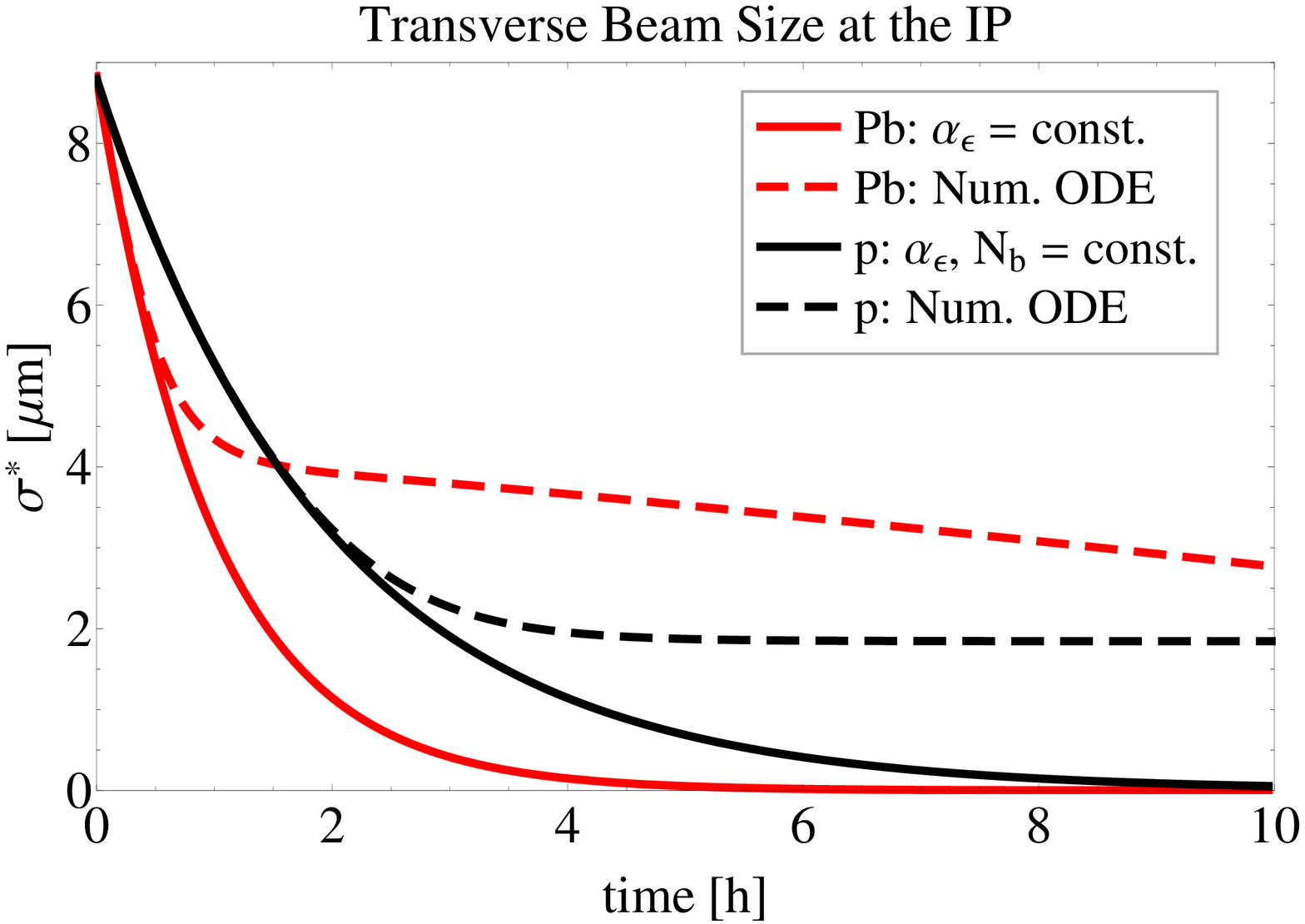}
		\end{minipage}
	\end{subfigure}

	\vspace*{0.5cm}
	\begin{subfigure}[h]{1\textwidth}
	\centering
		\begin{minipage}{0.45\textwidth}
		\includegraphics[width=1\textwidth]{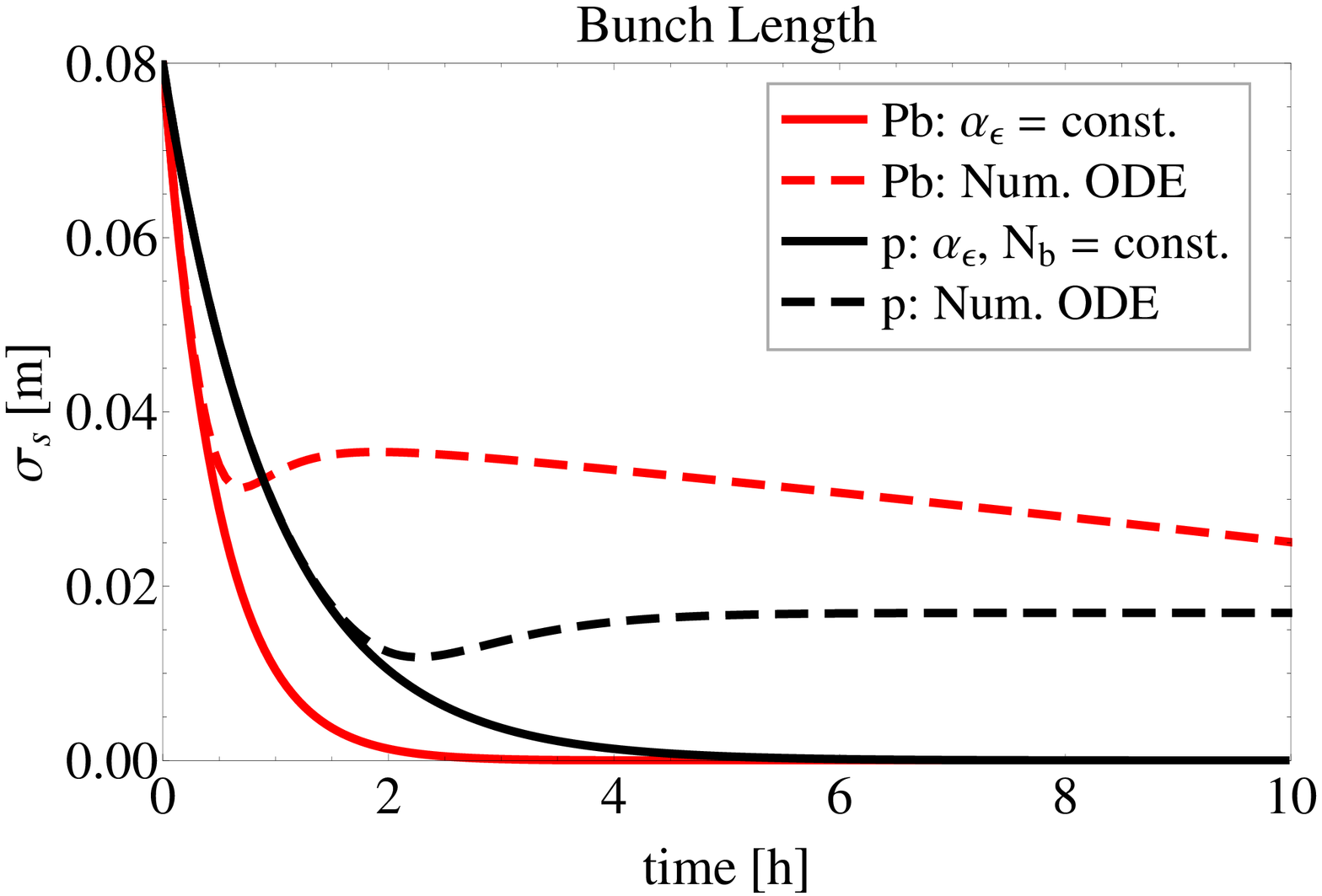}
		\end{minipage}
	\end{subfigure}
\caption[p-Pb beam and luminosity evolution.]{\label{f_fcc_ppbBeamEvolution}  p-Pb beam and luminosity evolution for one experiment in collisions. Top: instantaneous (left) and integrated (right) bunch luminosity, middle: intensity (left) and beam size at the IP (right), bottom: bunch length for the proton (black) and lead (red) beam. Approximated analytic calculations (solid lines), neglect the dynamically changing IBS, leading to unrealistically small beam sizes. The numerical ODE solution is shown as dashed lines, giving more realistic estimates.}
\end{figure*}

The  equations for the evolution of the emittance and intensity are inserted into Eq.~\eqref{eq_LuminosityPPb} to obtain the p-Pb luminosity evolution. Figure~\ref{f_fcc_ppbBeamEvolution} presents the results. The above derived analytical approximation is shown as the solid lines, while the dashed lines correspond to the numerical solution of the ODE system. The evolution of the intensity (middle left), beam size at the IP (middle right) and bunch length (bottom) are displayed in black for the proton and in red for the lead beam.

The peak luminosity  is shifted to later times compared to Pb-Pb operation, due to the slower radiation damping for protons, leading to longer fills. The Pb intensity burn-off is very fast, while the proton intensity hardly changes. This arises form the fact that in one collision one Pb nucleus is lost per proton.
A free knob to adjust the luminosity peak and evolution is the proton intensity. Increasing $\Nb(\text{p})$ would lead to higher initial and peak  rates followed by an even faster Pb burn-off and shorter fills. Decreasing $\Nb(\text{p})$ would distribute the achievable luminosity over a longer period with reduced peak rates.

The $1/e$-luminosity lifetime, extracted from the numerical solution of the ODE system shown in Fig.~\ref{f_fcc_ppbBeamEvolution},
determines to
\begin{eqnarray*}
\tau_{\mathcal{L}} = \qty{14.0}{h}.
\end{eqnarray*}

\subsection{Optimising the Integrated Luminosity}

Because of the weaker IBS for protons, their intensity losses at injection are smaller and the proton beam can wait in the machine without deteriorating significantly.  Therefore, the proton beam is injected first, followed by the lead. Depending on the number of injections, either both LHC rings are filled with the same species, or the filling is shared between the species and each LHC beam is injected in opposite directions in the FCC. In this way the number of particles surviving until top energy is maximised.

From the numerical solution of the ODE system, which provides the best estimate of the beam and luminosity evolution available today, the average luminosity per hour is determined. Similar to Pb-Pb, an expression for the total available p and Pb beam intensity in collision is derived, taking into account the different waiting times and loss rates at the injection plateau. The average luminosity per hour is then calculated as in Eq.~\eqref{eq_fcc_averageLumi} reduced by the factor 
\begin{equation}
(N_\text{beam}(\text{Pb})/n_\text{inj} \kb \Nb(\text{Pb}))\times (N_\text{beam}(\text{p})/n_\text{inj} \kb \Nb(\text{p}))
\end{equation}
for losses during injection.

\begin{figure}
	\begin{subfigure}[h]{0.45\textwidth}
		\includegraphics[width=1\textwidth]{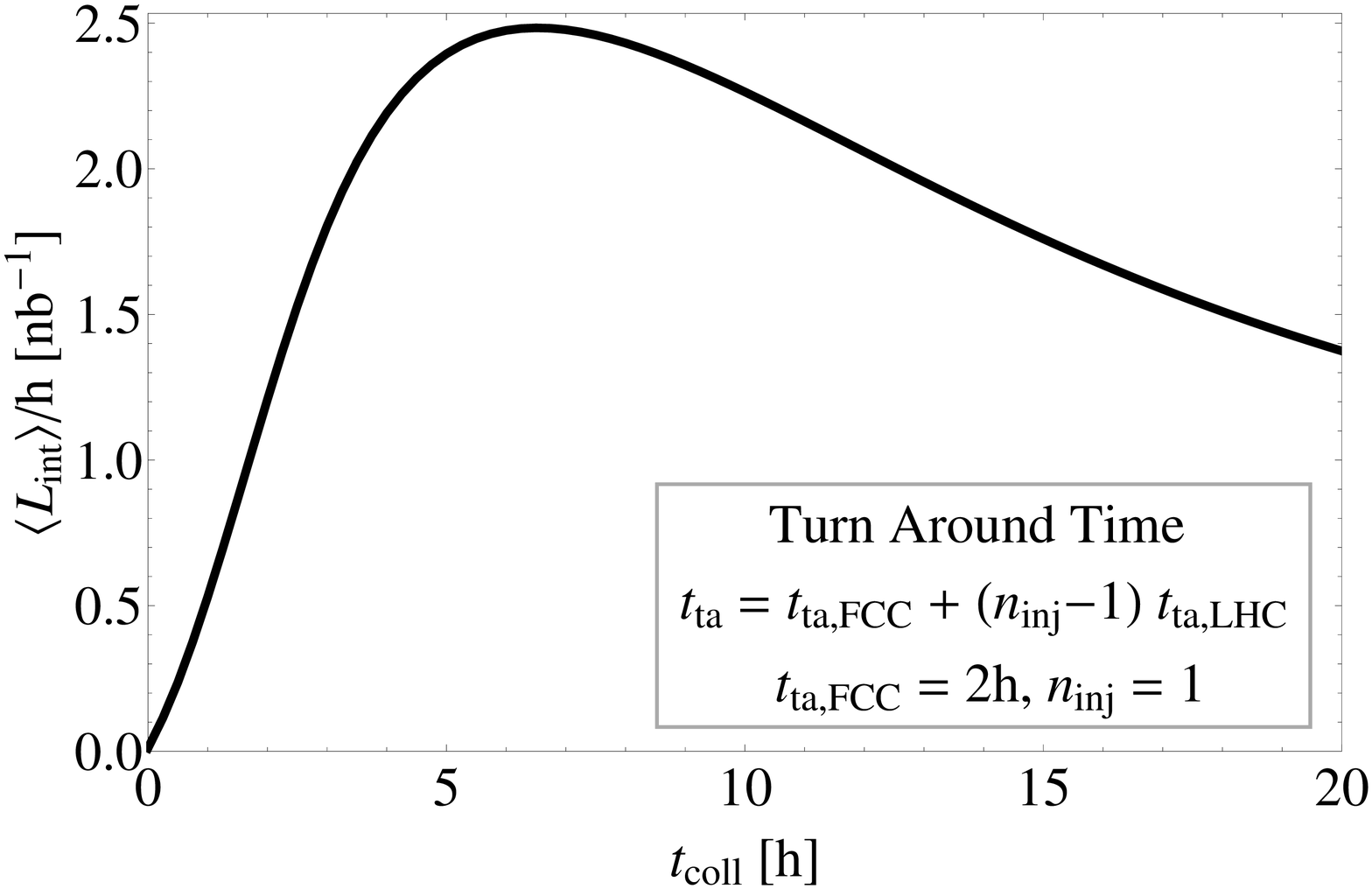}
	\caption{\label{f_fcc_optimisingLumiIntppbA} Average integrated luminosity per hour.}
	\vspace*{0.3cm}
	\end{subfigure}
	\begin{subfigure}[h]{0.45\textwidth}
		\includegraphics[width=1\textwidth]{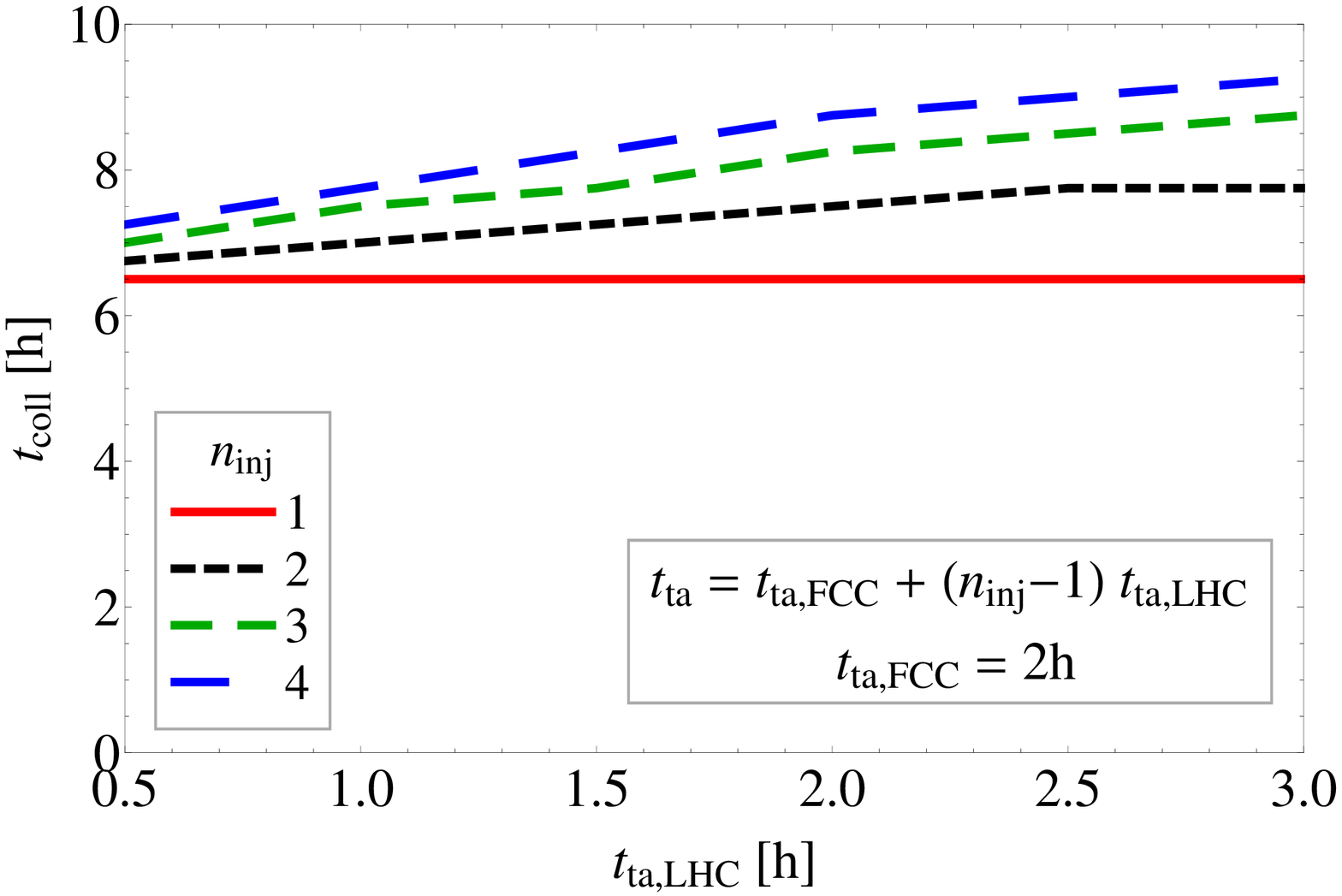}
	\caption{\label{f_fcc_optimisingLumiIntppbB} Optimal time in collision.}
	\vspace*{0.3cm}
	\end{subfigure}
	\begin{subfigure}[h]{0.45\textwidth}
	\centering
		\includegraphics[width=1\textwidth]{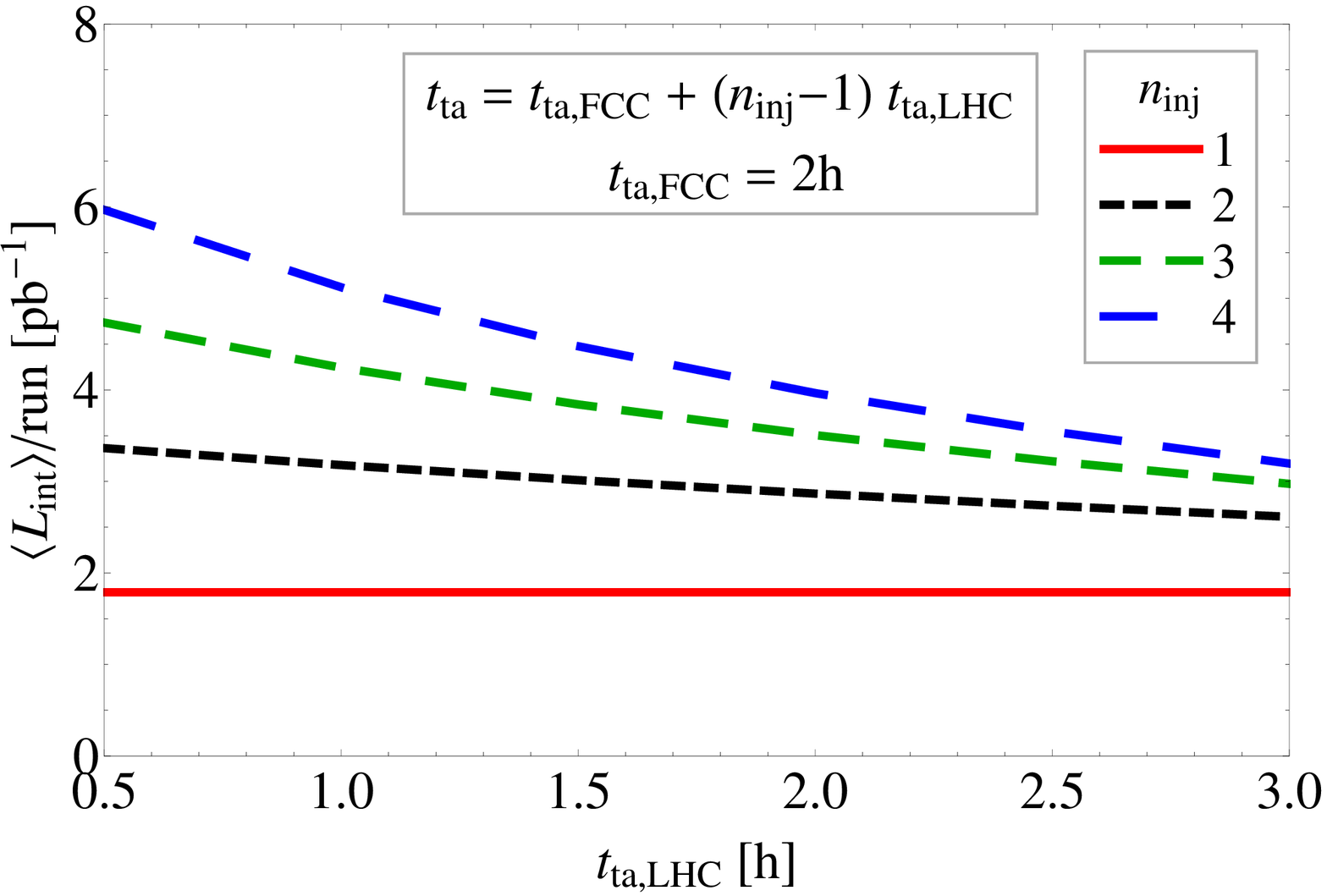}
	\caption{\label{f_fcc_optimisingLumiIntppbC} Optimised integrated luminosity.}
	\end{subfigure}
\caption[p-Pb optimising integrated luminosity.]{\label{f_fcc_optimisingLumiIntppb}  (\subref{f_fcc_optimisingLumiIntppbA}) Average integrated luminosity per hour,  (\subref{f_fcc_optimisingLumiIntppbB}) optimal time in collision,  (\subref{f_fcc_optimisingLumiIntppbC}) optimised integrated luminosity for a 30 days p-Pb run.}
\end{figure}

Figure~\ref{f_fcc_optimisingLumiIntppb} shows the results for the average integrated luminosity (\subref{f_fcc_optimisingLumiIntppbA}) and the corresponding time in collisions (\subref{f_fcc_optimisingLumiIntppbB}) to achieve the optimised integrated luminosity per \qty{30}{days} run  (\subref{f_fcc_optimisingLumiIntppbC}). For  $n_\text{inj}=1$ the maximal luminosity of \qty{1.7}{pb^{-1}}/run is reached for a fill length of \qty{6.5}{h}. This does not take into account any delays or early aborted fills changing the assumed optimised statistics. Again it becomes clear that the longer the LHC turnaround time, the less attractive it becomes to wait for more injections before colliding. The collectable luminosity decreases, due to particle losses on the lengthened injection plateau. Moreover, the optimal fill length becomes longer, enhancing the risk to be aborted ahead of schedule, potentially decreasing the predicted luminosity further.
Table~\ref{t_fcc_lumippb} summarises the initial, peak and integrated luminosity values in p-Pb operation.

\begin{table}
\begin{ruledtabular}
\caption[p-Pb luminosity.]{\label{t_fcc_lumippb} p-Pb luminosity. $\sigma_{c,tot} = \qty{2}{b}$ was used. The maximum integrated luminosity
per fill calculated with Eq.~\eqref{eq_maxLint} is $L_\text{int} = \qty{30}{nb^{-1}}$.}
\begin{tabular}{lccc}
   & Unit& per Bunch & \kb\ Bunches \\
\colrule
 $\mathcal{L}_\text{initial}$ & [Hz/mb] &  0.5  & 213  \\
 $\mathcal{L}_\text{peak}$ & [Hz/mb] &  2.8 &  1192 \\
 $L_\text{int,fill}$ & [$\mu$b$^{-1}$] & 48.7 & 21068  \\
 $L_\text{int,run}$ & [nb$^{-1}$] & 4.1 & 1784  \\
\end{tabular} 
\end{ruledtabular}
\end{table}

\subsection{Beam Current Lifetime}
As mentioned, the ion beam is naturally weak, while proton beams can be produced with much higher intensities. In the collision the lead beam loses $Z=82$ charges per lost proton. Thus, the ion beam will in general have the smaller beam current lifetime, i.e., faster intensity decay. Consequently, the ion beam lifetime determines the length of the fill in p-Pb operation.

The beam current lifetime is given by 
\begin{eqnarray*}
\frac{1}{\tau_N} =- \frac{1}{N}\frac{\text{d}N}{\text{d}t} = -\frac{1}{N} \sigma_{c,\text{tot}}\lumi,
\end{eqnarray*}
with $N=\kb \Nb$ and $N_b = N_b(\text{Pb})$. Inserting Eq.~\eqref{eq_LuminosityPPb} for the luminosity, the Pb beam current lifetime in p-Pb collisions is
\begin{eqnarray}
-\tau_N(\text{Pb},t) = \frac{ 2\pi \bstar (\epsilon(\text{p},t) + \epsilon(\text{Pb},t))}{\sigma_{c,\text{tot}} n_\text{exp} f_\text{rev} \Nb(\text{p})}.
\label{eq_beamcurrentlifetimePPb}
\end{eqnarray}
The first factor is constant for $\Nb(\text{p})\gg \Nb(\text{Pb})$. Hence, the lifetime only varies in time proportionally to the convoluted emittance of the two beams. As expected,  $\tau_N(\text{Pb})$ decreases with increasing proton intensity, because of the higher interaction probability.
The initial value evaluates to
\begin{eqnarray*}
\tau_N(\text{Pb}, t=0) = \qty{39.3}{h}
\end{eqnarray*}
for $n_\text{exp} =1$. Owing to the damping of the emittances, these values will decrease exponentially during the fill and lead to a much shorter fill durations. It is interesting to note that Eq.~\eqref{eq_beamcurrentlifetimePPb} is independent of the lead beam current.

\subsection{Beam-Beam Effects}
\subsubsection{Unequal Beam Sizes}
The initial beam sizes of the proton and lead beam in p-Pb operation is assumed to be equal. 
Because of the stronger radiation damping for Pb, the Pb beam size falls below the proton beam size in the first period of the fill, see Fig.~\ref{f_fcc_ppbBeamEvolution}.
After about one hour in collisions the Pb emittance reaches the balanced regime and does now change only slowly due to the intensity losses and the thus decreasing IBS rate.
Since the IBS is weaker for protons, the emittance is damped to a lower value. After about \qty{1.5}{h} in collisions the proton beam has become smaller than the Pb beam.
Over the duration of the fill, the ratio of the Pb to the proton beam size lies between \mbox{$0.8\leq \sigma_\text{Pb}/\sigma_\text{p}\leq 1.9$}. 

From experience at various past colliders, it is well known that beam lifetime can be significantly reduced when colliding beams have unequal sizes \cite{RBrinkmann_PAC93, KCornelis_SPSBB, MSyphers_BB}.
It is observed that the lifetime of the larger beam decreases with decreasing size of the opposite beam. 
This is because the particles in the beam with the larger size see more of the non-linear part of the beam-beam force exerted by the smaller beam, where they are affected by higher order resonances \cite{KCornelis_SPSBB}.

In case of the FCC p-Pb operation, the difference in beam size stays below 20\% over the first two hours, including a reversal of the rank. 
For the later part of the fill, the Pb beam is the larger and thus might suffer from a lifetime reduction. 
Considering the optimum fill length of \qty{6.5}{h} and the evolution of the luminosity in Fig.~\ref{f_fcc_ppbBeamEvolution}, the largest fraction of the luminosity is integrated in the first four hours of the fill. 
Comparing this to the discussed evolution of $\sigma_\text{Pb}/\sigma_\text{p}$, 
indicates that a potential reduction of the Pb beam lifetime due to unequal beam sizes would probably affect the luminosity only in the second half of the fill, when the collision rates have already past the maximum.

For comparison, the Tevatron ran with mismatched beam sizes between the proton and antiproton beam of around $\sigma_\text{p}/\sigma_{\bar{\text{p}}}\approx 3$ \cite{AccelPhysTevatron}.
In the first p-Pb run of the LHC in 2013, $\sigma_\text{Pb}/\sigma_\text{p}\approx 2$ was observed, while the beam lifetime was dominated by other effects \cite{JJowett_IPAC13}. 

\subsubsection{Tune Shift}
\begin{figure}
		\includegraphics[width=0.45\textwidth]{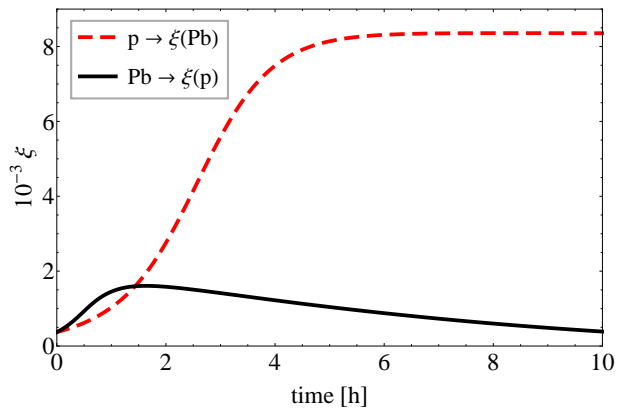}
\caption[p-Pb beam-beam tune shift.]{\label{f_fcc_BBtuneshiftppb}  p-Pb beam-beam tune shift for 1 IP in collision.}
\end{figure}

With Eq.~\eqref{eq_BeamBeamParameterLong} the beam-beam tune shift $\xi$ can be calculated for weak-strong beam-beam interactions as in the case of p-Pb collisions. The initial beam parameters in Table~\ref{t_fcc_beamparameters} are such that the number of charges and the beam sizes of both beams are approximately equal, resulting in the same tune shift, $\xi(\text{p}) \approx \xi(\text{Pb}) = 3.7 \times 10^{-4} $, at the beginning of the fill. However, the proton and lead beam properties evolve differently with time, changing the force exerted from one to the other during the fill. Figure~\ref{f_fcc_BBtuneshiftppb} shows the calculation of $\xi$ based on the numerical solution of the ODE system. The effect on the proton (black) beam is small ($\xi(\text{p})< 2 \times 10^{-3}$). The increase of $\xi(\text{p})$ due to the shrinking lead beam emittances is negated by the rapid Pb intensity losses. Owing to the almost constant proton intensity but damping emittances, the tune shift to the Pb beam becomes significant and approaches a value of $\xi(\text{Pb}) = 8.3 \times 10^{-3}$ in the regime where IBS and radiation damping start to balance each other. This value is close to the assumed beam-beam limit of $\xi = 0.01$ for p-p operation.

\section{Proton-Proton Operation}
In the following the tools used in the above analysis are applied to p-p operation in the FCC. In p-p operation two scenarios are under investigation, namely bunches spaced by 25 or \qty{5}{ns} with different beam properties. The proton beam parameters are listed in Table~\ref{t_fcc_pp_beamparameters}.

\begin{table}
\caption[Assumed beam parameters for p-p operation.]{\label{t_fcc_pp_beamparameters} Assumed beam parameters for proton-proton operation \cite{FCC1}. }
\begin{ruledtabular}
\begin{tabular}{rcccc}
 Parameter  & Symbol & Unit& \qty{25}{ns} & \qty{5}{ns} \\
\colrule
 No. of particles per bunch & \Nb & [$10^{11}$]&	1.0 & 0.2 \\
 Normalised transv. emittance & \emittn & [$\mu$m] & 2.2 & 0.44\\
 RMS bunch length & $\sigma_s$ & [m] & 0.08 & 0.08 \\
 No. of bunches per beam & \kb & - & 10600 & 53000 \\
 \bfunc\ at IP & \bstar & [m] & 1.1 & 1.1 \\
 Total cross section & $\sigma_{c,\text{tot}}$ &[mb] & 153 & 153 \\
 No. of main IPs &- & -& 2 & 2\\
 
\end{tabular} 
\end{ruledtabular}
\end{table}

Radiation damping is negligible for protons at injection energy. At \qty{50}{TeV} the transverse and longitudinal emittance radiation damping times are $1/\araddxy =\qty{1.0}{h}$ and $1/\araddxy =\qty{0.5}{h}$, respectively.  The horizontal equilibrium emittance from quantum excitation at top energy is in the order of \qty{10^{-2}}{\mu m}, which is still an order of magnitude smaller than the emittance ranges of the scenarios considered. 

\begin{figure}
	\begin{subfigure}[h]{0.45\textwidth}
		\includegraphics[width=1\textwidth]{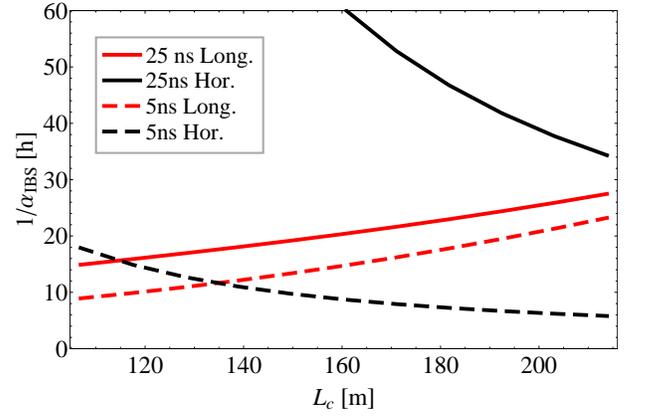}
	\caption{\label{f_fcc_pp_IBSinj} Initial IBS at injection.}
	\vspace*{0.3cm}
	\end{subfigure}
	\begin{subfigure}[h]{0.45\textwidth}
		\includegraphics[width=1\textwidth]{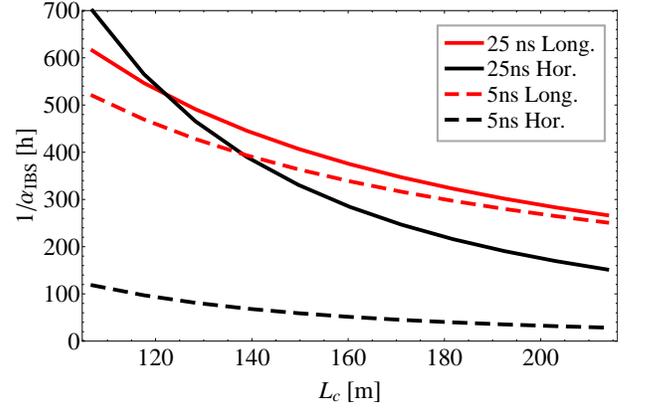}
	\caption{\label{f_fcc_pp_IBScoll} Initial IBS at top energy.}
	\end{subfigure}

\caption[p-p initial IBS growth times.]{\label{f_fcc_pp_IBS}  Initial IBS growth times and their dependence on the FODO cell length, $L_c$, at injection (\subref{f_fcc_pp_IBSinj}) and  top energy (\subref{f_fcc_pp_IBScoll}) for p-p operation.}
\end{figure}

As already explained, depending on the lattice choice, the IBS growth rates can be rather different. Figure~\ref{f_fcc_pp_IBS} shows the IBS growth times as a function of the FODO cell length, $L_c$, at (\subref{f_fcc_pp_IBSinj}) injection  and (\subref{f_fcc_pp_IBScoll}) top energy. The same behaviour as for Pb is observed, whereas the rates are lower. For the chosen baseline lattice with $L_c \approx \qty{203}{m}$  and $\gamma_T \approx 103$, the initial growth times calculated with Piwinski's algorithm are listed in Table~\ref{t_fcc_pp_ibsgrowthtimes}. IBS is in general small for the initial proton beam parameters. Only the horizontal growth time at injection is below \qty{10}{h}, which might lead to transverse emittance growth, if the time spent on the injection plateau becomes too long.

\begin{table}
\caption[p-p initial IBS growth times.]{\label{t_fcc_pp_ibsgrowthtimes} 
Initial IBS growth times for protons calculated with Piwinski's  formalism, assuming the baseline lattice ($L_c = \qty{203}{m}$). Assumption for momentum spread: injection $\sigp = \enum{1.5}{-4}$, collision (a) $\sigp = \enum{0.5}{-4}$ (obtained with $\gamma_T$ of baseline lattice), (b) $\sigp = \enum{1.1}{-4}$ (LHC design). }
\begin{ruledtabular}
\begin{tabular}{rccccccc}
 \multirow{3}{*}{Growth Times}   & \multirow{3}{*}{Unit}& 
\multicolumn{2}{c}{Injection} & \multicolumn{4}{c}{Collision}  \\

				&&	\multirow{2}{*}{\qty{25}{ns}} & \multirow{2}{*}{\qty{5}{ns}}	& \multicolumn{2}{c}{(a)} & \multicolumn{2}{c}{(b)} \\
				&& & & \qty{25}{ns} & \qty{5}{ns} &\qty{25}{ns} &\qty{5}{ns} \\
\colrule
 $1/\aibss$ & [h] & 25.9  &21.3 &	283.1  	& 264.7 &  1467 & 1534\\
 $1/\aibsx$ & [h] & 37.7 	& 6.2 & 169.5 	& 31.7 &265.4 & 55.5\\
 $1/\aibsy$ & [h] & $-10^5$ & $-10^4$& $-10^7$ &$-10^7$ & $-10^8$ & $-10^7$\\
\end{tabular} 
\end{ruledtabular}
\end{table}

\begin{figure*}
	\begin{subfigure}[h]{1\textwidth}
	\centering
		\begin{minipage}{0.45\textwidth}
		\includegraphics[width=1\textwidth]{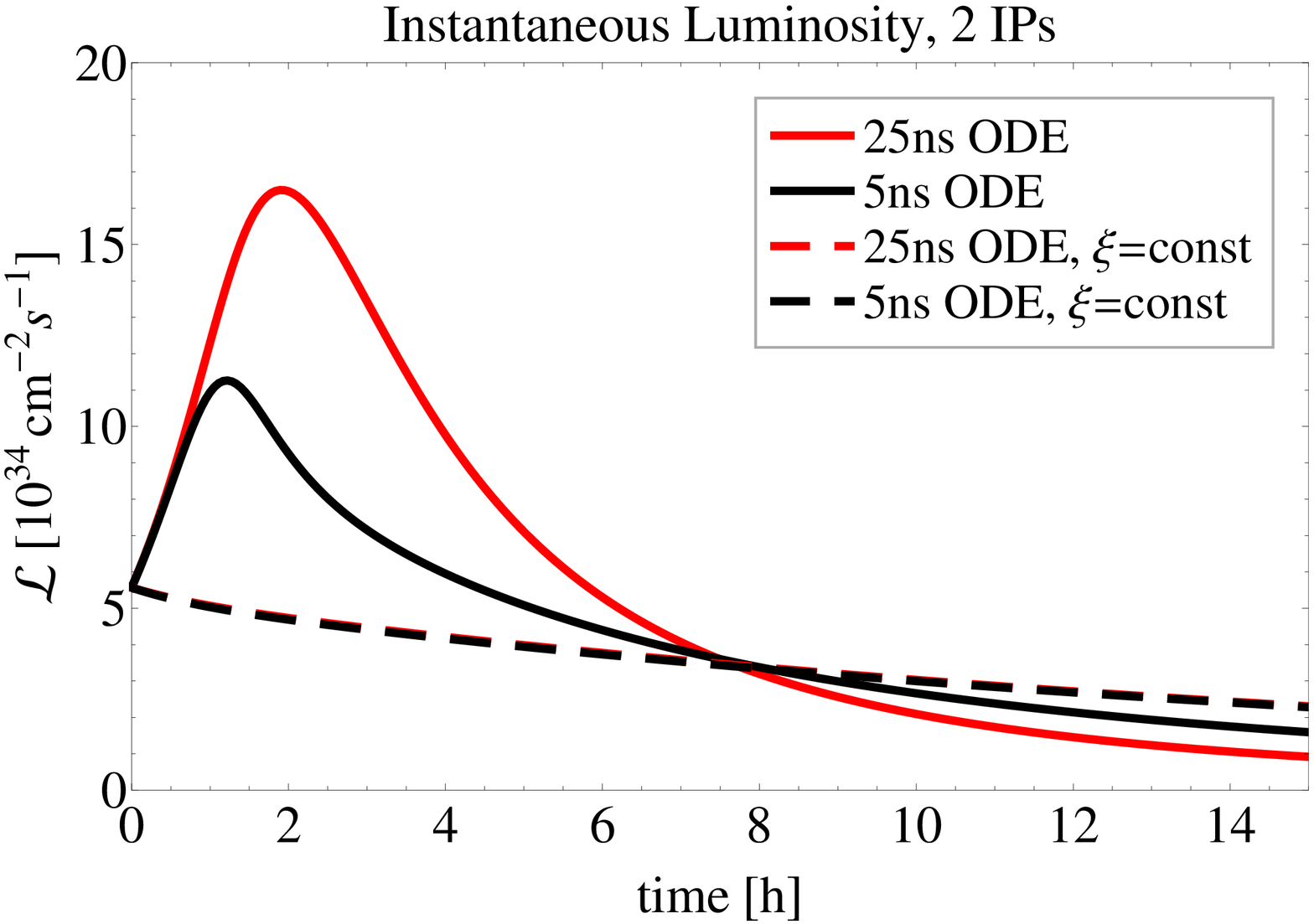}
		\end{minipage}
		\begin{minipage}{0.45\textwidth}
		\includegraphics[width=1\textwidth]{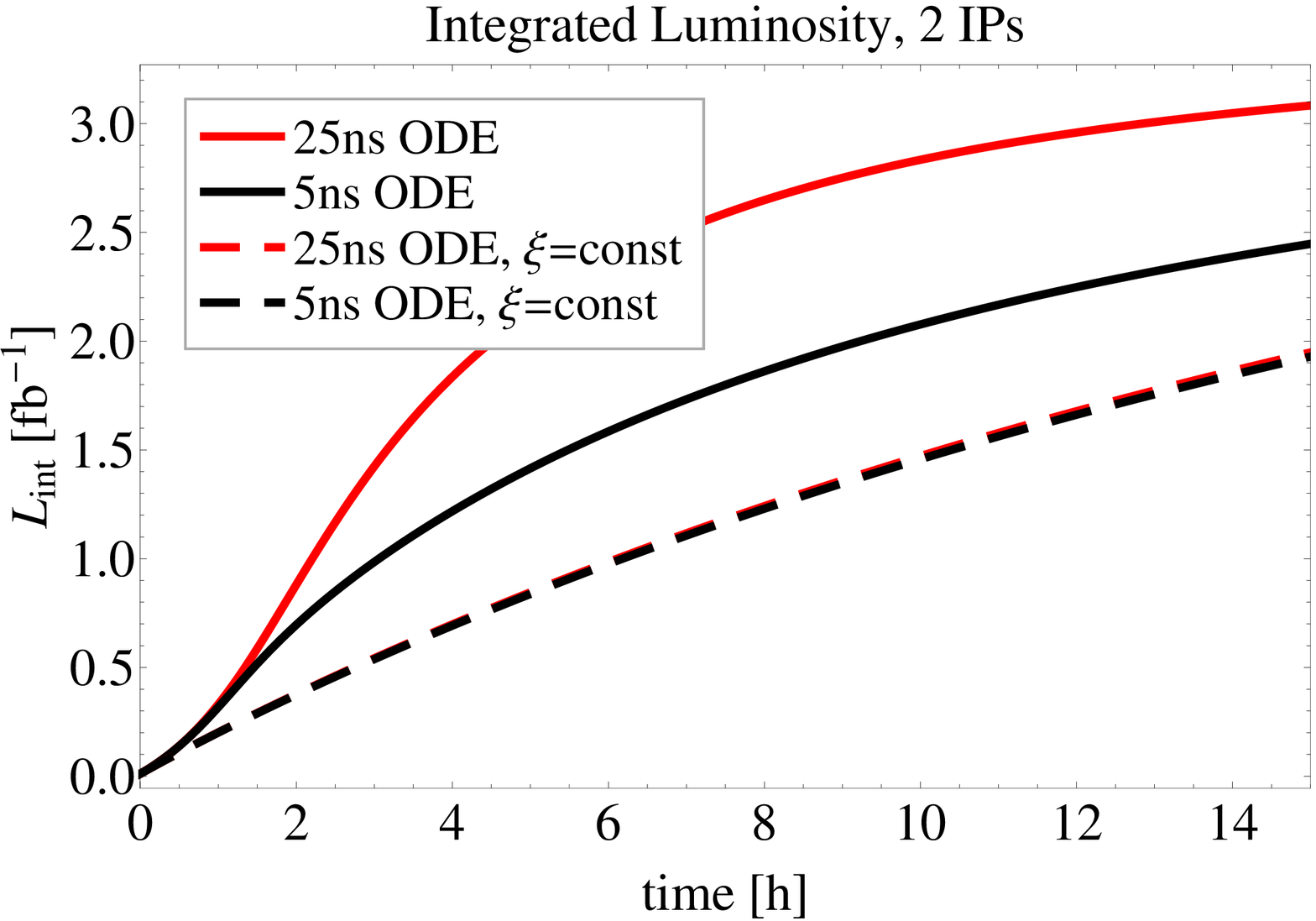}
		\end{minipage}
			\vspace*{0.5cm}
	\end{subfigure}

	\begin{subfigure}[h]{1\textwidth}
	\centering
		\begin{minipage}{0.45\textwidth}
		\includegraphics[width=1\textwidth]{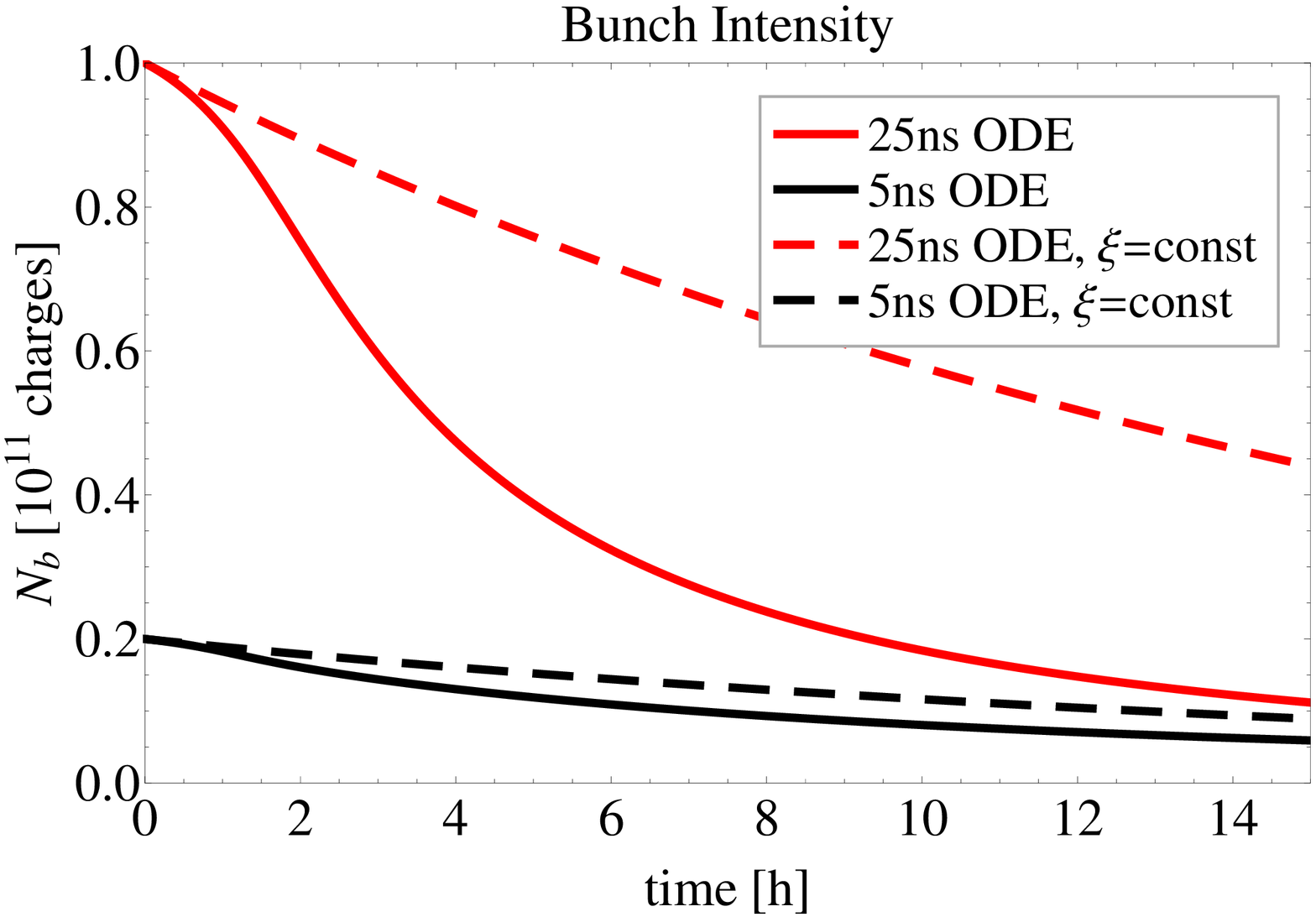}
		\end{minipage}
				\begin{minipage}{0.45\textwidth}
		\includegraphics[width=1\textwidth]{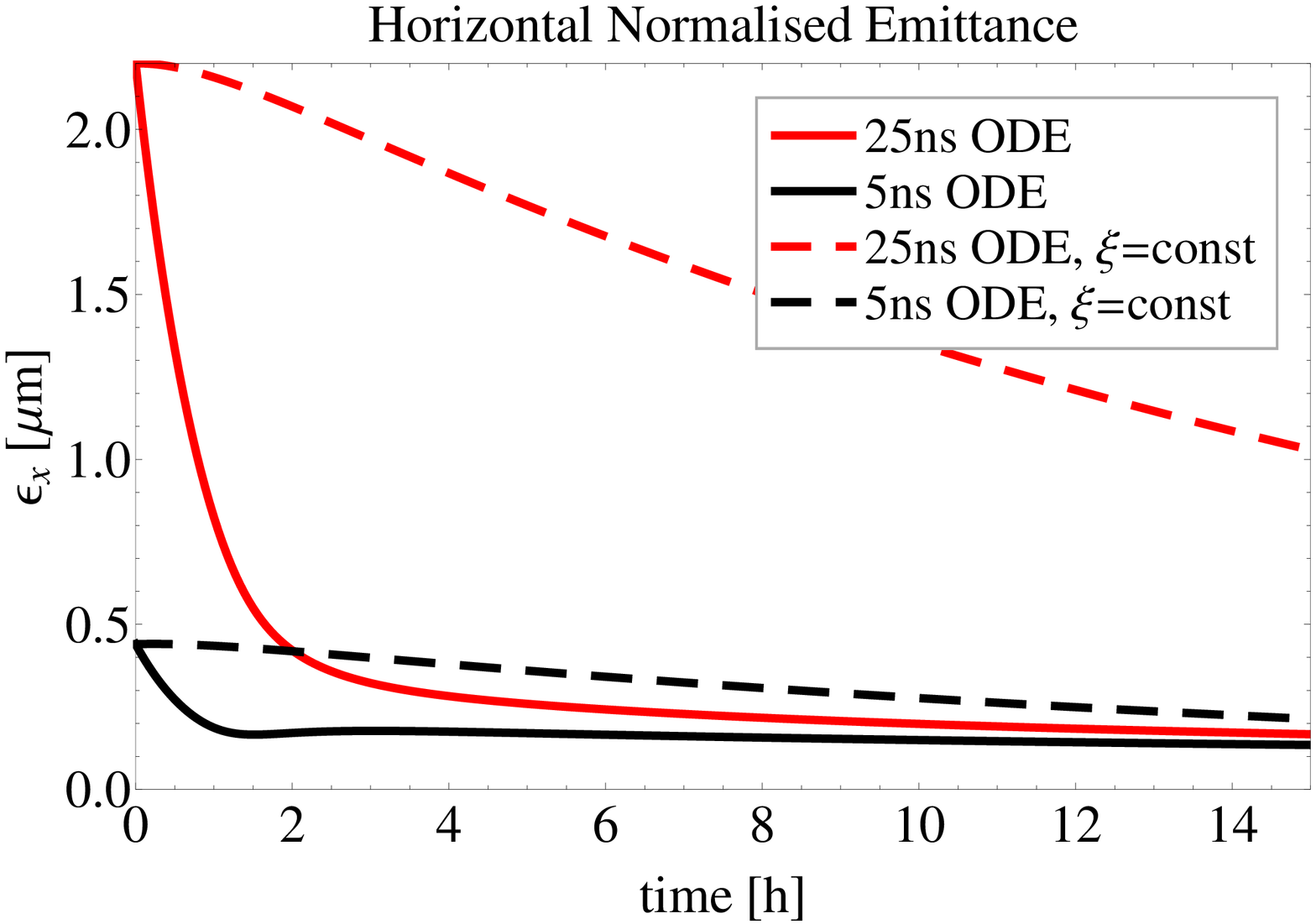}
		\end{minipage}
		\vspace*{0.5cm}
	\end{subfigure}
	
		\begin{subfigure}[h]{1\textwidth}
	\centering
		\begin{minipage}{0.45\textwidth}
		\includegraphics[width=1\textwidth]{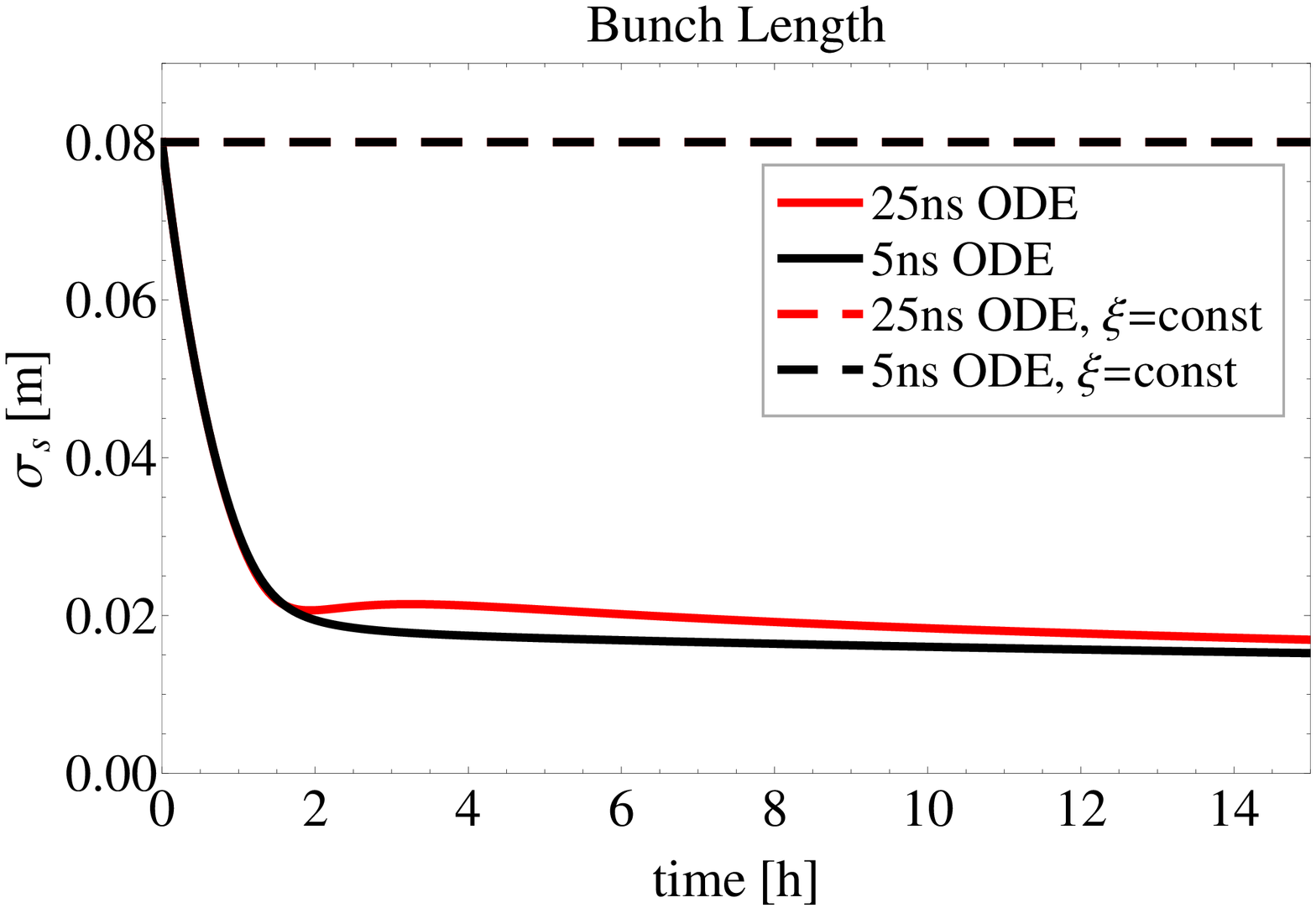}
		\end{minipage}
				\begin{minipage}{0.45\textwidth}
		\includegraphics[width=1\textwidth]{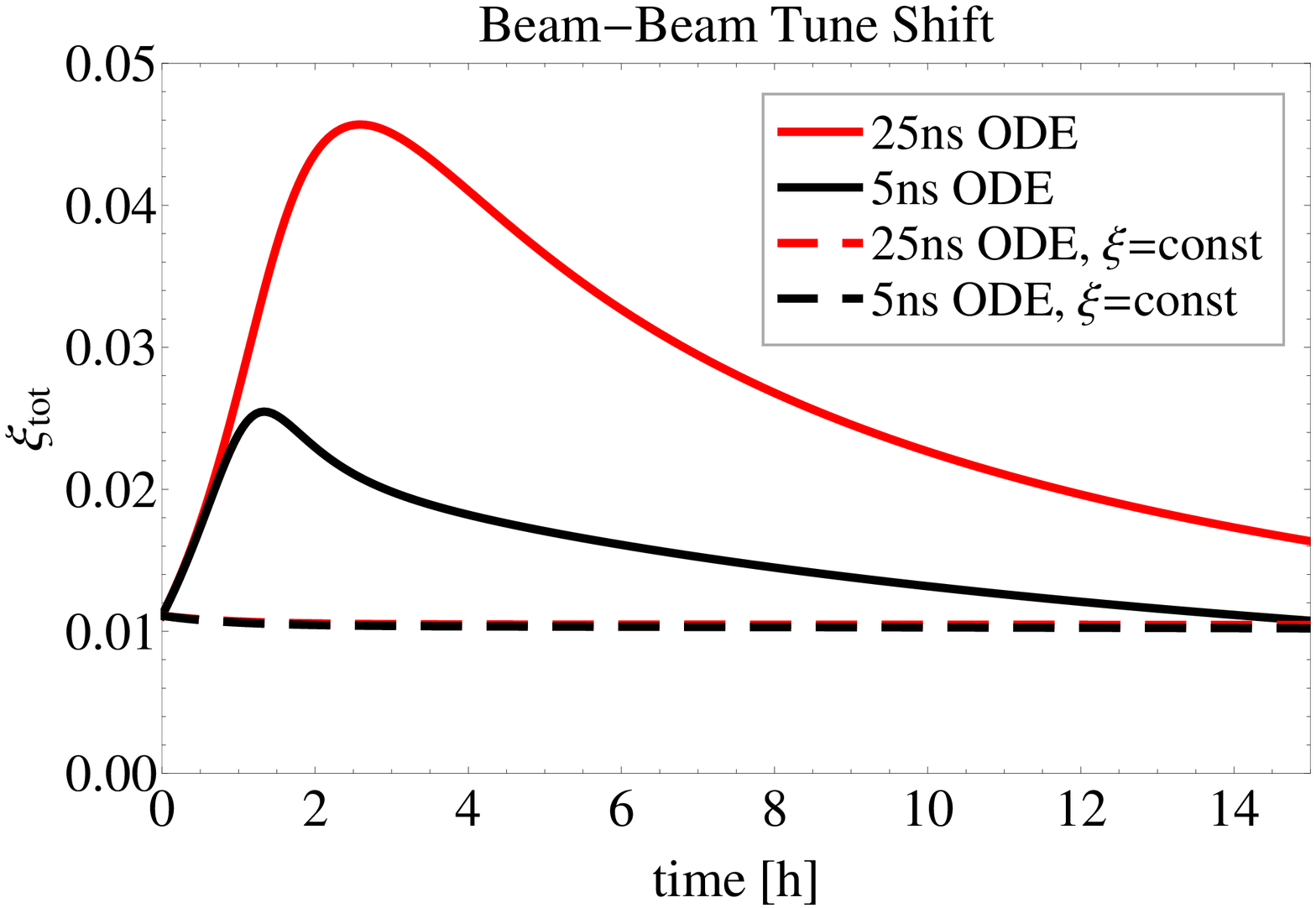}
		\end{minipage}
	\end{subfigure}
\caption[p-p beam and luminosity evolution.]{\label{f_fcc_pp_beamEvolution}  p-p beam and luminosity evolution for two experiment in collisions. Top: instantaneous (left) and integrated (right) luminosity, middle: intensity (left) and normalised emittance (right), bottom: bunch length (left) and total beam-beam tune shift (right). Solid lines show free beam evolution without artificial blow-up, dashed lines show situation with constant bunch length and transverse emittance blow-up such $\xi=\text{const.}$ Beams for 25 (red) and \qty{5}{ns} (black) bunch spacing are investigated. The instantaneous and integrated luminosity, the bunch length and tune shift evolution are very similar (overlapping lines) for both bunch spacings if $\xi=\text{const.}$}
\end{figure*}

In Fig.~\ref{f_fcc_pp_beamEvolution} the luminosity and beam evolution in p-p operation is displayed. The solid lines show the free beam evolution without any artificial blow-up, obtained by solving an ODE system of the form~\eqref{eq_intensityEvolutionDiff}-\eqref{eq_bunchlengthEvolutionDiff} for two experiments in collision. The dashed lines represent the solution of the following differential equations:
\begin{eqnarray*}
\frac{\text{d}\Nb}{\text{d}t} &=&  -\sigma_{c,\text{tot}} A \frac{\Nb^2 }{\epsilon}\\
\frac{\text{d}\epsilon}{\text{d}t}& =&  \aibsx~\epsilon -\araddx (\epsilon - \frac{\Nb}{N_{b0}} \epsilon_0) \\
\frac{\text{d}\sigs}{\text{d}t}& =&0,
\end{eqnarray*}
here a constant bunch length and a transverse emittance blow-up designed to keep the beam-beam parameter $\xi$ at its initial value is implemented. For the FCC study it is assumed that the peak luminosity is limited by a maximum beam-beam tune shift of $\xi = 0.01$, from which the initial beam parameters were derived. Leaving the beams to evolve freely leads to an increase of up to $\sim 5$ times this value, as shown in the bottom right plot of Fig.~\ref{f_fcc_pp_beamEvolution}. In this case, the bunch length shrinks to about \qty{2}{cm}, which is not acceptable for the experiments. The transverse normalised emittances balance around \qty{0.2}{\mu m}. The peak luminosity reaches
 \qty{16 \times 10^{34}}{cm^{-2}s^{-2}} for \qty{25}{ns} and to \qty{11 \times 10^{34}}{cm^{-2}s^{-2}} for the \qty{5}{ns} scenario. Since the beam-beam parameter is proportional to $\Nb/\epsilon$, the luminosity will decay exponentially, if $\xi =\text{const}$. This luminosity decay could be mitigated by \bstar-levelling. The minimum \bstar~is constrained by the aperture in the triplet, thus \bstar~could be lowered proportionally to the shrinking emittance, resulting in an about constant luminosity as long as the damping is strong enough.

\begin{figure}[tb]
		\includegraphics[width=0.45\textwidth]{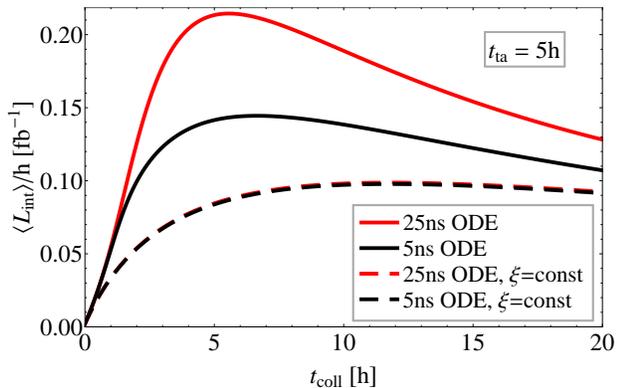}
\caption[p-p average integrated luminosity per hour.]{\label{f_fcc_pp_aveLint}  Average integrated luminosity per hour in p-p operation.}
\end{figure}

Figure~\ref{f_fcc_pp_aveLint} shows the average integrated luminosity as a function of the time in collisions, assuming a total turnaround time of $t_\text{ta}=\qty{5}{h}$ (as in \cite{FCC1}), evaluated with Eq.~\eqref{eq_fcc_averageLumi} and the results shown in the upper right plot of Fig.~\ref{f_fcc_pp_beamEvolution}. The four cases discussed in the previous paragraph are displayed.  The particle losses of proton bunches on the LHC injection plateau are small and thus neglected. The optimum time in collisions calculates to \qty{5.6}{h} and \qty{6.7}{h} for the \qty{25}{ns} and \qty{5}{ns} case of free beam evolution (solid lines), respectively. 
Under optimised conditions, \qty{5.1}{fb^{-1}} (\qty{25}{ns}) and \qty{3.4}{fb^{-1}} (\qty{5}{ns}) could be (on average) collected per day.
Considering $\xi=\text{const.}$, the two options are very similar. The integrated luminosity is maximised for \qty{11.8}{h} collision time, delivering on average \qty{2.3}{fb^{-1}/day}. 
If the beam-beam limit is higher than expected and the beams could be left to evolve freely, the luminosity outcome could potentially be doubled.

\section{Summary Table}

In Table~\ref{t_fcc_summary} calculated and assumed parameters for Pb-Pb,  p-Pb and p-p operation at \etev{50Z} in the FCC-hh are summarised. In case of p-Pb operation the Pb beam is assumed to be the same as for Pb-Pb, therefore the corresponding column only quotes the proton beam parameters. The Pb beam parameters at injection are listed as well as the LHC Pb-Pb and p-p design parameters \cite{LHCDesignReport}. The p-p luminosity parameters given are based on the case where the beam-beam tune shift is kept constant to its initial value. The "$/$" separates the results for two beam options.

\begin{table*}
 	  \caption[Summary table.]{ \label{t_fcc_summary}%
 	  Summary table.} 
 \begin{ruledtabular}
    \begin{tabular}{lc|cc|c|ccc}
    
    \multirow{2}[2]{*}{} &\multirow{2}[2]{*}{\textbf{Unit}} & \multicolumn{2}{c|}{\textbf{LHC}} & \textbf{FCC} & \multicolumn{3}{c}{ \textbf{FCC}} \\
          &        &  \multicolumn{2}{c|}{\textbf{Design} }  & \textbf{Injection} & \multicolumn{3}{c}{ \textbf{Collision}}\\
    \colrule
  	 Operation mode&-						&p-p	& Pb-Pb 				& Pb 	& Pb-Pb 	& p-Pb 		& p-p (\qty{25}{ns}/\qty{5}{ns})\\
    \colrule
 \multicolumn{8}{c}{\textbf{General storage ring parameters}} \\
    \colrule
     Circumference & [km] 					& \multicolumn{2}{c|}{26.659 }				& 100 	& \multicolumn{3}{c}{100}\\
     Field of main bends & [T] 				&\multicolumn{2}{c|}{ 8.33 }				& 1.0 	& \multicolumn{3}{c}{16}\\
     Bending radius & [m] 					&\multicolumn{2}{c|}{2803.95} 				& 10424	& \multicolumn{3}{c}{10424}\\
     Cell length & [m] 						&\multicolumn{2}{c|}{106.9}	 				& 203	& \multicolumn{3}{c}{203} \\
     Gamma transition $\gamma_T$&  			&\multicolumn{2}{c|}{55.7} 					& 103	& \multicolumn{3}{c}{103} \\
     Revolution frequency & [kHz]			&\multicolumn{2}{c|}{11.245} 				& 2.998	& \multicolumn{3}{c}{2.998} \\
     RF frequency & [MHz] 					&\multicolumn{2}{c|}{400.8} 				& 400.8	& \multicolumn{3}{c}{400.8} \\
     Harmonic number & 						&\multicolumn{2}{c|}{35640} 				& 133692& \multicolumn{3}{c}{133692} \\
     Total RF Voltage & [MV] 				&\multicolumn{2}{c|}{16} 					& 13 	& \multicolumn{3}{c}{32} \\
     Synchrotron frequency & [Hz] 			&\multicolumn{2}{c|}{23.0} 					& 8.4 	& \multicolumn{3}{c}{3.4} \\

  	\colrule
  	 \multicolumn{8}{c}{\textbf{General Beam Parameters}} \\
  	 \colrule

     Beam energy & [TeV] 					&7		& 574   				& 270  		& 4100  		& 50 		& 50  \\
     Relativistic $\gamma$-factor & 		&7461	& 2963.5  				& 1397  	& 21168 		& 53290 	& 53290 \\
     No. of bunches & -   					&2808	& 592   				& 432 		& 432 			& 432  		& 10600/53000 \\
     No. of particles per bunch &[$10^8$] 	&1150	& 0.7   				& 1.4 		& 1.4   		& 115  		& 1000/200\\
     Transv. norm. emittance &  [$\mu$m.rad]&3.75	& 1.5   				& 1.5 		& 1.5  			& 3.75 		& 2.2/0.44\\
     RMS bunch length &  [cm]  				&7.55	& 7.94  				& 10.0  	& \multicolumn{3}{c}{8.0}  \\
     RMS energy spread&  [$10^{-4}$] 		&1.129	& 1.1   				& 1.9   	& \multicolumn{3}{c}{0.6}    \\
     Long. emittance (4$\sigma$)&[eVs/charge]&2.5	& 2.5  				& 2.6   	& \multicolumn{3}{c}{10.1}     \\
     Circulating beam current& [mA]  		&584	& 6.12  				& 2.38		& 2.38			& 2.38 		&   509.14 \\
     Stored beam energy & [MJ]  			&362	& 3.8  					& 2.6   	& 39.8			& 39.8		&  8491.5  \\
    \colrule
     \multicolumn{8}{c}{\textbf{Intra Beam Scattering and Synchrotron Radiation}} \\
    \colrule
     Long. IBS emit. growth time & [h]  	&61		& 7.7   				& 6.2  		& 29.2  		& $4\times 10^3$ &  283.1/264.7\\
     Hor. IBS emit. growth time & [h]   	&80		& 13    				& 10.0  	& 30.0  		& $4\times 10^3$ & 169.5/31.7\\
     Long. emit. rad. damping time & [h]   	&13		& 6.3  					& 852   	& 0.24  		& 0.5		& 0.5 \\
     Hor. emit. rad. damping time  &[h]    	&26		& 12.6  				& 1704  	& 0.49  		& 1.0		& 1.0  \\
     Power loss per ion 	& [W] 		   &$1.8\times 10^{-11}$ 	& $2.0\times 10^{-9}$ 	& $1.1\times 10^{-11}$ & $5.7\times 10^{-7}$ 	& $3.4 \times 10^{-9}$	&$3.4 \times 10^{-9}$   \\
     Power loss per length in main bends &[W/m]&0.206	& 0.005 				& $1.0\times 10^{-5}$  & 0.53 & 0.26 	& 55.4   \\
     Energy loss per ion per turn & [MeV] & 0.007 	& 1.12  				& 0.01  	& 775.3 		& 4.7	 	& 4.7   \\
     Synch. radiation power per ring&[W] &$3.6\times 10^{3}$& 83.9  		& 0.7   	& 34389 		& 17016 	& $3.6\times10^6$  \\

  	\colrule
     \multicolumn{8}{c}{\textbf{Luminosity}} \\
  	\colrule
     $\beta$-function at the IP & [m] 		&0.55 	& 0.5        			& -			& \multicolumn{3}{c}{1.1} \\
     Initial RMS beam size at IP&  [$\mu$m] &16.7	& 15.9     				& -			& 8.8			& 8.8	 	& 6.7/3.0\\
	 Number of IPs in collision	& -			&2+2	& 1 					& -			&1				& 1			& 2 \\
	 Crossing-angle	& [$\mu$rad]			&$\pm$142.5& 0					& -			&\multicolumn{3}{c}{0}  \\
     Initial luminosity	&  [$10^{27}\text{cm}^{-2}\text{s}^{-1}$]&$10^7$& 1& -    		& 2.6  			& 213 		& $5.6\times 10^7$  \\
     Peak luminosity & [$10^{27}\text{cm}^{-2}\text{s}^{-1}$]&$10^7$& 1    & - 		& 7.3 	 		& 1192 		& $5.6\times 10^7$  \\
     Integrated luminosity per fill	& [$\mu$b$^{-1}$] &-	& $<$15     	& -		  	& 57.8  		& 21068 	&  $1.65\times 10^9$   \\
     Ave. Integrated luminosity/hour& [$\mu$b$^{-1}$]&- 	& - 			& - 		& 11.5   		& 2478  	&  $98.2\times 10^6$ \\
	 Optimum time in collision & [h]		&-		& -						& - 		& 3.0			& 6.5		& 11.8 \\
	 Assumed turnaround time& [h]			&-		& -						& - 		& 2.0			& 2.0		& 5.0 \\
     Initial bb tune shift per IP& [$10^{-4}$] &33	& 1.8				& -    		& 3.7 		 	& 3.7 		&  55.5  \\
     Total cross-section & [b]  		 	&0.1	& 515       			& -			& 597 	 		& 2  		&  0.153\\
     Peak BFPP beam power & [W]   			&0		& 26        			& - 		& 1705	 		& 0  		& 0 \\
     Luminosity lifetime ($\mathcal{L}_0/e$)& [h]&14.9	& $<$5.6 (2 exp.)	& -			& 6.2   		& 14.0 		&  17.0 \\
    
    \end{tabular}%
\end{ruledtabular}
\end{table*}%

\section{Conclusions}

The FCC will enter a new regime of hadron collider operation.  Strong radiation damping will lead to small emittances and very efficient intensity burn-off. The emittances and bunch length become so small that artificial blow-up might be necessary to avoid instabilities.
An  artificial blow-up might also be used as a way of luminosity levelling. Because of the small beam dimensions, the peak Pb-Pb luminosity can expected to be about 7 times the nominal LHC design value. 
The absolute integrated luminosity maximum per fill, when all particles are converted into luminosity, comes into reach, again because of the natural cooling from radiation damping. It is estimated that an integrated luminosity of about \qty{8}{nb^{-1}} could be expected per run of 30 days.

If the LHC is used as the last pre-accelerator, its cycle time has to be drastically improved. Otherwise, the time between two injections into the FCC will be in the same order as the expected time in collisions per fill. To optimise the  run time, the LHC could be re-filled in parallel to physics operation, maximising the time in physics and the integrated luminosity, while filling  only one fourth of the FCC.

In p-Pb operation, the fill length is determined by the burn-off of the lead beam. The longer radiation damping time and weaker IBS of the proton beam, lead to longer fills in p-Pb operation. However, by adjusting the proton beam intensity the luminosity peak and time distribution could be levelled. 

The formalisms developed for the heavy-ion operation have also been applied to p-p operation. First prediction of the p-p beam and luminosity evolution, under the assumption of constant bunch length and an emittance blow-up, designed to keep the beam-beam tune shift $\xi =\text{const.}$, have been presented. Furthermore, IBS calculations show that transverse emittance growth for long injection plateaus could become an issue for high intensity, low emittance protons.

\section{Acknowledgments}
It is a pleasure to thank John~M. Jowett for the revealing discussions and the support during the preparation of this work. 
I would also like to acknowledge Andrea Dainese, Silvia Masciocchi and Urs Wiedemann for the motivation of this study and the opportunity to present the results in their collaboration meetings, providing a discussion background on the physics potential of nuclear beams at the FCC. I am also grateful to Daniel Schulte for discussions.
This work is supported by the Wolfgang-Gentner-Programme of the Bundesministerium f\"ur Bildung und Forschung (BMBF), Germany.

\bibliography{FCC}

\begin{thebibliography}{29}%
\makeatletter
\providecommand \@ifxundefined [1]{%
 \@ifx{#1\undefined}
}%
\providecommand \@ifnum [1]{%
 \ifnum #1\expandafter \@firstoftwo
 \else \expandafter \@secondoftwo
 \fi
}%
\providecommand \@ifx [1]{%
 \ifx #1\expandafter \@firstoftwo
 \else \expandafter \@secondoftwo
 \fi
}%
\providecommand \natexlab [1]{#1}%
\providecommand \enquote  [1]{``#1''}%
\providecommand \bibnamefont  [1]{#1}%
\providecommand \bibfnamefont [1]{#1}%
\providecommand \citenamefont [1]{#1}%
\providecommand \href@noop [0]{\@secondoftwo}%
\providecommand \href [0]{\begingroup \@sanitize@url \@href}%
\providecommand \@href[1]{\@@startlink{#1}\@@href}%
\providecommand \@@href[1]{\endgroup#1\@@endlink}%
\providecommand \@sanitize@url [0]{\catcode `\\12\catcode `\$12\catcode
  `\&12\catcode `\#12\catcode `\^12\catcode `\_12\catcode `\%12\relax}%
\providecommand \@@startlink[1]{}%
\providecommand \@@endlink[0]{}%
\providecommand \url  [0]{\begingroup\@sanitize@url \@url }%
\providecommand \@url [1]{\endgroup\@href {#1}{\urlprefix }}%
\providecommand \urlprefix  [0]{URL }%
\providecommand \Eprint [0]{\href }%
\providecommand \doibase [0]{http://dx.doi.org/}%
\providecommand \selectlanguage [0]{\@gobble}%
\providecommand \bibinfo  [0]{\@secondoftwo}%
\providecommand \bibfield  [0]{\@secondoftwo}%
\providecommand \translation [1]{[#1]}%
\providecommand \BibitemOpen [0]{}%
\providecommand \bibitemStop [0]{}%
\providecommand \bibitemNoStop [0]{.\EOS\space}%
\providecommand \EOS [0]{\spacefactor3000\relax}%
\providecommand \BibitemShut  [1]{\csname bibitem#1\endcsname}%
\let\auto@bib@innerbib\@empty
\bibitem [{\citenamefont {Dominguez}\ and\ \citenamefont
  {Zimmermann}(2013)}]{VHELHC}%
  \BibitemOpen
  \bibfield  {author} {\bibinfo {author} {\bibfnamefont {O.}~\bibnamefont
  {Dominguez}}\ and\ \bibinfo {author} {\bibfnamefont {F.}~\bibnamefont
  {Zimmermann}},\ }in\ \href@noop {} {\emph {\bibinfo {booktitle} {Proceedings
  of European Coordination for Accelerator Research \& Development (EuCARD)
  Workshop 2013}}},\ \bibinfo {series and number} {\bibinfo {number}
  {EuCARD-CON-2013-010}}\ (\bibinfo {address} {Geneva, Switzerland},\ \bibinfo
  {year} {2013})\BibitemShut {NoStop}%
\bibitem [{\citenamefont {Armesto}\ \emph {et~al.}(2014)\citenamefont
  {Armesto}, \citenamefont {Dainese}, \citenamefont {d'Enterria}, \citenamefont
  {Masciocchi}, \citenamefont {Roland}, \citenamefont {Salgado}, \citenamefont
  {van Leeuwen},\ and\ \citenamefont {Wiedemann}}]{ADainese_QM2014}%
  \BibitemOpen
  \bibfield  {author} {\bibinfo {author} {\bibfnamefont {N.}~\bibnamefont
  {Armesto}}, \bibinfo {author} {\bibfnamefont {A.}~\bibnamefont {Dainese}},
  \bibinfo {author} {\bibfnamefont {D.}~\bibnamefont {d'Enterria}}, \bibinfo
  {author} {\bibfnamefont {S.}~\bibnamefont {Masciocchi}}, \bibinfo {author}
  {\bibfnamefont {C.}~\bibnamefont {Roland}}, \bibinfo {author} {\bibfnamefont
  {C.}~\bibnamefont {Salgado}}, \bibinfo {author} {\bibfnamefont
  {M.}~\bibnamefont {van Leeuwen}}, \ and\ \bibinfo {author} {\bibfnamefont
  {U.}~\bibnamefont {Wiedemann}},\ }in\ \href@noop {} {\emph {\bibinfo
  {booktitle} {Proceedings of Quark Matter 2014}}},\ \bibinfo {series and
  number} {\bibinfo {number} {arXiv:1407.7649}}\ (\bibinfo {address}
  {Darmstadt, Germany},\ \bibinfo {year} {2014})\ \bibinfo {note} {submitted to
  Nucl. Phys.}\BibitemShut {Stop}%
\bibitem [{\citenamefont {Bruce}\ \emph {et~al.}(2010)\citenamefont {Bruce},
  \citenamefont {Jowett}, \citenamefont {Blaskiewicz},\ and\ \citenamefont
  {Fischer}}]{cte}%
  \BibitemOpen
  \bibfield  {author} {\bibinfo {author} {\bibfnamefont {R.}~\bibnamefont
  {Bruce}}, \bibinfo {author} {\bibfnamefont {J.~M.}\ \bibnamefont {Jowett}},
  \bibinfo {author} {\bibfnamefont {M.}~\bibnamefont {Blaskiewicz}}, \ and\
  \bibinfo {author} {\bibfnamefont {W.}~\bibnamefont {Fischer}},\ }\href
  {\doibase 10.1103/PhysRevSTAB.13.091001} {\bibfield  {journal} {\bibinfo
  {journal} {Phys. Rev. ST Accel. Beams}\ }\textbf {\bibinfo {volume} {13}},\
  \bibinfo {pages} {091001} (\bibinfo {year} {2010})}\BibitemShut {NoStop}%
\bibitem [{\citenamefont {Ball}\ \emph {et~al.}(2014)\citenamefont {Ball} \emph
  {et~al.}}]{FCC1}%
  \BibitemOpen
  \bibfield  {author} {\bibinfo {author} {\bibfnamefont {A.}~\bibnamefont
  {Ball}} \emph {et~al.},\ }\href@noop {} {\emph {\bibinfo {title} {{Future
  Circular Collider Study, Hadron Collider Parameters}}}},\ \bibinfo {type}
  {Tech. Rep.}\ \bibinfo {number} {FCC-ACC-SPC-0001}\ (\bibinfo  {institution}
  {CERN},\ \bibinfo {address} {Geneva, Switzerland},\ \bibinfo {year}
  {2014})\BibitemShut {NoStop}%
\bibitem [{\citenamefont {Alemany~Fernandez}\ and\ \citenamefont
  {Holzer}(2014)}]{BHolzer_fcclattice}%
  \BibitemOpen
  \bibfield  {author} {\bibinfo {author} {\bibfnamefont {R.}~\bibnamefont
  {Alemany~Fernandez}}\ and\ \bibinfo {author} {\bibfnamefont {B.}~\bibnamefont
  {Holzer}},\ }\href@noop {} {\emph {\bibinfo {title} {{First Considerations on
  Beam Optics and Lattice Design for the Future Hadron-Hadron Collider
  FCC}}}},\ \bibinfo {type} {Tech. Rep.}\ \bibinfo {number}
  {CERN-ACC-NOTE-2014-0065}\ (\bibinfo  {institution} {CERN},\ \bibinfo
  {address} {Geneva, Switzerland},\ \bibinfo {year} {2014})\BibitemShut
  {NoStop}%
\bibitem [{\citenamefont {Chao}\ \emph {et~al.}(2013)\citenamefont {Chao},
  \citenamefont {Mess}, \citenamefont {Tigner},\ and\ \citenamefont
  {Zimmermann}}]{HandbookAccelPhys}%
  \BibitemOpen
  \bibinfo {editor} {\bibfnamefont {A.}~\bibnamefont {Chao}}, \bibinfo {editor}
  {\bibfnamefont {K.}~\bibnamefont {Mess}}, \bibinfo {editor} {\bibfnamefont
  {M.}~\bibnamefont {Tigner}}, \ and\ \bibinfo {editor} {\bibfnamefont
  {F.}~\bibnamefont {Zimmermann}},\ eds.,\ \href@noop {} {\emph {\bibinfo
  {title} {{Handbook of Accelerator Physics and Engineering}}}},\ \bibinfo
  {edition} {2nd}\ ed.\ (\bibinfo  {publisher} {World Scientific},\ \bibinfo
  {year} {2013})\BibitemShut {NoStop}%
\bibitem [{\citenamefont {Schaumann}\ and\ \citenamefont
  {Jowett}(2013)}]{IPAC13BbB}%
  \BibitemOpen
  \bibfield  {author} {\bibinfo {author} {\bibfnamefont {M.}~\bibnamefont
  {Schaumann}}\ and\ \bibinfo {author} {\bibfnamefont {J.~M.}\ \bibnamefont
  {Jowett}},\ }in\ \href@noop {} {\emph {\bibinfo {booktitle} {Proceedings of
  the 2013 International Particle Accelerator Conference}}},\ \bibinfo {series
  and number} {\bibinfo {number} {TUPFI025}}\ (\bibinfo {address} {Shanghai,
  China},\ \bibinfo {year} {2013})\ pp.\ \bibinfo {pages} {1391--1393},\
  \bibinfo {note} {{CERN-ACC-2013-0136}}\BibitemShut {NoStop}%
\bibitem [{\citenamefont {Jowett}\ \emph
  {et~al.}(2013{\natexlab{a}})\citenamefont {Jowett}, \citenamefont
  {Alemany-Fernandez}, \citenamefont {Baudrenghien}, \citenamefont {Jacquet},
  \citenamefont {Lamont}, \citenamefont {Manglunki}, \citenamefont {Redaelli},
  \citenamefont {Sapinski}, \citenamefont {Schaumann}, \citenamefont
  {Camillocci}, \citenamefont {Tom\'as}, \citenamefont {Uythoven},
  \citenamefont {Valuch}, \citenamefont {Versteegen},\ and\ \citenamefont
  {Wenninger}}]{JJowett_IPAC13}%
  \BibitemOpen
  \bibfield  {author} {\bibinfo {author} {\bibfnamefont {J.}~\bibnamefont
  {Jowett}}, \bibinfo {author} {\bibfnamefont {R.}~\bibnamefont
  {Alemany-Fernandez}}, \bibinfo {author} {\bibfnamefont {P.}~\bibnamefont
  {Baudrenghien}}, \bibinfo {author} {\bibfnamefont {D.}~\bibnamefont
  {Jacquet}}, \bibinfo {author} {\bibfnamefont {M.}~\bibnamefont {Lamont}},
  \bibinfo {author} {\bibfnamefont {D.}~\bibnamefont {Manglunki}}, \bibinfo
  {author} {\bibfnamefont {S.}~\bibnamefont {Redaelli}}, \bibinfo {author}
  {\bibfnamefont {M.}~\bibnamefont {Sapinski}}, \bibinfo {author}
  {\bibfnamefont {M.}~\bibnamefont {Schaumann}}, \bibinfo {author}
  {\bibfnamefont {M.~S.}\ \bibnamefont {Camillocci}}, \bibinfo {author}
  {\bibfnamefont {R.}~\bibnamefont {Tom\'as}}, \bibinfo {author} {\bibfnamefont
  {J.}~\bibnamefont {Uythoven}}, \bibinfo {author} {\bibfnamefont
  {D.}~\bibnamefont {Valuch}}, \bibinfo {author} {\bibfnamefont
  {R.}~\bibnamefont {Versteegen}}, \ and\ \bibinfo {author} {\bibfnamefont
  {J.}~\bibnamefont {Wenninger}},\ }in\ \href@noop {} {\emph {\bibinfo
  {booktitle} {Proceedings of the 2013 International Particle Accelerator
  Conference}}},\ \bibinfo {series and number} {\bibinfo {number} {MOODB201}}\
  (\bibinfo {address} {Shanghai, China},\ \bibinfo {year} {2013})\ pp.\
  \bibinfo {pages} {49--51},\ \bibinfo {note}
  {{CERN-ACC-2013-0057}}\BibitemShut {NoStop}%
\bibitem [{\citenamefont {Schaumann}(shed)}]{MSchaumann_phd}%
  \BibitemOpen
  \bibfield  {author} {\bibinfo {author} {\bibfnamefont {M.}~\bibnamefont
  {Schaumann}},\ }\emph {\bibinfo {title} {{Maximising Luminosity in Heavy-Ion
  Operation of the LHC and Future Colliders}}},\ \href@noop {} {Ph.D. thesis},\
  \bibinfo  {school} {RWTH Aachen University}, \bibinfo {address} {Aachen,
  Germany} (\bibinfo {year} {to be published})\BibitemShut {NoStop}%
\bibitem [{\citenamefont {Schaumann}(2014)}]{IPAC14Model}%
  \BibitemOpen
  \bibfield  {author} {\bibinfo {author} {\bibfnamefont {M.}~\bibnamefont
  {Schaumann}},\ }in\ \href@noop {} {\emph {\bibinfo {booktitle} {Proceedings
  of the 2014 International Particle Accelerator Conference}}},\ \bibinfo
  {series and number} {\bibinfo {number} {TUPRO014}}\ (\bibinfo {address}
  {Dresden, Germany},\ \bibinfo {year} {2014})\ pp.\ \bibinfo {pages}
  {1033--1035},\ \bibinfo {note} {{CERN-ACC-2014-0084}}\BibitemShut {NoStop}%
\bibitem [{LHC(2004)}]{LHCDesignReport}%
  \BibitemOpen
  \href@noop {} {\enquote {\bibinfo {title} {{LHC Design Report Volume I - The
  LHC main ring}},}\ }\bibinfo {howpublished} {CERN-2004-003-V-1},\ \bibinfo
  {address} {Geneva, Switzerland} (\bibinfo {year} {2004})\BibitemShut
  {NoStop}%
\bibitem [{\citenamefont {Piwinski}(1974)}]{APiwinski_IBS}%
  \BibitemOpen
  \bibfield  {author} {\bibinfo {author} {\bibfnamefont {A.}~\bibnamefont
  {Piwinski}},\ }in\ \href@noop {} {\emph {\bibinfo {booktitle} {{Proc. 9th
  Int. Conf. on High Energy Accelerators}}}}\ (\bibinfo {address} {Stanford},\
  \bibinfo {year} {1974})\ p.\ \bibinfo {pages} {405}\BibitemShut {NoStop}%
\bibitem [{\citenamefont {Bjorken}\ and\ \citenamefont
  {Mtingwa}(1983)}]{JBjorken_IBS}%
  \BibitemOpen
  \bibfield  {author} {\bibinfo {author} {\bibfnamefont {J.}~\bibnamefont
  {Bjorken}}\ and\ \bibinfo {author} {\bibfnamefont {S.}~\bibnamefont
  {Mtingwa}},\ }\href@noop {} {\bibfield  {journal} {\bibinfo  {journal} {Part.
  Accel.}\ }\textbf {\bibinfo {volume} {13}},\ \bibinfo {pages} {115} (\bibinfo
  {year} {1983})}\BibitemShut {NoStop}%
\bibitem [{\citenamefont {Bane}\ \emph {et~al.}(2002)\citenamefont {Bane},
  \citenamefont {Hayano}, \citenamefont {Kubo}, \citenamefont {Naito},
  \citenamefont {Okugi},\ and\ \citenamefont {Urakawa}}]{KBane_IBS}%
  \BibitemOpen
  \bibfield  {author} {\bibinfo {author} {\bibfnamefont {K.}~\bibnamefont
  {Bane}}, \bibinfo {author} {\bibfnamefont {H.}~\bibnamefont {Hayano}},
  \bibinfo {author} {\bibfnamefont {K.}~\bibnamefont {Kubo}}, \bibinfo {author}
  {\bibfnamefont {T.}~\bibnamefont {Naito}}, \bibinfo {author} {\bibfnamefont
  {T.}~\bibnamefont {Okugi}}, \ and\ \bibinfo {author} {\bibfnamefont
  {J.}~\bibnamefont {Urakawa}},\ }\href {\doibase 10.1103/PhysRevSTAB.5.084403}
  {\bibfield  {journal} {\bibinfo  {journal} {Phys. Rev. ST Accel. Beams}\
  }\textbf {\bibinfo {volume} {5}},\ \bibinfo {pages} {084403} (\bibinfo {year}
  {2002})}\BibitemShut {NoStop}%
\bibitem [{\citenamefont {Nagaitsev}(2005)}]{SNagaitsev_IBS}%
  \BibitemOpen
  \bibfield  {author} {\bibinfo {author} {\bibfnamefont {S.}~\bibnamefont
  {Nagaitsev}},\ }\href@noop {} {\bibfield  {journal} {\bibinfo  {journal}
  {Phys. Rev. ST Accel. Beams}\ }\textbf {\bibinfo {volume} {8}},\ \bibinfo
  {pages} {064403} (\bibinfo {year} {2005})}\BibitemShut {NoStop}%
\bibitem [{\citenamefont {Wei}(1993)}]{JWei_IBS}%
  \BibitemOpen
  \bibfield  {author} {\bibinfo {author} {\bibfnamefont {J.}~\bibnamefont
  {Wei}},\ }in\ \href@noop {} {\emph {\bibinfo {booktitle} {Proceedings of the
  1993 Particle Accelerator Conference}}}\ (\bibinfo {address} {Washington,
  DC},\ \bibinfo {year} {1993})\ pp.\ \bibinfo {pages} {3651--3653}\BibitemShut
  {NoStop}%
\bibitem [{\citenamefont {Piwinski}(1985)}]{APiwinski_IBS1}%
  \BibitemOpen
  \bibfield  {author} {\bibinfo {author} {\bibfnamefont {A.}~\bibnamefont
  {Piwinski}},\ }in\ \href@noop {} {\emph {\bibinfo {booktitle} {CAS - CERN
  Accelerator School: Accelerator Physics}}},\ \bibinfo {series and number}
  {\bibinfo {number} {CERN-1987-003-V-1.402}}\ (\bibinfo {address} {Oxford,
  UK},\ \bibinfo {year} {1985})\ pp.\ \bibinfo {pages} {402--415}\BibitemShut
  {NoStop}%
\bibitem [{\citenamefont {Meier}\ \emph {et~al.}(2001)\citenamefont {Meier},
  \citenamefont {Halabuka}, \citenamefont {Hencken}, \citenamefont
  {Trautmann},\ and\ \citenamefont {Baur}}]{HMeier_BFPP}%
  \BibitemOpen
  \bibfield  {author} {\bibinfo {author} {\bibfnamefont {H.}~\bibnamefont
  {Meier}}, \bibinfo {author} {\bibfnamefont {Z.}~\bibnamefont {Halabuka}},
  \bibinfo {author} {\bibfnamefont {K.}~\bibnamefont {Hencken}}, \bibinfo
  {author} {\bibfnamefont {D.}~\bibnamefont {Trautmann}}, \ and\ \bibinfo
  {author} {\bibfnamefont {G.}~\bibnamefont {Baur}},\ }\href@noop {} {\bibfield
   {journal} {\bibinfo  {journal} {Phys. Rev. A}\ }\textbf {\bibinfo {volume}
  {63}},\ \bibinfo {pages} {032713} (\bibinfo {year} {2001})}\BibitemShut
  {NoStop}%
\bibitem [{\citenamefont {Braun}\ \emph {et~al.}(2014)\citenamefont {Braun},
  \citenamefont {Fass\`o}, \citenamefont {Ferrari}, \citenamefont {Jowett},
  \citenamefont {Sala},\ and\ \citenamefont {Smirnov}}]{JJowett_EDM}%
  \BibitemOpen
  \bibfield  {author} {\bibinfo {author} {\bibfnamefont {H.~H.}\ \bibnamefont
  {Braun}}, \bibinfo {author} {\bibfnamefont {A.}~\bibnamefont {Fass\`o}},
  \bibinfo {author} {\bibfnamefont {A.}~\bibnamefont {Ferrari}}, \bibinfo
  {author} {\bibfnamefont {J.~M.}\ \bibnamefont {Jowett}}, \bibinfo {author}
  {\bibfnamefont {P.~R.}\ \bibnamefont {Sala}}, \ and\ \bibinfo {author}
  {\bibfnamefont {G.~I.}\ \bibnamefont {Smirnov}},\ }\href {\doibase
  10.1103/PhysRevSTAB.17.021006} {\bibfield  {journal} {\bibinfo  {journal}
  {Phys. Rev. ST Accel. Beams}\ }\textbf {\bibinfo {volume} {17}},\ \bibinfo
  {pages} {021006} (\bibinfo {year} {2014})}\BibitemShut {NoStop}%
\bibitem [{\citenamefont {Fischer}\ \emph {et~al.}(2014)\citenamefont
  {Fischer}, \citenamefont {Baltz}, \citenamefont {Blaskiewicz}, \citenamefont
  {Gassner}, \citenamefont {Drees}, \citenamefont {Luo}, \citenamefont {Minty},
  \citenamefont {Thieberger}, \citenamefont {Wilinski},\ and\ \citenamefont
  {Pshenichnov}}]{WFischer_UUcrosssection}%
  \BibitemOpen
  \bibfield  {author} {\bibinfo {author} {\bibfnamefont {W.}~\bibnamefont
  {Fischer}}, \bibinfo {author} {\bibfnamefont {A.~J.}\ \bibnamefont {Baltz}},
  \bibinfo {author} {\bibfnamefont {M.}~\bibnamefont {Blaskiewicz}}, \bibinfo
  {author} {\bibfnamefont {D.}~\bibnamefont {Gassner}}, \bibinfo {author}
  {\bibfnamefont {K.~A.}\ \bibnamefont {Drees}}, \bibinfo {author}
  {\bibfnamefont {Y.}~\bibnamefont {Luo}}, \bibinfo {author} {\bibfnamefont
  {M.}~\bibnamefont {Minty}}, \bibinfo {author} {\bibfnamefont
  {P.}~\bibnamefont {Thieberger}}, \bibinfo {author} {\bibfnamefont
  {M.}~\bibnamefont {Wilinski}}, \ and\ \bibinfo {author} {\bibfnamefont
  {I.~A.}\ \bibnamefont {Pshenichnov}},\ }\href {\doibase
  10.1103/PhysRevC.89.014906} {\bibfield  {journal} {\bibinfo  {journal} {Phys.
  Rev. C}\ }\textbf {\bibinfo {volume} {89}},\ \bibinfo {pages} {014906}
  (\bibinfo {year} {2014})}\BibitemShut {NoStop}%
\bibitem [{\citenamefont {Schaumann}(2011)}]{MSchaumann_MasterThesis}%
  \BibitemOpen
  \bibfield  {author} {\bibinfo {author} {\bibfnamefont {M.}~\bibnamefont
  {Schaumann}},\ }\emph {\bibinfo {title} {{Beam-Beam Interaction Studies at
  LHC}}},\ \href@noop {} {Master's thesis},\ \bibinfo  {school} {RWTH Aachen
  University}, \bibinfo {address} {Aachen, Germany} (\bibinfo {year} {2011}),\
  \bibinfo {note} {{CERN-THESIS-2011-138}}\BibitemShut {NoStop}%
\bibitem [{\citenamefont {Baltz}\ \emph {et~al.}(1996)\citenamefont {Baltz},
  \citenamefont {Rhoades-Brown},\ and\ \citenamefont
  {Weneser}}]{ABaltz_EMinteractions}%
  \BibitemOpen
  \bibfield  {author} {\bibinfo {author} {\bibfnamefont {A.~J.}\ \bibnamefont
  {Baltz}}, \bibinfo {author} {\bibfnamefont {M.~J.}\ \bibnamefont
  {Rhoades-Brown}}, \ and\ \bibinfo {author} {\bibfnamefont {J.}~\bibnamefont
  {Weneser}},\ }\href {\doibase 10.1103/PhysRevE.54.4233} {\bibfield  {journal}
  {\bibinfo  {journal} {Phys. Rev. E}\ }\textbf {\bibinfo {volume} {54}},\
  \bibinfo {pages} {4233} (\bibinfo {year} {1996})}\BibitemShut {NoStop}%
\bibitem [{\citenamefont {Bruce}\ \emph {et~al.}(2009)\citenamefont {Bruce},
  \citenamefont {Bocian}, \citenamefont {Gilardoni},\ and\ \citenamefont
  {Jowett}}]{RBruce_BFPP}%
  \BibitemOpen
  \bibfield  {author} {\bibinfo {author} {\bibfnamefont {R.}~\bibnamefont
  {Bruce}}, \bibinfo {author} {\bibfnamefont {D.}~\bibnamefont {Bocian}},
  \bibinfo {author} {\bibfnamefont {S.}~\bibnamefont {Gilardoni}}, \ and\
  \bibinfo {author} {\bibfnamefont {J.~M.}\ \bibnamefont {Jowett}},\ }\href
  {\doibase 10.1103/PhysRevSTAB.12.071002} {\bibfield  {journal} {\bibinfo
  {journal} {Phys. Rev. ST Accel. Beams}\ }\textbf {\bibinfo {volume} {12}},\
  \bibinfo {pages} {071002} (\bibinfo {year} {2009})}\BibitemShut {NoStop}%
\bibitem [{\citenamefont {Jowett}\ \emph
  {et~al.}(2013{\natexlab{b}})\citenamefont {Jowett}, \citenamefont
  {Schaumann},\ and\ \citenamefont {Versteegen}}]{JJowett_RLIUP}%
  \BibitemOpen
  \bibfield  {author} {\bibinfo {author} {\bibfnamefont {J.}~\bibnamefont
  {Jowett}}, \bibinfo {author} {\bibfnamefont {M.}~\bibnamefont {Schaumann}}, \
  and\ \bibinfo {author} {\bibfnamefont {R.}~\bibnamefont {Versteegen}},\ }in\
  \href
  {{https://indico.cern.ch/event/260492/session/5/contribution/28/material/paper/0.pdf}}
  {\emph {\bibinfo {booktitle} {Proc. of RLIUP: Review of LHC and
  Injector Upgrade Plans}}},\ \bibinfo {series and number} {\bibinfo {number}
  {CERN-2014-006}}\ (\bibinfo {address} {Archamps, France},\ \bibinfo {year}
  {2013})\BibitemShut {NoStop}%
\bibitem [{\citenamefont {Zlobin}\ \emph {et~al.}(2014)\citenamefont {Zlobin}
  \emph {et~al.}}]{AZlobin_11Tstatus}%
  \BibitemOpen
  \bibfield  {author} {\bibinfo {author} {\bibfnamefont {A.}~\bibnamefont
  {Zlobin}} \emph {et~al.},\ }in\ \href@noop {} {\emph {\bibinfo {booktitle}
  {Proceedings of the 2014 International Particle Accelerator Conference}}},\
  \bibinfo {series and number} {\bibinfo {number} {WEPRI097}}\ (\bibinfo
  {address} {Dresden, Germany},\ \bibinfo {year} {2014})\ pp.\ \bibinfo {pages}
  {2722--2724}\BibitemShut {NoStop}%
\bibitem [{\citenamefont {Brinkmann}\ and\ \citenamefont
  {Willeke}(1993)}]{RBrinkmann_PAC93}%
  \BibitemOpen
  \bibfield  {author} {\bibinfo {author} {\bibfnamefont {R.}~\bibnamefont
  {Brinkmann}}\ and\ \bibinfo {author} {\bibfnamefont {F.}~\bibnamefont
  {Willeke}}\ }(\bibinfo {address} {Washington, DC},\ \bibinfo {year} {1993})\
  p.\ \bibinfo {pages} {3742–3744}\BibitemShut {NoStop}%
\bibitem [{\citenamefont {Cornelis}\ \emph {et~al.}(1990)\citenamefont
  {Cornelis}, \citenamefont {Meddahi},\ and\ \citenamefont
  {Schmidt}}]{KCornelis_SPSBB}%
  \BibitemOpen
  \bibfield  {author} {\bibinfo {author} {\bibfnamefont {K.}~\bibnamefont
  {Cornelis}}, \bibinfo {author} {\bibfnamefont {M.}~\bibnamefont {Meddahi}}, \
  and\ \bibinfo {author} {\bibfnamefont {R.}~\bibnamefont {Schmidt}},\
  }\href@noop {} {\emph {\bibinfo {title} {{The beam-beam effect in the SPS
  proton anti-proton collider for beams with unequal emittances}}}},\ \bibinfo
  {type} {Tech. Rep.}\ \bibinfo {number} {{CERN SL/90-73 (AP)}}\ (\bibinfo
  {institution} {{CERN}},\ \bibinfo {address} {Geneva, Switzerland},\ \bibinfo
  {year} {1990})\BibitemShut {NoStop}%
\bibitem [{\citenamefont {Syphers}(2008)}]{MSyphers_BB}%
  \BibitemOpen
  \bibfield  {author} {\bibinfo {author} {\bibfnamefont {M.}~\bibnamefont
  {Syphers}},\ }\href@noop {} {\emph {\bibinfo {title} {{Beam-beam Tune
  Distribution with Differing Beam Sizes}}}},\ \bibinfo {type} {Tech. Rep.}\
  \bibinfo {number} {{Fermilab-AD, Beams-doc-3031-v1}}\ (\bibinfo
  {institution} {{Fermilab}},\ \bibinfo {address} {Geneva, Switzerland},\
  \bibinfo {year} {2008})\BibitemShut {NoStop}%
\bibitem [{\citenamefont {Lebedev}\ and\ \citenamefont
  {Shiltsev}(2014)}]{AccelPhysTevatron}%
  \BibitemOpen
  \bibinfo {editor} {\bibfnamefont {V.}~\bibnamefont {Lebedev}}\ and\ \bibinfo
  {editor} {\bibfnamefont {V.}~\bibnamefont {Shiltsev}},\ eds.,\ \href@noop {}
  {\emph {\bibinfo {title} {{Accelerator physics at the Tevatron collider}}}}\
  (\bibinfo  {publisher} {Springer},\ \bibinfo {year} {2014})\BibitemShut
  {NoStop}%
\end{thebibliography}%

\end{document}